\documentclass[3p]{elsarticle}
\usepackage{amsmath}
\usepackage{amssymb}
\usepackage{amsthm}
\usepackage{bm}
\usepackage{subfigure}
\usepackage{color}
\usepackage{enumerate}
\usepackage{geometry}
\usepackage{hyperref}
\usepackage[capitalize,nameinlink]{cleveref}

\renewcommand{\Re}{\operatorname{Re}}
\renewcommand{\Im}{\operatorname{Im}}
\renewcommand{\i}{\mathrm{i}}
\DeclareMathOperator{\sech}{sech}
\newtheorem{theorem}{Theorem}[section]

\newtheorem{lemma}[theorem]{Lemma}

\begin{document}

\title{Soliton solutions to the coupled Sasa-Satsuma-mKdV equation}
\author[1]{Changyan Shi}
\author[1]{Bao-Feng Feng\corref{cor1}}
\ead{baofeng.feng@utrgv.edu}
\cortext[cor1]{Corresponding author}
\affiliation[1]{organization={School of Mathematical and Statistical Sciences, The University of Texas Rio Grande Valley},
postcode={78539},
city={Edinburg, TX},
country={USA}}

\begin{abstract}
    We consider the soliton solutions of a recently proposed coupled Sasa-Satsuma-mKdV equation using the Kadomtsev-Petviashvili reduction method. The system consists of a complex-valued component coupled with a real-valued one. Under zero or nonzero boundary conditions, we derive four distinct classes of soliton solutions: bright-bright, dark-dark, bright-dark, and dark-bright. These solutions are derived from the vector Hirota equation, for which the bright, dark, and bright-dark soliton solutions are provided in the Appendix. We perform asymptotic analysis of soliton collisions for each class of solutions, in which inelastic collisions are observed between bright-bright solitons. In the dark-dark case, we identify soliton profiles similar to the Sasa-Satsuma equation, including double-hole, Mexican hat, and anti-Mexican hat solutions; this study further explores the collisions between these structures and hyperbolic tangent shaped kink solitons. Regarding the bright-dark case, beyond the expected soliton-kink interactions, we report and analyze a notable collision occurring between kink solitons. 
\end{abstract}


\maketitle

\section{Introduction}
The study of the generalized nonlinear Schr\"odinger equation by Kodama and Hasegawa \cite{kodama1987nonlinear}
\begin{equation}\label{gnls}
    \i q_\xi + \alpha_1 q_{\tau \tau} + \alpha_2 |q|^2 q + \i \left(\beta_1 q_{\tau \tau \tau} + \beta_2 |q|^2 q_\tau + \beta_3 q \left(|q|^2\right)_\tau\right) = 0,
\end{equation}
represents an important step in the research of higher-order extensions to the nonlinear Schr\"odinger (NLS) equation. Here the terms associated with \(\alpha_1\), \(\alpha_2\) are from the original NLS equation, which standing for group velocity dispersion and self-phase modulation \cite{agrawal2000nonlinear}. While \(\beta_1\), \(\beta_2\), \(\beta_3\) describe the third-order dispersion, self-frequency shift, and self-steepening, respectively \cite{trippenbach1998effects}. In particular, in the case \((\alpha_1,\alpha_2)=(\frac 12, 1)\), \((\beta_1, \beta_2, \beta_3) = (1,6,3)\), together with the transformation
\[
u(x,t) = q(\tau,\xi) \exp\left\{-\frac{\i}{6}\left(\tau - \frac{\xi}{18}\right)\right\}
\]
where \(x = \tau - \frac{\xi}{12}, t = -\xi\), we have the following completely integrable system \cite{sasa1991new}
\begin{align}\label{ss}
    u_t = u_{xxx} + 6|u|^2 u_x + 3 u \left(|u|^2\right)_x.
\end{align}
Eq.\eqref{ss} is known as the Sasa-Satsuma (SS) equation. As an integrable higher-order extension to the physically significant NLS equation \cite{hasegawa1973transmission,fibich2015nonlinear,dalfovo1999theory,pitaevskii2003bose,zakharov1972collapse,kato2005nonlinear,benney1967propagation}, \cref{ss} was comprehensively studied through different approaches, including inverse scattering transform \cite{sasa1991new,mihalache1993inverse}, Hirota's bilinear method  and Kadomtsev-Petviashvili (KP) reduction \cite{jiang2013bright,gilson2003sasa,ohta2010dark,shi2025general}, Darboux transformation \cite{chen2013twisted,xu2014soliton}, Riemann-Hilbert approach \cite{xu2018initial,wen2023sasa}. Similar to the NLS equation, the SS equation admits bright \cite{gilson2003sasa,jiang2013bright,shi2025general} and dark solitons \cite{ohta2010dark,xu2014soliton,shi2025general}, as well as breather \cite{xu2014soliton,wu2022multi} and rogue wave solutions \cite{bandelow2012sasa,bandelow2012persistence,chen2013twisted,akhmediev2015rogue,feng2022higher,wu2024general}. Beyond these, the SS equation features additional solutions than the NLS model, such as double-hump bright solitons \cite{jiang2013bright,shi2025general}, double-hole dark solitons \cite{ohta2010dark,shi2025general}, (anti-)Mexican hat solitons \cite{xu2015anti,shi2025general}, and twisted rogue pairs \cite{chen2013twisted,wu2022multi}.

Multi-component generalizations to the NLS equation, such as the Manakov system \cite{manakov1974theory}, are required for studying different light polarizations \cite{gelash2023vector} in optical pulse propagation within birefringent fibers \cite{agrawal2000nonlinear,maimistov2013nonlinear}. Similarly, the integrable multi-component SS equation has also been extensively investigated, leading to the proposal and analysis of various multi-component extensions \cite{gilson2003sasa,sakovich2000symmetrically,wang2020riemann}.
The coupled Sasa-Satsuma (cSS) equation
\begin{subequations}
    \begin{align}
    &u_{1t}=u_{1xxx} - 6\left(c_1|u_1|^2+c_2|u_2|^2\right)u_{1x} - 3u_1\left(c_1|u_1|^2+c_2|u_2|^2\right)_x,\label{case3a}\\
    &u_{2t}=u_{2xxx} - 6\left(c_1|u_1|^2+c_2|u_2|^2\right)u_{2x} - 3u_2\left(c_1|u_1|^2+c_2|u_2|^2\right)_x,\label{case3b}
    \end{align}
\end{subequations}
is one of the such multi-component extension to the SS equation firstly studied by Porsezian et al.\ in Ref.~\cite{porsezian1994coupled}. 
Studies find \eqref{case3a}-\eqref{case3b} possesses various exact solutions including bright-bright \cite{lu2014bright,xu2013single,liu2018vector,liu2023riemann} and bright-dark soliton solutions \cite{liu2018dark,shi2026soliton}, as well as dark-dark soliton \cite{zhang2025dark}, breather \cite{zhang2025dark} and rogue wave solutions \cite{zhao2014localized,zhang2025rogue}. 

Another coupled extension to \cref{ss}, also known as the coupled Hirota (cHirota) equation \cite{tasgal1992soliton, porsezian1994coupled, gilson2003sasa} is of the form
\begin{subequations}
    \begin{align}
    &u_{1t}=u_{1xxx}-3\left(c_1 |u_1|^2 + c_2 |u_2|^2\right)u_{1x}-3 u_1(c_1 u_1^* u_{1x}+c_2 u_2^* u_{2x}),\label{case4a}\\
    &u_{2t}=u_{2xxx}-3\left(c_1 |u_1|^2 + c_2 |u_2|^2\right)u_{2x}-3 u_2(c_1 u_1^* u_{1x}+c_2 u_2^* u_{2x}),\label{case4b}
    \end{align}
\end{subequations}
where \(*\) denotes the complex conjugate. If \(u_2 = u_1^*, c_2 = c_1 = -1\), above system reduces to the SS equation \eqref{ss}. Prior researches have derived the bright-bright soliton \cite{kang2019construction, xie2020elastic,wang2021analytical,shi2025general}, multiple higher-order poles \cite{zhao2025nonlinear}, dark-dark soliton \cite{bindu2001dark, jiang2023asymptotic,shi2025general}, breather \cite{xu2018breathers,chai2019localized, pan2024super} and rogue wave \cite{chen2013rogue, chen2014dark, huang2016rational, chan2017rogue} solutions to \eqref{case4a}-\eqref{case4b}. Moreover, in the mixed boundary conditions, bright-dark soliton \cite{wang2014generalized, liu2016mixed,shi2025general}, bright-dark rogue wave \cite{wang2014rogue} solutions to the cHirota equation have been derived.

In addition to above cSS and cHirota equations, this study focuses on another two-component generalization of the SS equation, which was recently introduced by Wang et al.\ in Ref.~\cite{wang2020riemann}:
\begin{subequations}
    \begin{align}
        &u_{t}=u_{xxx} - 6 \varepsilon_1 |u|^2 u_{x} - 3 \varepsilon_1 u \left(|u|^2\right)_x - 3 \varepsilon_2 v \left(u v\right)_x, \label{ss_mkdv_1}\\
        &v_{t}=v_{xxx} - 6 \varepsilon_1 |u|^2 v_{x} - 3 \varepsilon_1 v \left(|u|^2\right)_x - 6 \varepsilon_2 v^2v_{x},  \label{ss_mkdv_2}
    \end{align}
\end{subequations}
where \(u\) is a complex-valued function and \(v\) is real-valued. It reduces to the SS equation \eqref{ss} for \(v = 0\) and to the modified Korteweg-de Vries (mKdV) equation for \(u = 0\). This equation is named the coupled Sasa-Satsuma-mKdV (SS-mKdV) equation. Since the introduction of the SS-mKdV equation \eqref{ss_mkdv_1}-\eqref{ss_mkdv_2}, various aspects of this model have been extensively investigated. For instance, bright-bright soliton and oscillated soliton solutions were constructed via the Riemann-Hilbert approach \cite{wang2020riemann} and the Darboux transformation \cite{geng2021darboux}, alongside rogue wave \cite{geng2021darboux} and multiple pole solutions \cite{zhao2025dynamics}. Beyond exact solutions, researchers have also explored the initial-boundary value problems \cite{hu2022initial} and the long-time asymptotic behavior \cite{zhao2024two}. However, existing studies on the solutions to the SS-mKdV equation have primarily focused on zero boundary condition, resulting in bright-bright and multiple-pole solitons; 
Soliton solutions under nonzero and mixed boundary conditions remain unexplored. In this paper, we aim to derive soliton solutions using the KP reduction method, which differs from the approaches used in previous studies. 
In particular, our approach yields solutions to \eqref{ss_mkdv_1}-\eqref{ss_mkdv_2} under various boundary conditions, such that components \(u,v\) satisfy one of the following
\begin{enumerate}
    \item Zero boundary condition: the functions \(u,v\) vanish as \(x \to \pm \infty\). Soliton solutions in this category are referred to as bright solitons.
    \item Nonzero boundary condition: the magnitudes \(|u|,|v|\) approach positive constants \(\rho_1, \rho_2\) as \(x \to \pm \infty\). These are classified as dark solitons.
\end{enumerate}
Since \(u\) is complex-valued and \(v\) is real-valued, four distinct combinations of boundary conditions can be investigated, leading to bright-bright, dark-dark, bright-dark, and dark-bright soliton solutions. Furthermore, considering the special dynamical behavior in SS equation such as double-hump bright soliton, double-hole dark soliton, and (anti-)Mexican-hat dark soliton solutions, we are motivated to explore whether similar phenomena occur in the SS-mKdV equation.

It should be noted that above coupled extensions of SS equation \eqref{case3a}-\eqref{case3b}, \eqref{case4a}-\eqref{case4b}, and \eqref{ss_mkdv_1}-\eqref{ss_mkdv_2}, are all special cases of the following vector Hirota equation 
\begin{align}\label{m-Hirota}
    u_{k,t}=u_{k,xxx} - 3 \left(\sum_{l=1}^M \varepsilon_l |u_l|^2\right) u_{k,x} - 3 u_k \sum_{l=1}^M \varepsilon_l u_l^*u_{l,x}.
\end{align}
This vector equation, first studied in Refs.~\cite{kim1998conservation,nakkeeran2000exact}, is completely integrable with a \((N+1)\times (N+1)\) Lax pair. 
Its bright soliton \cite{kang2019construction,shi2024study}, dark soliton \cite{zhang2017general,shi2024study}, rational rogue waves \cite{weng2021rational} and multiple poles \cite{wei2022vector} solutions were derived. 
To derive the SS-mKdV equation, one sets \(M = 3\) in \eqref{m-Hirota} and employs the complex conjugate reduction
\begin{align}
    u = u_1 = u_3^*, \quad v = u_2 = u_2^*, \quad c_1 = \varepsilon_1 = \varepsilon_3, \quad c_2 = \varepsilon_2.
\end{align}
Similarly, \eqref{case3a}-\eqref{case3b} is obtained from \eqref{m-Hirota} with \(M=4\) and the reduction \(u_1 = u_3^*, u_2 = u_4^*, c_1 = c_3, c_2 = c_4\) (see Ref.~\cite{zhang2025dark}), while \eqref{case4a}-\eqref{case4b} corresponds to the case \(M=2\) (see Ref.~\cite{shi2025general}). 
Although the general \(N\)-bright and \(N\)-dark soliton solutions to \cref{m-Hirota} were constructed in ~\cite{shi2024study}, \(N\)-bright-dark soliton solution under mixed boundary conditions remains open. Our goal in this paper is to derive and study the general soliton solutions to the SS-mKdV equation \eqref{ss_mkdv_1}-\eqref{ss_mkdv_2} by solving the vector Hirota equation \eqref{m-Hirota} via the method of KP reduction.

The present paper is organized as follows. In \cref{section:BL_ss_mkdv}, we derive the bilinear forms of the SS-mKdV equation under four different kinds of boundary conditions. In \cref{section:solution_ss_mkdv}, we present the general \(N\) bright-bright, dark-dark, bright-dark, and dark-bright soliton solutions. The dynamical behaviors of the aforementioned soliton solutions are presented in \cref{section:dynmics_bb}-\cref{section:dynmics_db}. Finally, we provide bright, dark, and bright-dark soliton solutions to the vector Hirota equation in \ref{section:append_a}, with the corresponding bilinear equations from the KP-Toda hierarchy presented in \ref{section:append_b}.

\section{Bilinearization of the coupled Sasa-Satsuma-mKdV equation}\label{section:BL_ss_mkdv}
This section we list the result of bilinear forms to \eqref{ss_mkdv_1}-\eqref{ss_mkdv_2} under different boundary conditions.
\begin{enumerate}
    \item Zero boundary condition, i.e., \(u,v \to 0\) as \(x\to \pm \infty\). In this case, the transformation
    \begin{equation}
        u = \frac{g_1}{f}, \quad v = \frac{g_2}{f}. \label{bb_transfer}
    \end{equation}
    converts Eq. \eqref{ss_mkdv_1}-\eqref{ss_mkdv_2} into    \begin{align*}
        &f^2\left(D_x^3 - D_t\right) g_1 \cdot f - 3 D_x g_1 \cdot f \left(D_x^2 f\cdot f + 4 c_1 |g_1|^2  + 2c_2 g_2^2\right) + 3c_2 g_2 f D_x g_1 \cdot g_2 \\
        &\quad + 3c_1 g_1 f D_x g_1 \cdot g_1^* = 0,\\
        &f^2\left(D_x^3 - D_t\right) g_2 \cdot f - 3 D_x g_2 \cdot f \left(D_x^2 f\cdot f + 4 c_1 |g_1|^2  + 2c_2 g_2^2\right) - 3c_2 g_1^* f D_x g_1 \cdot g_2 \\
        &\quad + 3c_1 g_1 f D_x g_2 \cdot g_1^* = 0.
    \end{align*}
    where the identity \(a D_x b\cdot c - b D_x a \cdot c = c D_x b \cdot a\) is utilized. Introducing auxiliary functions \(s_{12}, s_{13}\) and \(s_{23}\) by
    \[
    D_x g_1 \cdot g_2 = s_{12} f, \quad D_x g_1 \cdot g_1^* = s_{13} f, \quad D_x g_2 \cdot g_1^* = s_{23} f,
    \]
    and set
    \[
        D_x^2 f\cdot f + 4 c_1 |g_1|^2  + 2c_2 g_2^2 = 0,
    \]
    we have the following lemma.
    \begin{lemma}
    Under transformation \eqref{bb_transfer}, equation \eqref{ss_mkdv_1}-\eqref{ss_mkdv_2} is bilinearized into
        \begin{align}
        &(D_x^3 - D_t)g_1 \cdot f= - 3 c_2 g_2 s_{12} - 3 c_1 g_1 s_{13},\label{ssmkdv_bl_bb_1}\\
        &(D_x^3 - D_t)g_2 \cdot f= 3 c_2 g_1^* s_{12} - 3 c_1 g_1 s_{23},\label{ssmkdv_bl_bb_2}\\
        &D_x^2 f\cdot f + 4 c_1 |g_1|^2  + 2c_2 g_2^2 = 0,\label{ssmkdv_bl_bb_3}\\
        &D_x g_1 \cdot g_2 = s_{12} f, \label{ssmkdv_bl_bb_4}\\
        &D_x g_1 \cdot g_1^* = s_{13} f, \label{ssmkdv_bl_bb_5}\\
        &D_x g_2 \cdot g_1^* = s_{23} f. \label{ssmkdv_bl_bb_6}
        \end{align}
    \end{lemma}

    \item Nonzero boundary condition, i.e., \(|u| \to \rho_1, |v| \to \rho_2\) as \(x\to \pm \infty\), where \(\rho_1,\rho_2 > 0\). In this case, the transformation
    \begin{equation}
        u = \rho_1\frac{h_1}{f}e^{\i \left(\alpha x - \left(\alpha^3 + 3\alpha\left(2 c_1\rho_1^2 + c_2\rho_2^2\right) \right) t\right)}, \quad v = \rho_2\frac{h_2}{f}. \label{dd_transfer}
    \end{equation}
converts Eq. Eq. \eqref{ss_mkdv_1}-\eqref{ss_mkdv_2} into    \begin{align*}
       &f^2 \left[D_x^3 - D_t + 3\i \alpha D_x^2 - 3\left(\alpha^2 + 4c_1 \rho_1^2 + 2c_2 \rho_2^2\right)D_x - 6 \i c_1 \alpha \rho_1^2 - 3\i c_2\alpha \rho_2^2\right]h_1 \cdot f \\
       &\quad -3(D_x h_1 \cdot f + \i \alpha h_1 f) \left[\left(D_x^2 - 4c_1 \rho_1^2 - 2c_2\rho_2^2\right)f\cdot f + 4c_1 \rho_1^2 |h_1|^2 + 2c_2 \rho_2^2 h_2^2\right] \\
       &\quad +3c_2 \rho_2^2 h_2 f\left(D_x h_1 \cdot h_2 + \i \alpha h_1 h_2 \right) + 3c_1 \rho_1^2 h_1 f\left(D_x h_1 \cdot h_1^* +2 \i \alpha |h_1|^2\right) = 0,\\
       &f^2 \left[D_x^3 - D_t - 3\left(4c_1 \rho_1^2 + 2c_2 \rho_2^2\right)D_x \right]h_2 \cdot f \\
       &\quad -3(D_x h_2 \cdot f) \left[\left(D_x^2 - 4c_1 \rho_1^2 - 2c_2\rho_2^2\right)f\cdot f + 4c_1 \rho_1^2 |h_1|^2 + 2c_2 \rho_2^2 h_2^2\right] \\
       &\quad -3c_1 \rho_1^2 h_1^* f\left(D_x h_1 \cdot h_2 + \i \alpha h_1 h_2\right) + 3c_1 \rho_1^2 h_1 f\left(D_x h_2 \cdot h_1^* + \i \alpha h_2 h_1^*\right) = 0.
    \end{align*}
    Introducing auxiliary functions \(r_{12}, r_{13}, r_{23}\) by
    \[
        D_x h_1 \cdot h_2 + \i \alpha h_1 h_2 = \i \alpha r_{12} f, \quad D_x h_1 \cdot h_1^* +2 \i \alpha |h_1|^2 = 2\i \alpha r_{13} f, \quad D_x h_2 \cdot h_1^* + \i \alpha h_2 h_1^* = \i \alpha r_{23} f,
    \]
    and set
    \[
        \left(D_x^2 - 4c_1 \rho_1^2 - 2c_2\rho_2^2\right)f\cdot f + 4c_1 \rho_1^2 |h_1|^2 + 2c_2 \rho_2^2 h_2^2 = 0,
    \]
    we have the following lemma.
    \begin{lemma}
    Under transformation \eqref{dd_transfer}, equation \eqref{ss_mkdv_1}-\eqref{ss_mkdv_2} is bilinearized into
        \begin{align}
        &\left[D_x^3 - D_t + 3\i \alpha D_x^2 - 3\left(\alpha^2 + 4c_1 \rho_1^2 + 2c_2 \rho_2^2\right)D_x - 6 \i c_1 \alpha \rho_1^2 - 3\i c_2\alpha \rho_2^2\right]h_1 \cdot f \nonumber \\
        &\quad = -3\i \alpha c_2 \rho_2^2 h_2 r_{12} - 6\i \alpha \rho_1^2 c_1 h_1 r_{13},\label{ssmkdv_bl_dd_1}\\
        &\left[D_x^3 - D_t - 3\left(4c_1 \rho_1^2 + 2c_2 \rho_2^2\right)D_x \right]h_2 \cdot f = 3 \i \alpha c_1 \rho_1^2 h_1^* r_{12} - 3\i \alpha c_1 \rho_1^2 h_1 r_{23}, \label{ssmkdv_bl_dd_2}\\
        &\left(D_x^2 - 4c_1 \rho_1^2 - 2c_2\rho_2^2\right)f\cdot f + 4c_1 \rho_1^2 |h_1|^2 + 2c_2 \rho_2^2 h_2^2 = 0,\label{ssmkdv_bl_dd_3}\\
        &D_x h_1 \cdot h_2 + \i \alpha h_1 h_2 = \i \alpha r_{12} f, \label{ssmkdv_bl_dd_4}\\
        &D_x h_1 \cdot h_1^* +2 \i \alpha |h_1|^2 = 2\i \alpha r_{13} f, \label{ssmkdv_bl_dd_5}\\
        &D_x h_2 \cdot h_1^* + \i \alpha h_2 h_1^* = \i \alpha r_{23} f. \label{ssmkdv_bl_dd_6}
        \end{align}
    \end{lemma}

    \item 
    Mixed boundary condition (i): \(u \to 0, |v| \to \rho_2\) as \(x\to \pm \infty\), where \(\rho_2 > 0\). In this case, the bilinearization process is similar to the above cases. Thus we have the transformation
    \begin{lemma}
    Under the transformation 
     \begin{align}
        u = \frac{g_1}{f},\quad v = \rho_2\frac{h_2}{f}. \label{bd_transfer}
    \end{align}
    Eq. \eqref{ss_mkdv_1}-\eqref{ss_mkdv_2} is bilinearized into
        \begin{align}
        &(D_x^3 - D_t - 6c_2 \rho_2^2)g_1 \cdot f= - 3 c_1 g_1 s_{13},\label{ssmkdv_bl_bd_1}\\
        &(D_x^3 - D_t - 6c_2 \rho_2^2)h_2 \cdot f= 0,\label{ssmkdv_bl_bd_2}\\
        &\left(D_x^2 - 2c_2 \rho_2^2\right) f\cdot f + 4 c_1 |g_1|^2  + 2c_2\rho_2^2 g_2^2 = 0,\label{ssmkdv_bl_bd_3}\\
        &D_x g_1 \cdot h_2 = 0, \label{ssmkdv_bl_bd_4}\\
        &D_x g_1 \cdot g_1^* = s_{13} f, \label{ssmkdv_bl_bd_5}\\
        &D_x h_2 \cdot g_1^* = 0. \label{ssmkdv_bl_bd_6}
        \end{align}
    \end{lemma}

    \item 
 Mixed boundary condition (ii): \(|u| \to \rho_1, v \to 0\) as \(x\to \pm \infty\), where \(\rho_1 > 0\). 
In this case, we have the following lemma
    \begin{lemma}
    Under the transformation 
        \begin{align}
        u = \rho_1\frac{h_1}{f}e^{\i \left(\alpha x - \left(\alpha^3 + 6c_1\alpha\rho_1^2\right) t\right)},\quad v = \frac{g_2}{f}. \label{db_transfer}
    \end{align}
    Eq. \eqref{ss_mkdv_1}-\eqref{ss_mkdv_2} is bilinearized into
        \begin{align}
        &\left[D_x^3 - D_t + 3\i \alpha D_x^2 - 3\left(\alpha^2 + 4c_1 \rho_1^2\right)D_x - 6 \i c_1 \alpha \rho_1^2\right]h_1 \cdot f \nonumber \\
        &\quad = -3\i \alpha c_2 g_2 r_{12} - 6\i \alpha \rho_1^2 c_1 h_1 r_{13},\label{ssmkdv_bl_db_1}\\
        &\left[D_x^3 - D_t - 12c_1 \rho_1^2 D_x \right]g_2 \cdot f = 3 \i \alpha c_1 \rho_1^2 h_1^* r_{12} - 3\i \alpha c_1 \rho_1^2 h_1 r_{23}, \label{ssmkdv_bl_db_2}\\
        &\left(D_x^2 - 4c_1 \rho_1^2\right)f\cdot f + 4c_1 \rho_1^2 |h_1|^2 + 2c_2 g_2^2 = 0,\label{ssmkdv_bl_db_3}\\
        &D_x h_1 \cdot g_2 + \i \alpha h_1 g_2 = \i \alpha r_{12} f, \label{ssmkdv_bl_db_4}\\
        &D_x h_1 \cdot h_1^* +2 \i \alpha |h_1|^2 = 2\i \alpha r_{13} f, \label{ssmkdv_bl_db_5}\\
        &D_x g_2 \cdot h_1^* + \i \alpha g_2 h_1^* = \i \alpha r_{23} f. \label{ssmkdv_bl_db_6}
        \end{align}
    \end{lemma}
\end{enumerate}

\section{Soliton solutions to the coupled Sasa-Satsuma-mKdV equation}\label{section:solution_ss_mkdv}
Theorems in this section are derived from the soliton solution to the three-component Hirota equation in \ref{section:append_a}. The detailed reduction process is similar to our previous researches \cite{zhang2025dark,shi2025general,shi2026soliton,shi2024study}. 
\begin{theorem}\label{thm:b-b}
    Equation \eqref{ss_mkdv_1}-\eqref{ss_mkdv_2} admits the bright soliton solutions given by \(u=g_1/f,\ v=g_2/f\) with \(f,\ g_1, \ g_2\) defined as
    \begin{align}\label{def_bright_f_g}
        f=|M|,\quad
        g_1=\begin{vmatrix}
            M & \Phi \\ -\left(\Psi\right)^T & 0
        \end{vmatrix},\quad
        g_2=\begin{vmatrix}
            M & \Phi \\ -\left(\Upsilon\right)^T & 0
        \end{vmatrix},
    \end{align}
    where \(M\) is an \(N\times N\) matrix, \(\Phi \), \(\bar{\Psi} \), are \(N\)-component row vectors whose elements are defined respectively as
    \begin{align}
        &m_{ij}=\frac{1}{p_i+p_j^*}\left(e^{\xi_i+\xi_j^*}+c_{i,j}\right),\quad \xi_i=p_i x + p_i^3 t+\xi_{i0},\\
        &c_{i,j} = - c_1 \left(C_i\right)^* C_j-c_1 C_{N+1-i} \left(C_{N+1-j}\right)^* - c_2 D_i^* D_j,   \\
        &\Phi = \left( e^{\xi _{1}},e^{\xi _2},\ldots, e^{\xi _{N}}\right)^T,\quad
        \Psi=\left(C_1, C_2,\ldots, C_N\right)^T, \quad \Upsilon=\left(D_1, D_2,\ldots, D_N\right)^T,
    \end{align}
    Here, \(p_i\), \(\xi_{i0}\), \(C_i\), \(D_i\) are complex parameters which satisfy the following restrictions
    \begin{align}\label{cond_bb}
        p_{N+1-i}^* = p_i, \quad \xi_{N+1-i,0}^* = \xi_{i,0}, \quad \left(D_i\right)^* = D_{N+1-i}.
    \end{align}
\end{theorem}

\begin{theorem}\label{thm:d-d}
Equation \eqref{ss_mkdv_1}-\eqref{ss_mkdv_2} admits the dark soliton solutions given by
\begin{equation}\label{def_dark_u}
    u=\rho_1 \frac{h_1}{f} e^{\i \left(\alpha x - \left(\alpha^3 + 3\alpha\left(2c_1\rho_1^2 + c_2\rho_2^2\right) \right) t\right)}, \quad v=\rho_2 \frac{h_2}{f},
\end{equation}
and \(f,\ h_1, \ h_2\) are defined as
\begin{equation}
    f=\tau_{0,0}, \quad h_1=\tau_{1,0}, \quad h_2=\tau_{0,1},
\end{equation}
where \(\tau_{k,l}\) is an \(N\times N\) determinant defined as
\begin{equation}
    \tau_{k,l}=\det\left(\delta_{ij}d_i e^{-\xi_i-\eta_j} + \frac{1}{p_i+q_j} \left(-\frac{p_i - \i \alpha}{q_j + \i \alpha}\right)^k\left(-\frac{p_i}{q_j}\right)^l\right),
\end{equation}
with \(\xi_i=p_i (x - 3(2c_1\rho_1^2 + c_2\rho_2^2) t) + p_i^3 t + \xi_{i 0}\), \(\eta_i=q_i (x - 3(2c_1\rho_1^2 + c_2\rho_2^2) t) + q_i^3 t + \xi_{i 0}\). Here \(\alpha,\ \alpha_2,\ \rho_1,\ \rho_2\) are real parameters, the parameters \(d_i,\ \xi_{i,0},\ p_i,\ q_i\) satisfy the following complex conjugate relation for each \(h = 0, 1, \ldots, \lfloor N/2 \rfloor\)
\begin{align}
        \begin{split}
            d_i = d_{N+1-i}\in \mathbb{R},\quad \xi_{i,0} = \xi_{N+1-i,0}\in \mathbb{R}, & \quad \text{for } i = 1, 2, \ldots, N,\\
            p_i = p_{N+1-i}^* = q_i^* = q_{N+1-i},& \quad \text{for } i \in \{\mathbb{Z} | 1 \leq i \leq h\},\\
            p_i = q_{N+1-i} \in \mathbb{R},\quad p_{N+1-i} = q_i \in \mathbb{R},& \quad \text{for } i \in  \{\mathbb{Z} | h+1 \leq i \leq \lceil N/2 \rceil\},
        \end{split}
\end{align}
Moreover, these parameters need to satisfy the constraint \(G(p_i,q_i) = 0\), for \(i = 1, 2, \ldots, N\), where \(G(p,q)\) defined as
\begin{align}\label{Gpq}
    G(p,q)&=\frac{c_1\rho_1^2}{(p_i-\i \alpha)(q_i+\i \alpha)}+\frac{c_1\rho_1^2}{(p_i+\i \alpha)(q_i-\i \alpha)}+\frac{c_2\rho_2^2}{p_i q_i}-1.
\end{align}
\end{theorem}

\begin{theorem}\label{thm:b-d}
Equation \eqref{ss_mkdv_1}-\eqref{ss_mkdv_2}  admits the following bright-dark soliton solution,
\begin{align}
    u = \frac{g_1}{f},\quad v =\rho_2\frac{h_2}{f}
\end{align}
and \(f, g_1, h_2\) are determinants defined as
\begin{equation}
    f = \left|M_0\right|,\quad g_1 = \begin{vmatrix}
        M_0 & \Phi \\
        -\left(\Psi\right)^T & 0
    \end{vmatrix},\quad h_2 = \left|M_1\right|,
\end{equation}
where \(M_q\) is \(N\times N\) matrix, \(\Phi\) and \(\bar{\Psi}^{(k)}\) are \(N\)-component vectors whose elements are defined as
\begin{align}
    &\left(M_q\right)_{ij}=\frac{1}{p_i+p_j^*}\left(\left(-\frac{p_i}{p_j^*}\right)^q e^{\xi_i+\xi_j^*}+ c_{i,j}\right),\\
    &c_{i,j} = \frac{c_1 \left(C_i\right)^* C_j+c_1 C_{N+1-i} \left(C_{N+1-j}\right)^* }{(c_2 \rho_2^2)/(p_i p_j^*) - 1},\label{bd_cij}\\
    &\xi_i=p_i \left(x - 3c_2\rho_2^2 t\right) + p_i^3 t +\xi_{i0}, \\
    &\Phi = \left(e^{\xi _{1}}, e^{\xi _2},\ldots, e^{\xi _{N}}\right)^T, \quad \Psi = \left(C_1, C_2,\ldots, C_N\right)^T.
\end{align}
Here, \(p_i, \xi_{i0}, C_i\) are complex parameters and
\begin{align}\label{cond_bd}
    p_{N+1-i}^* = p_i, \quad \xi_{N+1-i,0}^* = \xi_{i,0}.
\end{align}
\end{theorem}

\begin{theorem}\label{thm:d-b}
Equation \eqref{ss_mkdv_1}-\eqref{ss_mkdv_2}  admits the following dark-bright soliton solution,
\begin{align}
    u = \rho_1\frac{h_1}{f}e^{\i \left(\alpha x - \left(\alpha^3 + 6c_1\alpha\rho_1^2\right) t\right)},\quad v = \frac{g_2}{f}
\end{align}
and \(f, h_1, g_2\) are determinants defined as
\begin{equation}
    f = \left|M_0\right|,\quad h_1 = \left|M_1\right|,
    \quad g_2 = \begin{vmatrix}
        M_0 & \Phi \\
        -\left(\Psi\right)^T & 0
    \end{vmatrix},
\end{equation}
where \(M_q\) is \(N\times N\) matrix, \(\Phi\) and \(\bar{\Psi}\) are \(N\)-component vectors whose elements are defined as
\begin{align}
    &\left(M_q\right)_{ij}=\frac{1}{p_i+p_j^*}\left(\left(-\frac{p_i - \i \alpha}{p_j^* + \i \alpha}\right)^q e^{\xi_i+\xi_j^*}+ c_{i,j}\right),\\
    &c_{i,j} = \frac{c_2 \left(D_i\right)^* D_j}{\dfrac{2c_1 \rho_1^2 (p_i p_j^* + \alpha^2)}{(p_i^2 + \alpha^2)((p_j^*)^2 + \alpha^2)} - 1},\\
    &\xi_i=p_i \left(x - 6c_1\rho_1^2 t\right) + p_i^3 t +\xi_{i,0}, \\
    &\Phi = \left(e^{\xi _{1}}, e^{\xi _2},\ldots, e^{\xi _{N}}\right)^T, \quad \Psi = \left(D_1, D_2,\ldots, D_N\right)^T.
\end{align}
Here, \(p_i, \xi_{i,0}\) are complex parameters and \(\alpha_l\) is a real number and
\begin{align}
    p_{N+1-i}^* = p_i, \quad \xi_{N+1-i,0}^* = \xi_{i,0}, \quad \left(D_i\right)^* = D_{N+1-i}.
\end{align}
\end{theorem}

\section{Dynamics of bright-bright solitons}\label{section:dynmics_bb}
For the following analysis, we take \(c_1 = c_2 = -1\) for simplicity.
By taking \(N=1\) in \cref{thm:b-b}, we have the one-bright soliton solution
\begin{align}\label{bright_n=1}
    u &= \frac{2C_1 p_1}{\sqrt{2\left|C_1\right|^2 + \left|D_1\right|^2}} \sech\left(p_1^3 t + p_1 x + \xi_{1,0} - \log\sqrt{2\left|C_1\right|^2 + \left|D_1\right|^2}\right),\\
    v &= \frac{2D_1 p_1}{\sqrt{2\left|C_1\right|^2 + \left|D_1\right|^2}} \sech\left(p_1^3 t + p_1 x + \xi_{1,0} - \log\sqrt{2\left|C_1\right|^2 + \left|D_1\right|^2}\right),
\end{align}
where \(\ p_1, \xi_{1,0}, D_1 \in \mathbb{R}\) by \eqref{cond_bb}. The coupled equations \eqref{ss_mkdv_1}-\eqref{ss_mkdv_2} can be decoupled as follows: setting \(C_1 = 0\) results in \(u = 0\), reducing the system to an equation where \(v\) satisfies the mKdV equation. Conversely, choosing \(D_1\) leads to \(u\) being the solution to the Sasa-Satsuma equation.
For component \(v\), if \(\Re(p_1) C_3^{(1)} < 0\) anti-bright soliton is obtained, if \(\Re(p_1) C_3^{(1)} > 0\), bright soliton is obtained. The intensity of above solution is
\begin{align*}
    N(u) = \int_{-\infty}^{+\infty} |u|^2 \, dx = \frac{2|\Re (p_1)|\left|C_1\right|^2}{2\left|C_1\right|^2 + \left|D_1\right|^2}, \quad N(v) = \int_{-\infty}^{+\infty} |v|^2 \, dx = \frac{2|\Re (p_1)|\left|D_1\right|^2}{2\left|C_1\right|^2 + \left|D_1\right|^2}
\end{align*}
The total intensity for all components is \(N = 2N(u) + N(v) = 2 \Re(p_1)\).

\begin{figure}[!ht]
    \centering
    \subfigure[]{\label{fig:SS_mKdV_bright_N=2(a)}
        \includegraphics[width=60mm]{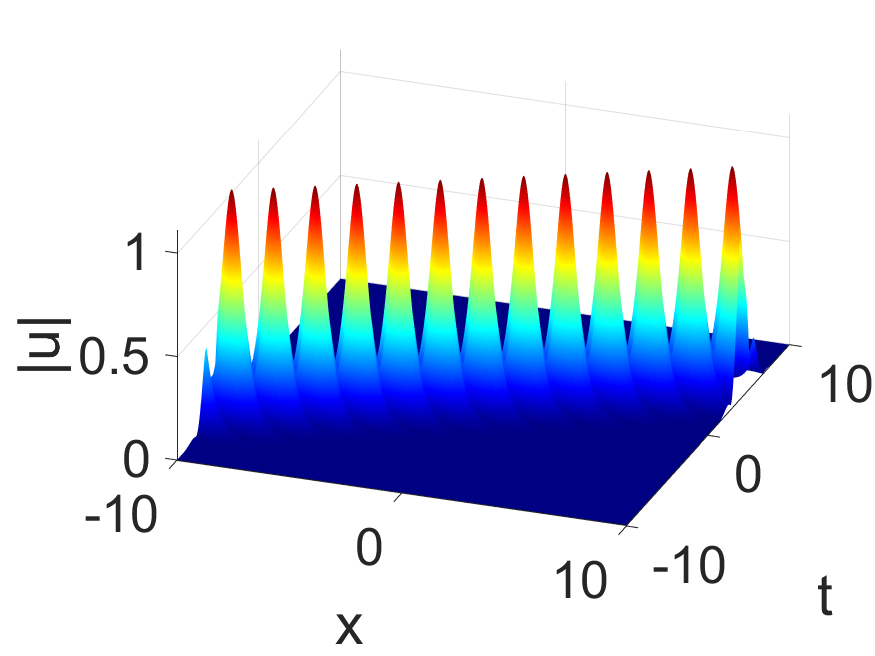}
    }
    \subfigure[]{\label{fig:SS_mKdV_bright_N=2(b)}
        \includegraphics[width=60mm]{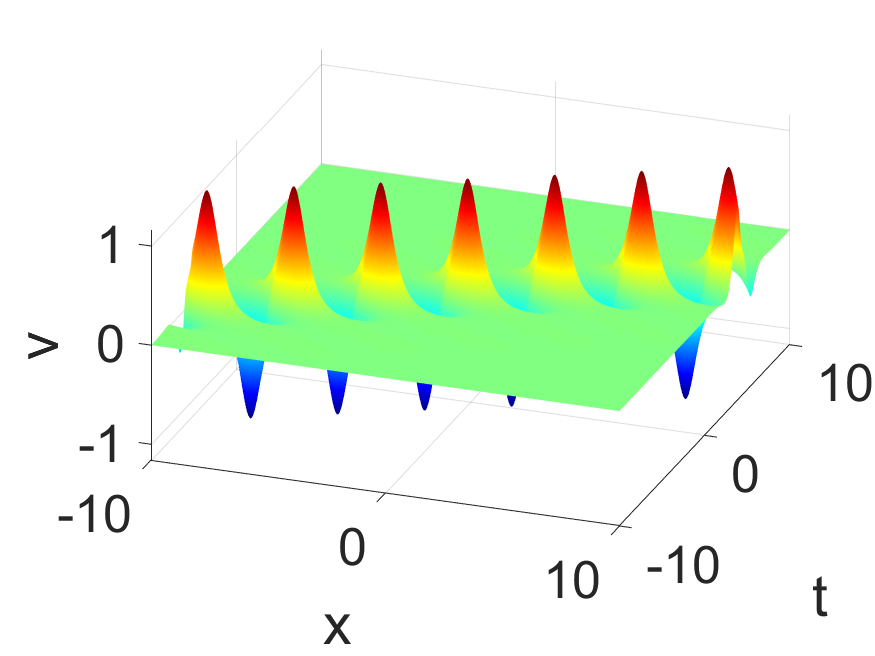}
    }
    \caption{One-oscillated soliton solution to Eq.~\eqref{ss_mkdv_1}-\eqref{ss_mkdv_2} with parameters \(p_1= 1+\i, C_1 = 1 + 1\i, C_2 = 1 - 2\i, D_1 = 1+2\i, \xi_{1,0} = 0\).}
    \label{fig:SS_mKdV_bright_N=2}
\end{figure}

The second order bright soliton solution takes the form \(u = g_1/f, \ v = g_2/f\) with
\begin{align*}
    f ={}& \left(\dfrac{1}{p_1 + p_1^*} \left(e^{\xi_1 + \xi_1^*} + c_{1,1}\right)\right)^2 - \left|\frac{1}{2p_1} \left(e^{2\xi_1} + c_{1,2}\right)\right|^2,\\
    g_1 ={}& \frac{C_1}{2 p_1 (p_1 + p_1^*)} \left(2 p_1 c_{1,1} \exp(\xi_1) - c_{1,2} (p_1 + p_1^*) \exp(\xi_1^*) + \exp(2\xi_1 + \xi_1^*) (p_1-p_1^*)\right)\\
    & + \frac{C_2}{2 p_1 (p_1 + p_1^*)} \left(2 p_1^* c_{1,1} \exp(\xi_1^*) - c_{1,2}^* (p_1 + p_1^*) \exp(\xi_1) + \exp(\xi_1 + 2\xi_1^*) (p_1^*-p_1)\right)\\
    g_2 ={}& 2 \Re \left(\frac{D_1}{2 p_1 (p_1 + p_1^*)} \left(2 p_1 c_{1,1} \exp(\xi_1) - c_{1,2} (p_1 + p_1^*) \exp(\xi_1^*) + \exp(2\xi_1 + \xi_1^*) (p_1-p_1^*)\right)\right),
\end{align*}
where \(\xi_1 = p_1 x + p_1^3 t + \xi_{1,0}\). It is noted that when \(\Im(p_1) \neq 0\)
the oscillated soliton solution is obtained (see \cref{fig:SS_mKdV_bright_N=2}). When \(\Im(p_1) = 0\), it reduces to the one-bright soliton solution. 
However, we cannot obtain double-hump solution to \eqref{ss_mkdv_1}-\eqref{ss_mkdv_2}, in spite of the fact that double-hump solution exists to the Sasa-Satsuma equation.
This is because the conditions
\[
    c_{1,2} = 2\left(C_1^* C_2\right) + \left(D_1^*\right)^2 = 0, \quad C_1 C_2 = 0,
\]
lead to \(D_1 = 0\) and \(v = 0\), which implies the SS equation.

\begin{figure}[!ht]
    \centering
    \subfigure[]{\label{fig:SS_mKdV_bright_N=1(a)}
        \includegraphics[width=50mm]{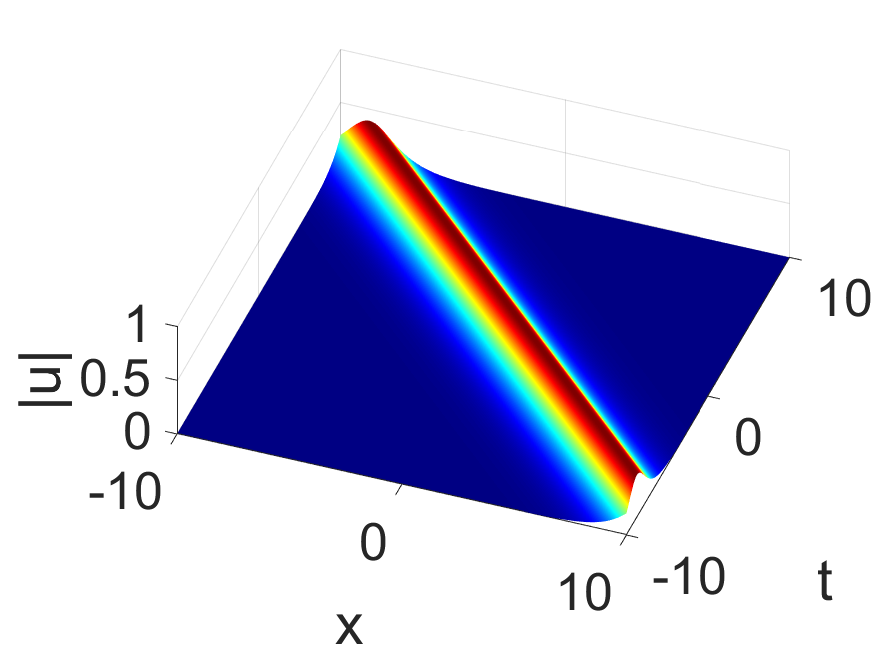}
    }
    \subfigure[]{\label{fig:SS_mKdV_bright_N=1(b)}
        \includegraphics[width=50mm]{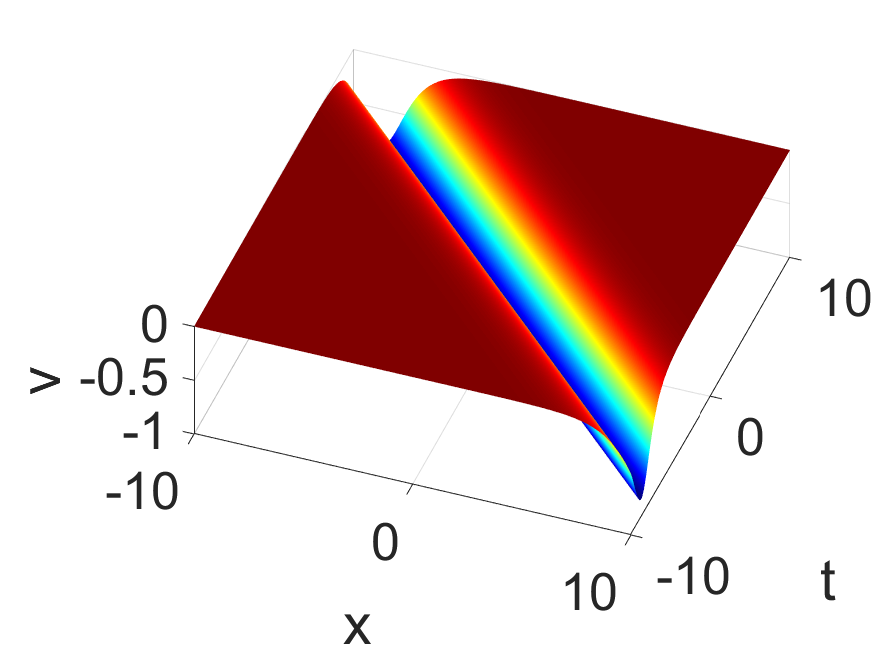}
    }
    \subfigure[]{\label{fig:SS_mKdV_bright_N=1(c)}
        \includegraphics[width=50mm]{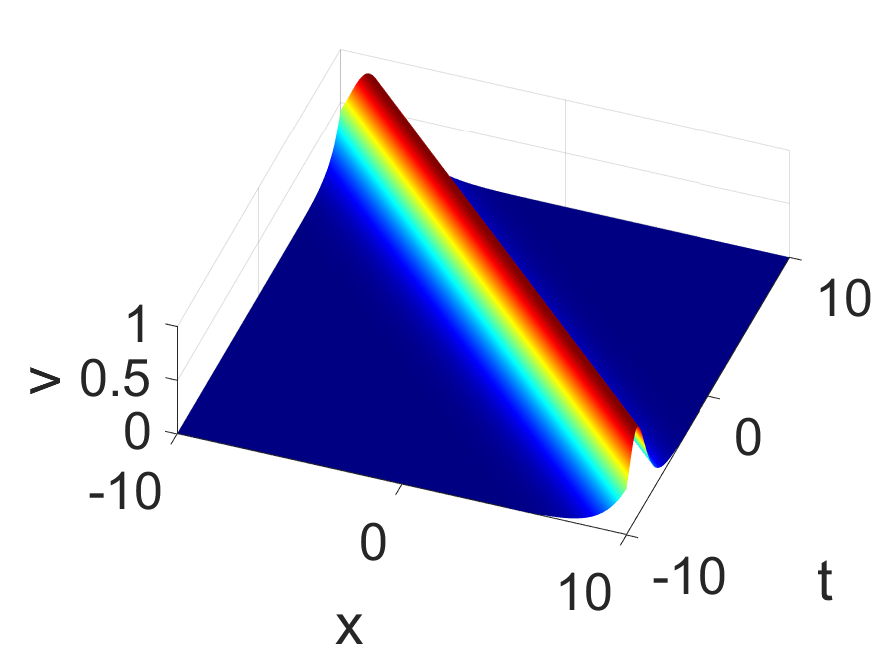}
    }
    \caption{One-bright-one-anti-bright and one-bright-one-bright soliton solution to Eq.~\eqref{ss_mkdv_1}-\eqref{ss_mkdv_2} with parameters (a-b) \(p_1= 1, C_1 = 1 + \i, D_1 = -3, \xi_{1,0} = 0\), (c) \(p_1= 1, C_1 = 1 + \i, D_1 = 3, \xi_{1,0} = 0\).}
    \label{fig:SS_mKdV_bright_N=1}
\end{figure}

\begin{figure}[!ht]
    \centering
    \subfigure[]{
        \includegraphics[width=60mm]{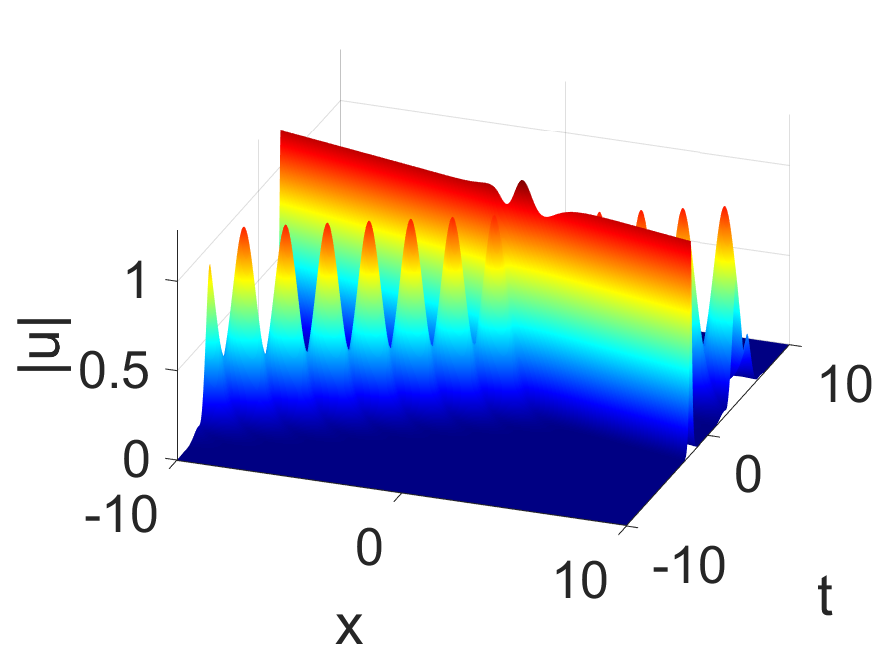}
    }
    \subfigure[]{
        \includegraphics[width=60mm]{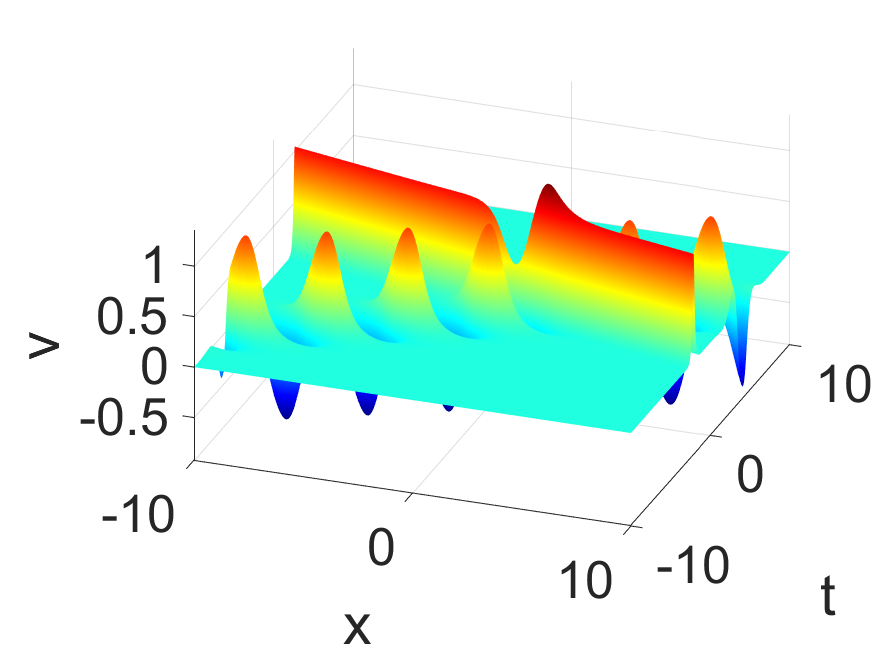}
    }
    \caption{Soliton solution to Eq.~\eqref{ss_mkdv_1}-\eqref{ss_mkdv_2} with collision between oscillated soliton and traveling soliton solution under parameters \(p_1= 1+\i, p_2 = 2, C_1 = 1 + \i, C_2 = 1 - 2\i, C_3 = 2 + 2\i, D_1 = 1+2\i, D_2 = 2, \xi_{1,0} = \xi_{2,0} = 0\).}
    \label{fig:SS_mKdV_bright_N=3}
\end{figure}

\begin{figure}[!ht]
    \centering
    \subfigure[]{
        \includegraphics[width=60mm]{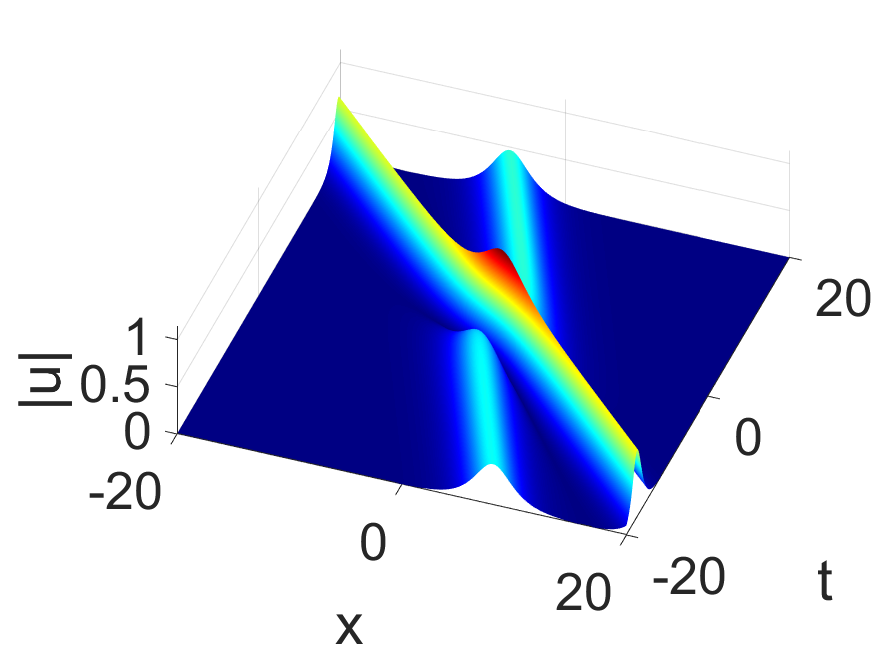}
    }
    \subfigure[]{
        \includegraphics[width=60mm]{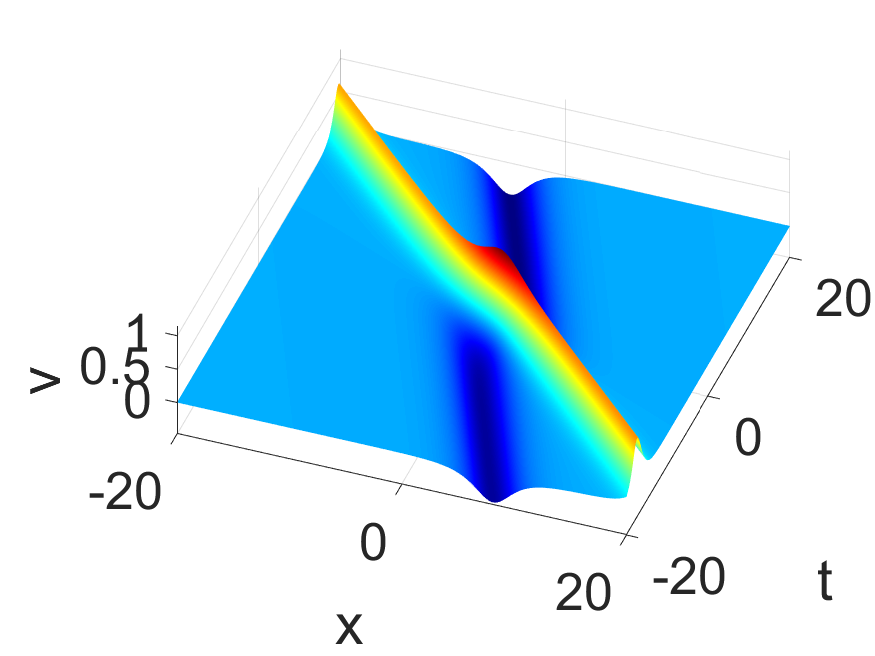}
    }
    \caption{Solution to Eq.~\eqref{ss_mkdv_1}-\eqref{ss_mkdv_2} with collision between traveling solitons under parameters \(p_1= \frac{2}{3}, p_2 = 1, C_1 = C_2 = C_3 = 2, D_1 = 1 + \i, D_2 = 1,  \xi_{1,0} = \xi_{2,0} = 0\).}
    \label{fig:SS_mKdV_bright_N=3_two_soliton}
\end{figure}

\begin{figure}[!ht]
    \centering
    \subfigure[]{
        \includegraphics[width=60mm]{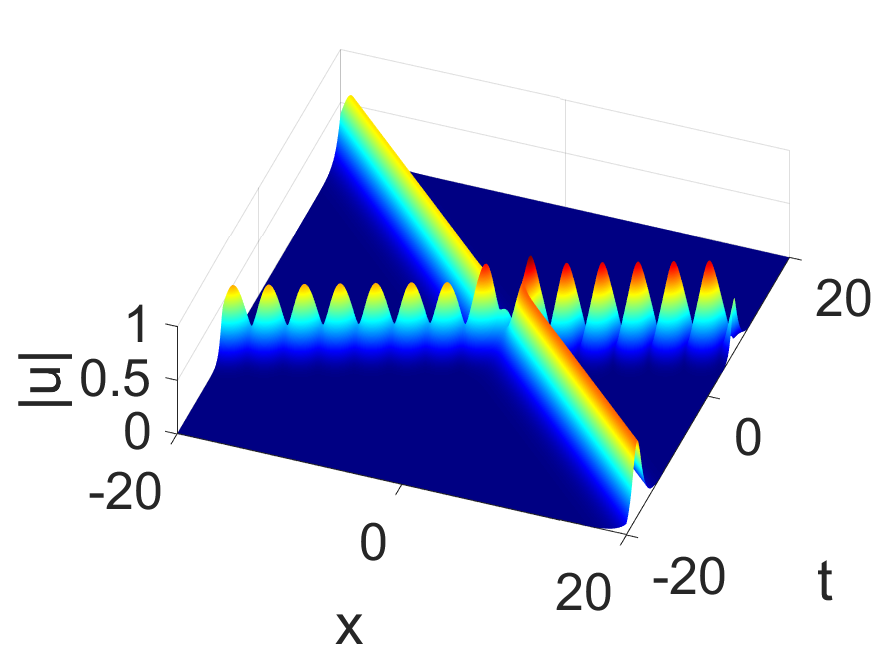}
    }
    \subfigure[]{
        \includegraphics[width=60mm]{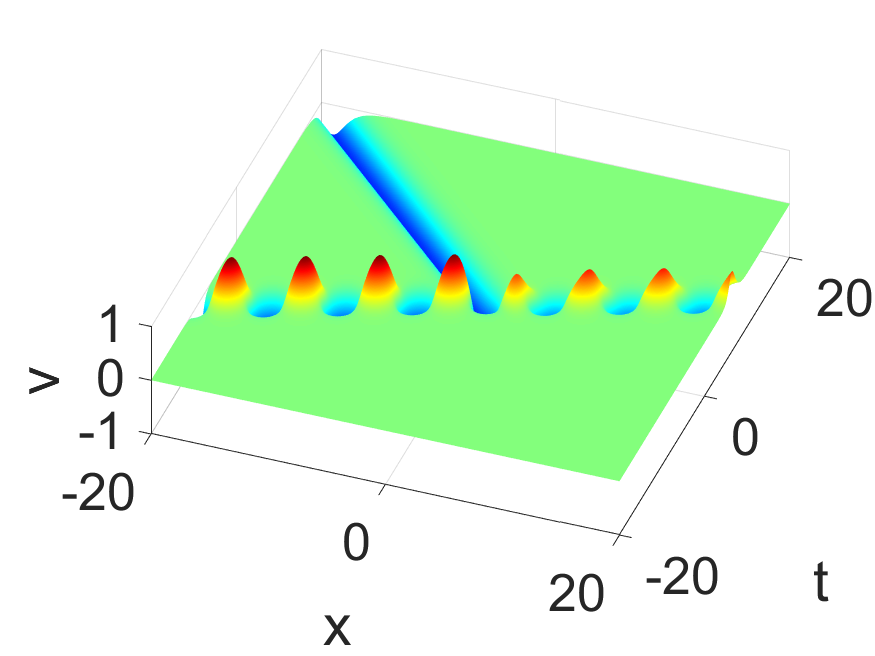}
    }
    \caption{Y-shaped Solution to Eq.~\eqref{ss_mkdv_1}-\eqref{ss_mkdv_2} under parameters \(p_1= \frac{2}{3}+\i, p_2 = 1, C_1 = 1+2\i, C_2 = 2+2\i, C_3 = 3-\i, D_1 = 1 + 2\i, D_2 = 0,  \xi_{1,0} = \xi_{2,0} = 0\).}
    \label{fig:SS_mKdV_bright_N=3_Y_shape}
\end{figure}

\begin{figure}[!ht]
    \centering
    \subfigure[]{
        \includegraphics[width=60mm]{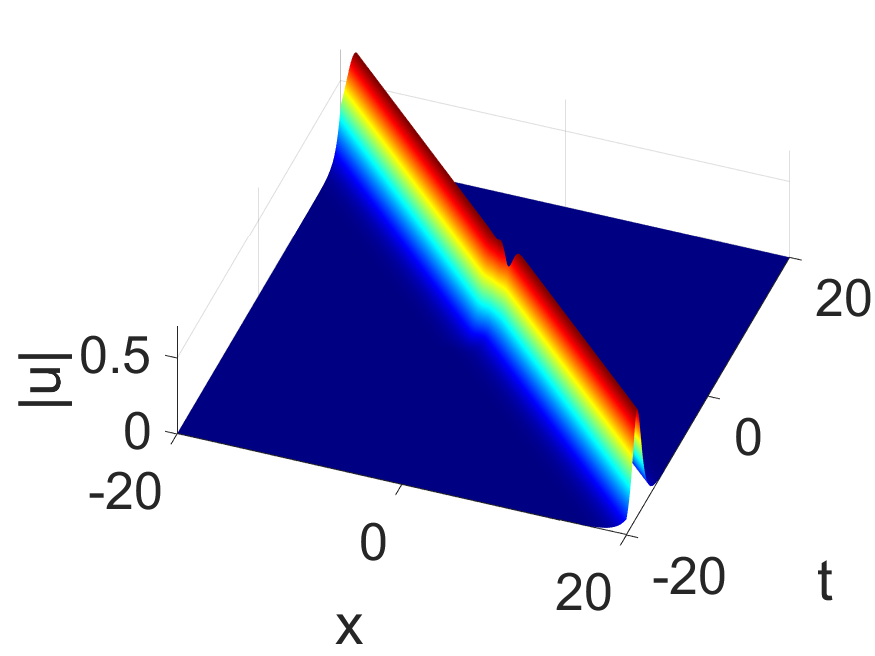}
    }
    \subfigure[]{
        \includegraphics[width=60mm]{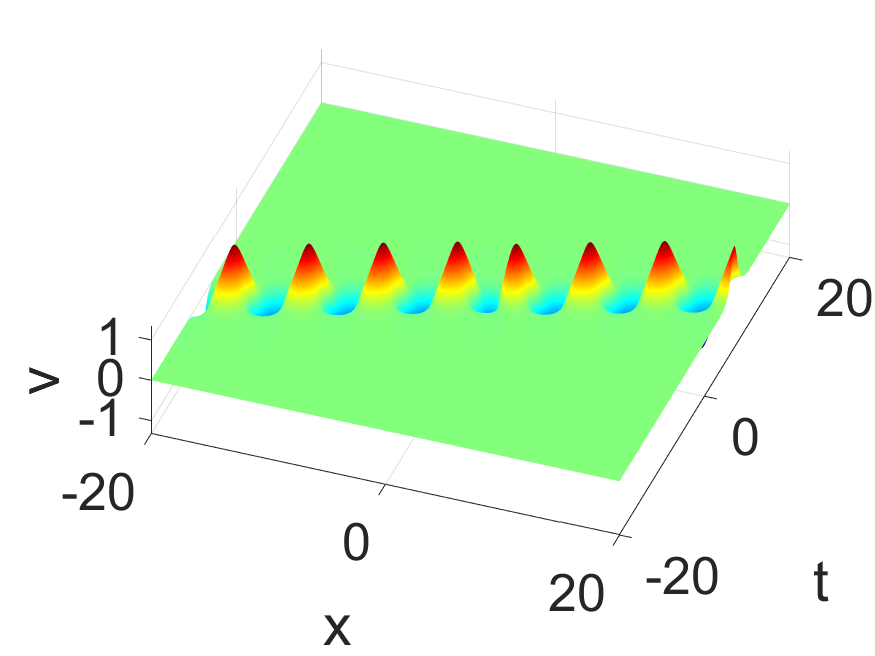}
    }
    \caption{Bright soliton solution to Eq.~\eqref{ss_mkdv_1}-\eqref{ss_mkdv_2} under parameters \(p_1= \frac{2}{3}+\i, p_2 = 1, C_1 = C_3 = 0, C_2 = 2+2\i, D_1 = 1 + 2\i, D_2 = 0,  \xi_{1,0} = \xi_{2,0} = 0\).}
    \label{fig:SS_mKdV_bright_N=3_alt}
\end{figure}

\begin{figure}[!ht]
    \centering
    \subfigure[]{
        \includegraphics[width=60mm]{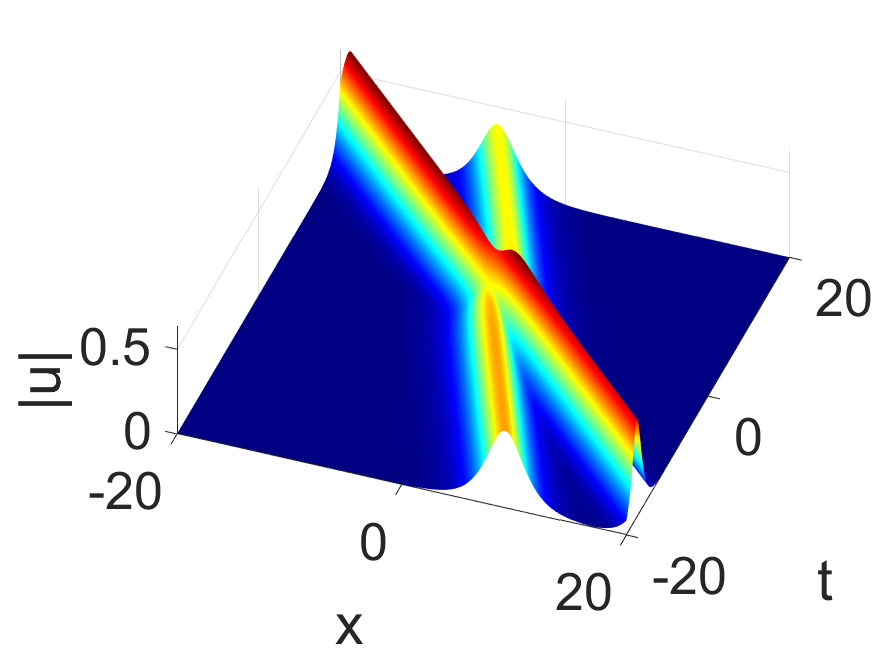}
    }
    \subfigure[]{
        \includegraphics[width=60mm]{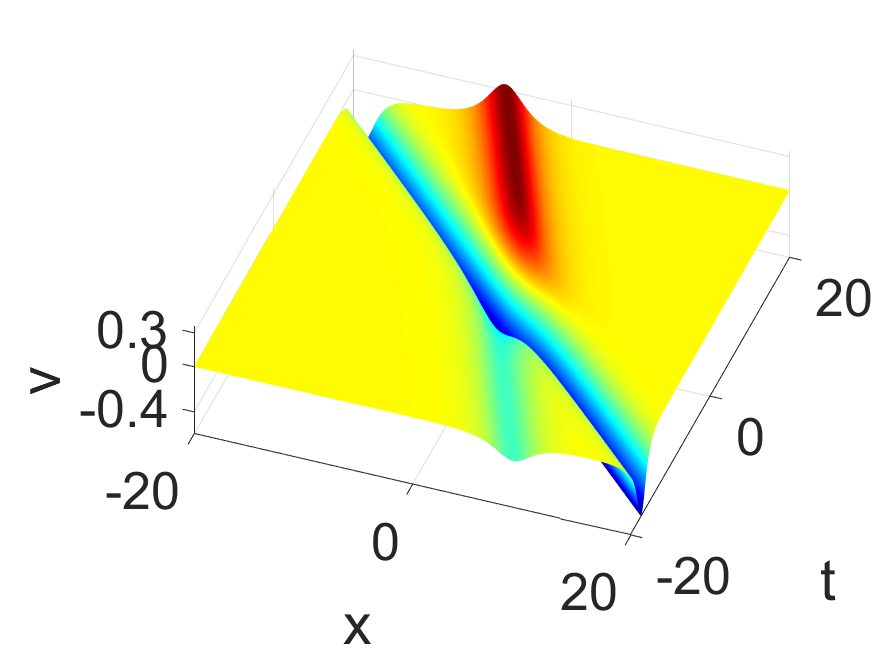}
    }
    \caption{Bright soliton solution to Eq.~\eqref{ss_mkdv_1}-\eqref{ss_mkdv_2} under parameters \(p_1= \frac{2}{3}, p_2 = 1, C_1 = 1+2\i, C_2 = 2+2\i, C_3 = 3-\i, D_1 = 1 + 2\i, D_2 = 0,  \xi_{1,0} = \xi_{2,0} = 0\).}
    \label{fig:SS_mKdV_bright_N=3_real_notY}
\end{figure}

We may observe collision behavior between multi-soliton or multi-oscillated soliton in the higher order cases. For \(N=3\) case, the condition \eqref{cond_bb} requires
\begin{equation}
    p_3 = p_1^*, \quad p_2 \in \mathbb{R}, \quad D_3 = D_1^*, \quad D_2 \in \mathbb{R}.
\end{equation}
If \(\Im(p_1) \neq 0\), the collision between oscillated soliton and traveling soliton is obtained, see \cref{fig:SS_mKdV_bright_N=3}. On the other hand, if \(\Im(p_1) = 0\), collision between two traveling solitons is obtained, see \cref{fig:SS_mKdV_bright_N=3_two_soliton}. We aim to examine the asymptotic behavior of the third-order soliton solutions. To begin, we denote soliton 1 as the one corresponding to \(\xi_1 = 0\) and soliton 2 as \(\xi_2 = 0\). Note that \(\xi_2 = p_2 x + p_2^3 t + \xi_{2,0}\in \mathbb{R}\), if we take \(p_2 > 0\) without loss of generality, we have \(\xi_2 \to \pm\infty\) as \(t \to \pm\infty\). In the following analysis, we focus on the case \(\Im(p_1) \neq 0\).
\begin{enumerate}[(1)]
    \item Before collision, i.e., \(t \rightarrow - \infty\)

    Soliton 1 (\(\xi_1 + \xi_1^* \approx 0,\ \xi_2\rightarrow - \infty\))
    \begin{align*}
        f &\simeq \det(M), \quad  g_1 \simeq \begin{vmatrix}
        M & \begin{matrix} \exp(\xi_1) \\ 0 \\ \exp(\xi_1^*) \end{matrix} \\
        \begin{matrix} -C_1 & -C_2 & -C_3 \end{matrix} &  0
        \end{vmatrix},\quad g_2 \simeq \begin{vmatrix}
            M & \begin{matrix} \exp(\xi_1) \\ 0 \\ \exp(\xi_1^*) \end{matrix} \\
            \begin{matrix} -D_1 & -D_2 & -D_1^* \end{matrix} &  0
        \end{vmatrix}, \\
        M & = \begin{pmatrix}
        \dfrac{1}{p_1 + p_1^*} \left(e^{\xi_1+\xi_1^*} + c_{1, 1}\right) &
        \dfrac{c_{1, 2}}{p_1 + p_2} &
        \dfrac{1}{2p_1} \left(e^{2\xi_1} + c_{1, 3}\right)  \\
        \dfrac{c_{1, 2}^*}{p_1^* + p_2} &
        \dfrac{c_{2, 2}}{2p_2}  &
        \dfrac{c_{2, 3}}{p_1+p_2} \\
        \dfrac{1}{2p_1^*} \left(e^{2\xi_1^*} + c_{1, 3}^*\right)&
        \dfrac{c_{2, 3}^*}{p_1^*+p_2}  &
        \dfrac{1}{p_1 + p_1^*} \left(e^{\xi_1+\xi_1^*} + c_{3, 3}\right)
        \end{pmatrix}.
    \end{align*}

    Soliton 2 (\(\xi_2\approx 0,\ \xi_1 + \xi_1^* \rightarrow + \infty\))
    \begin{align}
    \begin{split}\label{bb_s2_before}
        f &\simeq \det(M), \quad  g_1 \simeq \begin{vmatrix}
        M & \begin{matrix} 1 \\ \exp(\xi_2) \\ 1 \end{matrix} \\
        \begin{matrix} 0 & -C_2 & 0 \end{matrix} &  0
        \end{vmatrix},\quad g_2 \simeq \begin{vmatrix}
            M & \begin{matrix} 1 \\ \exp(\xi_2) \\ 1 \end{matrix} \\
            \begin{matrix} 0 & -D_2 & 0 \end{matrix} &  0
        \end{vmatrix}, \\
        M & = \begin{pmatrix}
        \dfrac{1}{p_1 + p_1^*} &
        \dfrac{1}{p_1 + p_2} e^{\xi_2} &
        \dfrac{1}{2p_1}  \\
        \dfrac{1}{p_1^* + p_2}  e^{\xi_2} &
        \dfrac{1}{2p_2} \left( e^{2\xi_2} + c_{2, 2} \right) &
        \dfrac{1}{p_1+p_2}  e^{\xi_2}\\
        \dfrac{1}{2p_1^*} &
        \dfrac{1}{p_1^*+p_2} e^{\xi_2} &
        \dfrac{1}{p_1 + p_1^*}
        \end{pmatrix}.
    \end{split}
    \end{align}

    \item After collision, i.e., \(t \rightarrow + \infty\)

    Soliton 1 (\(\xi_1 + \xi_1^* \approx 0,\ \xi_2\rightarrow + \infty\))
    \begin{align*}
        f &\simeq \det(M), \quad  g_1 \simeq \begin{vmatrix}
        M & \begin{matrix} \exp(\xi_1) \\ 1 \\ \exp(\xi_1^*) \end{matrix} \\
        \begin{matrix} -C_1 & 0 & -C_3 \end{matrix} &  0
        \end{vmatrix},\quad g_2 \simeq \begin{vmatrix}
            M & \begin{matrix} \exp(\xi_1) \\ 1 \\ \exp(\xi_1^*) \end{matrix} \\
            \begin{matrix} -D_1 & 0 & -D_1^* \end{matrix} &  0
        \end{vmatrix}, \\
        M & = \begin{pmatrix}
        \dfrac{1}{p_1 + p_1^*} \left(e^{\xi_1 + \xi_1^*} + c_{1,1}\right) &
        \dfrac{1}{p_1 + p_2} e^{\xi_1} &
        \dfrac{1}{2p_1} \left(e^{2\xi_1} + c_{1,3}\right) \\
        \dfrac{1}{p_1^* + p_2} e^{\xi_1^*} &
        \dfrac{1}{2p_2} &
        \dfrac{1}{p_1+p_2} e^{\xi_1} \\
        \dfrac{1}{2p_1^*} \left(e^{2\xi_1^*} + c_{1,3}^*\right) &
        \dfrac{1}{p_1^*+p_2} e^{\xi_1^*} &
        \dfrac{1}{p_1 + p_1^*} \left(e^{\xi_1 + \xi_1^*} + c_{3,3}\right)
        \end{pmatrix}.
    \end{align*}

    Soliton 2 (\(\xi_2\approx 0,\ \xi_1 + \xi_1^* \rightarrow - \infty\))
    \begin{align*}
        f &\simeq \det(M), \quad  g_1 \simeq \begin{vmatrix}
        M & \begin{matrix} 0 \\ \exp(\xi_2) \\ 0 \end{matrix} \\
        \begin{matrix} -C_1 & -C_2 & -C_3 \end{matrix} &  0
        \end{vmatrix},\quad g_2 \simeq \begin{vmatrix}
            M & \begin{matrix} 0 \\ \exp(\xi_2) \\ 0 \end{matrix} \\
            \begin{matrix} -D_1 & -D_2 & -D_1^* \end{matrix} &  0
        \end{vmatrix}, \\
        M & = \begin{pmatrix}
        \dfrac{c_{1, 1}}{p_1 + p_1^*} &
        \dfrac{c_{1, 2}}{p_1 + p_2} &
        \dfrac{c_{1, 3}}{2p_1}  \\
        \dfrac{c_{1, 2}^*}{p_1^* + p_2} &
        \dfrac{1}{2p_2} \left(e^{2\xi_2} + c_{2, 2}\right) &
        \dfrac{c_{2, 3}}{p_1+p_2} \\
        \dfrac{c_{1, 3}^*}{2p_1^*} &
        \dfrac{c_{2, 3}^*}{p_1^*+p_2}  &
        \dfrac{c_{3, 3}}{p_1 + p_1^*}
        \end{pmatrix}.
    \end{align*}
\end{enumerate}
These limiting results suggest that imposing the following parameter constraints may lead to different outcomes:
\begin{equation}\label{n3_bb_asym_conds}
    \text{(i) } C_1 = C_3 = 0, \quad \text{(ii) } C_2 = 0, \quad \text{(iii) } D_1 = 0, \quad \text{(iv) } D_2 = 0.
\end{equation}
The nontrivial cases include:
\begin{itemize}
    \item Choosing any one of the conditions in \eqref{n3_bb_asym_conds}, i.e., (i), (ii), (iii), or (iv). In this case, one can verify that one of the solitons in \(u\) or \(v\) vanishes either before or after the collision, leading to a Y-shaped dynamic behavior. An example under condition (iv) is illustrated in \cref{fig:SS_mKdV_bright_N=3_Y_shape}: here, setting \(D_2 = 0\) ensures that soliton 2 asymptotically approaches \(0\) in the \(v\) component before the collision, i.e., \(g_2 \simeq 0\) as \(t\to -\infty, \xi_1 \to +\infty\), as shown in \eqref{bb_s2_before}.
    \item Choosing any of the following pairs: (i,iv), (ii,iii), (i,iii), or (ii,iv) in which (i) and (ii) cannot occur simultaneously, as this would cause component \(u\) to vanish, effectively decoupling the system. A similar argument shows (iii) and (iv) cannot occur simultaneously. An example under condition (i,iv) is shown in \cref{fig:SS_mKdV_bright_N=3_alt}.
\end{itemize}
Next, we examine the case where \(p_1 \in \mathbb{R}\) for \(N=3\). Imposing the parameter constraints from \eqref{n3_bb_asym_conds} does not always result in a Y-shaped collision. For instance, setting \(p_1 = 2/3\) while keeping all other parameters the same as in \cref{fig:SS_mKdV_bright_N=3_Y_shape} does not yield a Y-shaped solution (see \cref{fig:SS_mKdV_bright_N=3_real_notY}). As analyzed earlier, we have \(g_2 \simeq 0\) as \(t\to -\infty, \xi_1 \to +\infty\), but when \(p_1 \in \mathbb{R}\), we also find that \(f \simeq 0\) in this case. This implies that taking the limits of the numerator and denominator separately is not valid; instead, we must consider the limit of \(g_2 / f\).
To interpret \cref{fig:SS_mKdV_bright_N=3_real_notY}, we compute:
\begin{equation*}
    \lim_{\xi_1 \to +\infty} \frac{g_2}{f} = -\frac{192 e^{\xi_2}}{13 e^{2 \xi_2}+3472}.
\end{equation*}
This explains the absence of a Y-shaped solution under these parameter choices.

\begin{figure}[!ht]
    \centering
    \subfigure[]{
        \includegraphics[width=60mm]{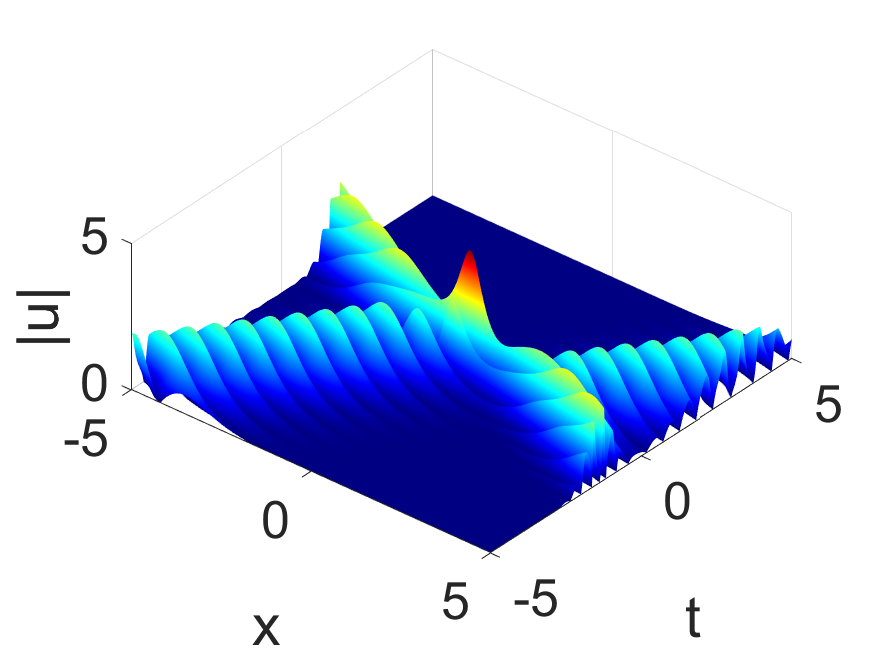}
    }
    \subfigure[]{
        \includegraphics[width=60mm]{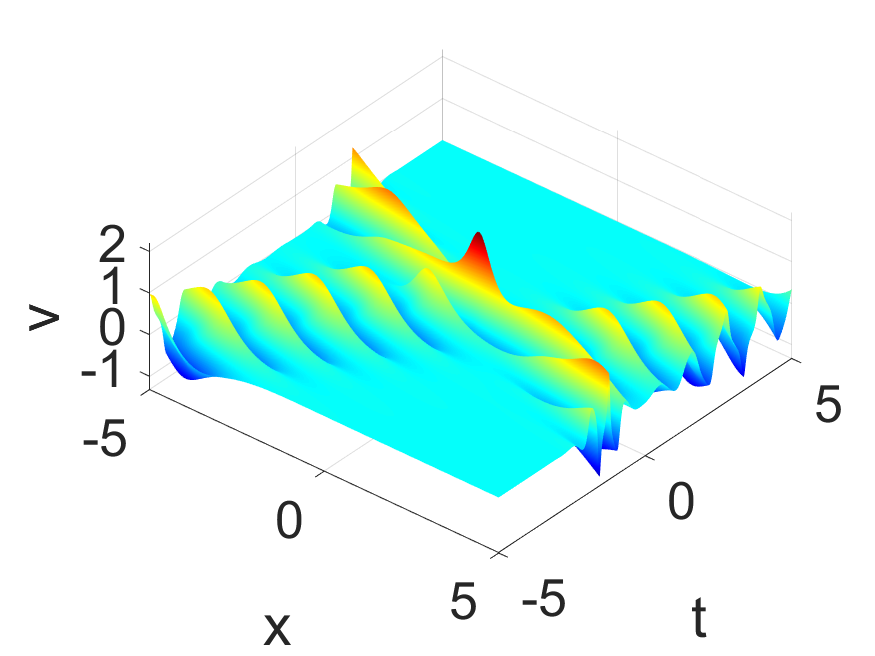}
    }
    \caption{Solution to Eq.~\eqref{ss_mkdv_1}-\eqref{ss_mkdv_2} with collision between oscillated solitons under parameters \(p_1= \frac{3}{2} + \i, p_2 = 2 - \frac{3}{4}\i, C_1 = C_2 = C_3 = 2, D_1 = D_2 = 1,  \xi_{1,0} = \xi_{2,0} = 0\).}
    \label{fig:SS_mKdV_bright_N=4}
\end{figure}

\(N=4\) would give us not only the collision between oscillated soliton and traveling soliton, but also the collision between two oscillated solions, see \cref{fig:SS_mKdV_bright_N=4}.

\section{Dynamics of dark-dark solitons}\label{section:dynmics_dd}
\begin{figure}[!ht]
    \centering
    \subfigure[]{
        \includegraphics[width=60mm]{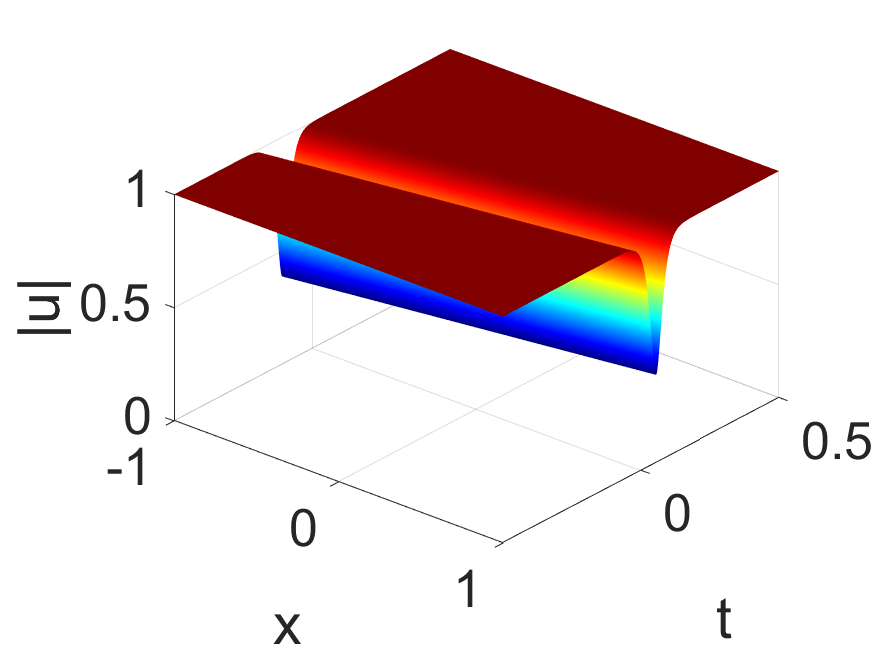}
    }
    \subfigure[]{
        \includegraphics[width=60mm]{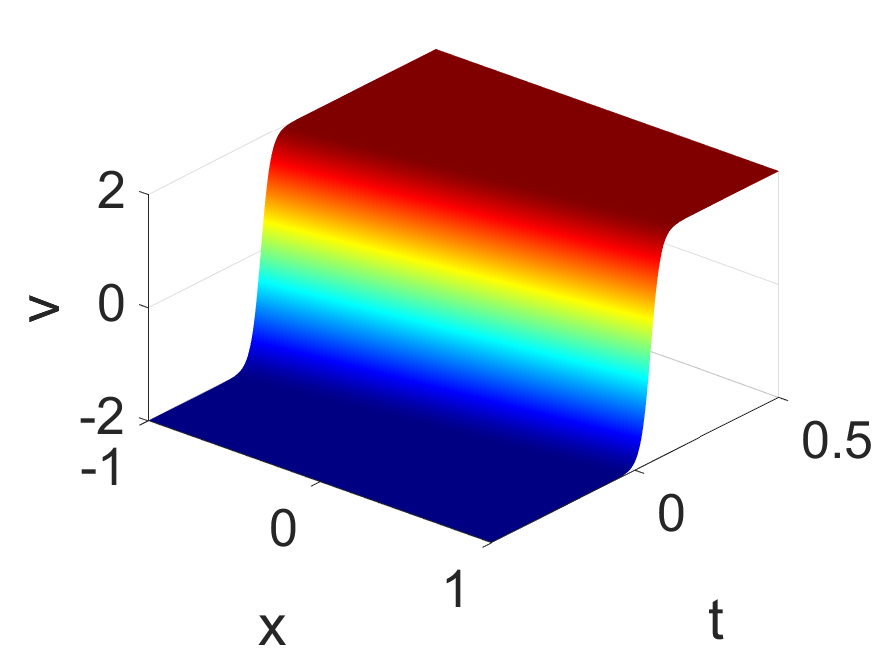}
    }
    \caption{One-dark soliton solution to Eq.~\eqref{ss_mkdv_1}-\eqref{ss_mkdv_2} with parameters \(p_1= \sqrt{\frac{1}{2}(5+\sqrt{41})}, d_1 = 1, c_1 = c_2 = 1, \alpha = 1, \rho_1 = 1, \rho_2 = 2, \xi_{1,0} = 0\).}
    \label{fig:SS_mKdV_dark_N=1}
\end{figure}

The dark soliton solution is obtained from \cref{thm:d-d}. For the first order case, taking \(h=0\) or \(h=1\) give the same one-dark soliton solution as
\begin{align*}
    u &= \frac{\rho_1 \exp(\i \theta_1)}{p_1 + \i \alpha} \biggl(\i \alpha - p_1\tanh\Bigl(p_1 (x - 3(2c_1\rho_1^2 + c_2\rho_2^2) t) + p_1^3 t + \xi_{1 0} - \log{\sqrt{2 d_1 p_1}}\Bigr) \biggr) ,\\
    v &= -\rho_2\tanh\Bigl(p_1 (x - 3(2c_1\rho_1^2 + c_2\rho_2^2) t) + p_1^3 t + \xi_{1 0} - \log{\sqrt{2 d_1 p_1}}\Bigr),
\end{align*}
and \(|u|^2\) can be further simplified as
\begin{align*}
    |u|^2 &= \rho_1^2 \left(1 - \frac{p_1^2}{|p_1 + \i \alpha|^2}\sech^2\Bigl(p_1 (x - 3(2c_1\rho_1^2 + c_2\rho_2^2) t) + p_1^3 t + \xi_{1 0} - \log{\sqrt{2 d_1 p_1}}\Bigr) \right).
\end{align*}
Here, \(\xi_1 = p_1 (x - 3(2c_1\rho_1^2 + c_2\rho_2^2) t) + p_1^3 t + \xi_{1 0}\), \(p_1 = q_1 \in \mathbb{R}, d_1 \in \mathbb{R}, \xi_{1,0} = \eta_{1,0} \in \mathbb{R}\), and \(\exp(\i \theta_1)\) is the plane wave solution for component \(u\) where \(\theta_1 = \alpha x - \left(\alpha^3 + 3\alpha \left(2c_1\rho_1^2 + c_2\rho_2^2\right)\right)t\). Moreover, the parameters need to satisfy
\begin{align*}
    \frac{c_1\rho_1^2}{|p_1-\i \alpha|^2} + \frac{c_1\rho_1^2}{|p_1+\i \alpha|^2} + \frac{c_2\rho_1^2}{p_1^2}=1.
\end{align*}
As seen in \cref{fig:SS_mKdV_dark_N=1}, the shape of the one-dark solution for component \(|u|\) resembles a regular dark soliton, while the shape the solution for component \(v\) is a kink determined by a hyperbolic tangent function. 
The background intensity to above solution is calculated as
\begin{align*}
    N(u) = \int_{-\infty}^{+\infty} \left(|u|^2 - \rho_1^2\right) \, dx= -\frac{2p_1}{|p_1 - \i \alpha|^2}, \quad N(v) = \int_{-\infty}^{+\infty} \left(v^2 - \rho_2^2\right) \, dx= -\frac{2}{p_1}.
\end{align*}

\begin{figure}[!ht]
    \centering
    \subfigure[]{
        \includegraphics[width=60mm]{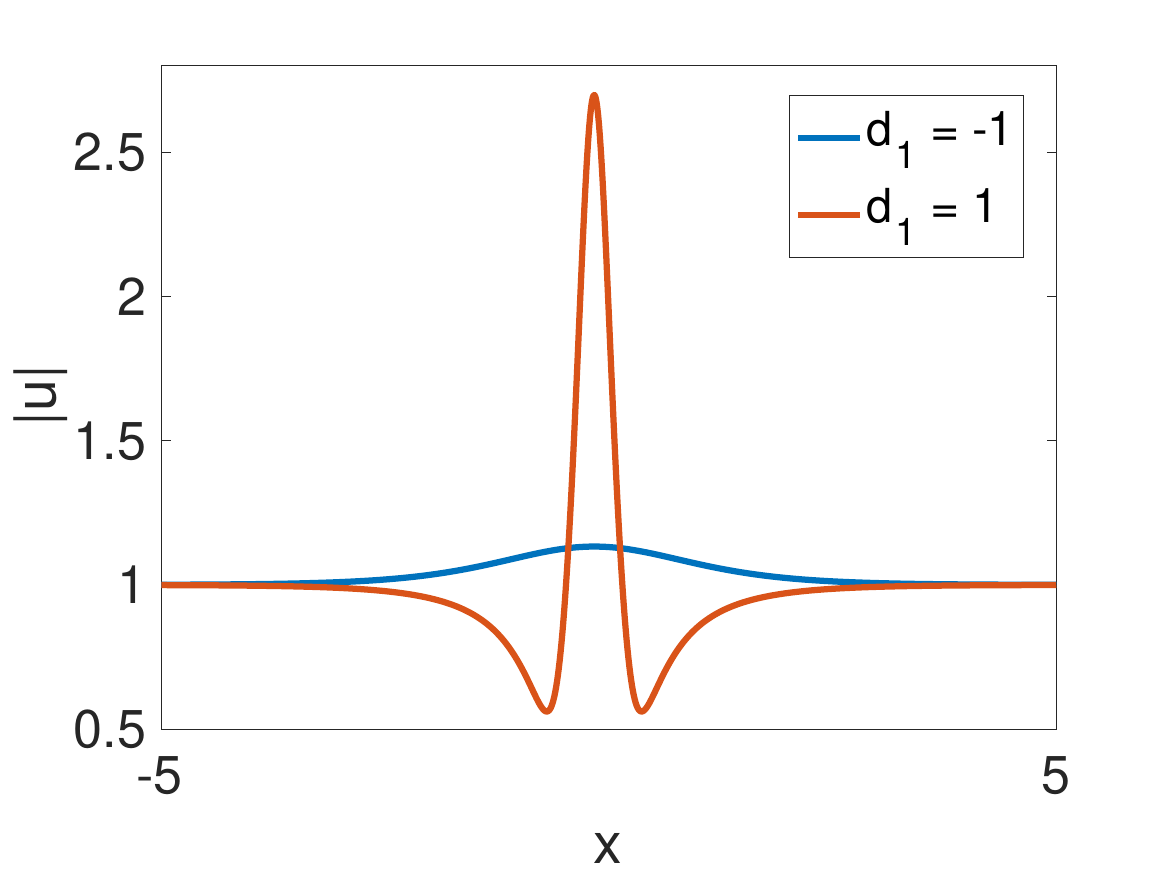}
    }
    \subfigure[]{
        \includegraphics[width=60mm]{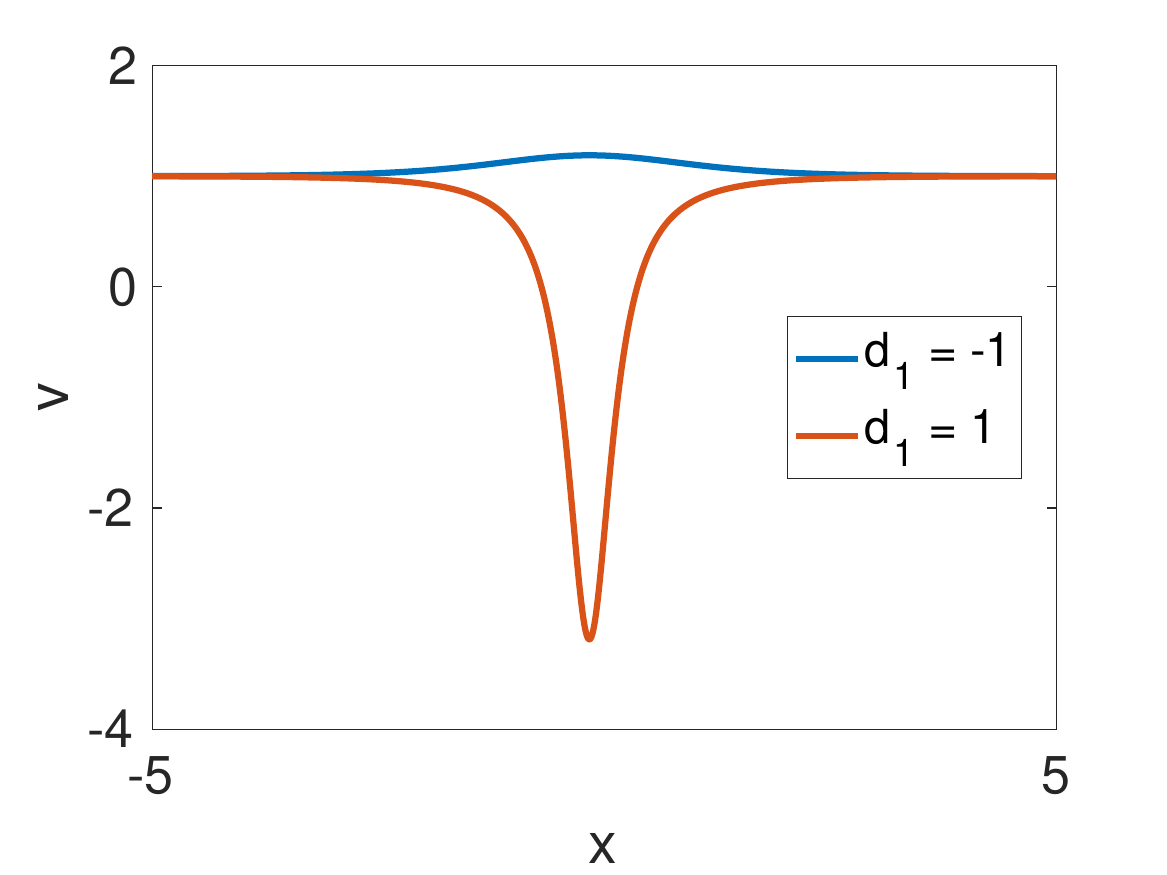}
    }
    \caption{Dark soliton solution to Eq.~\eqref{ss_mkdv_1}-\eqref{ss_mkdv_2} with parameters \(p_1= 1, p_2 \approx -2.37026, c_1 = c_2 = -1, \alpha = 0.5, \rho_1 = 1, \rho_2 = 1, \xi_{1,0} = 0\) and the value of \(d_1\) is shown on the legend.}
    \label{fig:SS_mKdV_dark_N=2_mh}
\end{figure}

\begin{figure}[!ht]
    \centering
    \subfigure[]{
        \includegraphics[width=60mm]{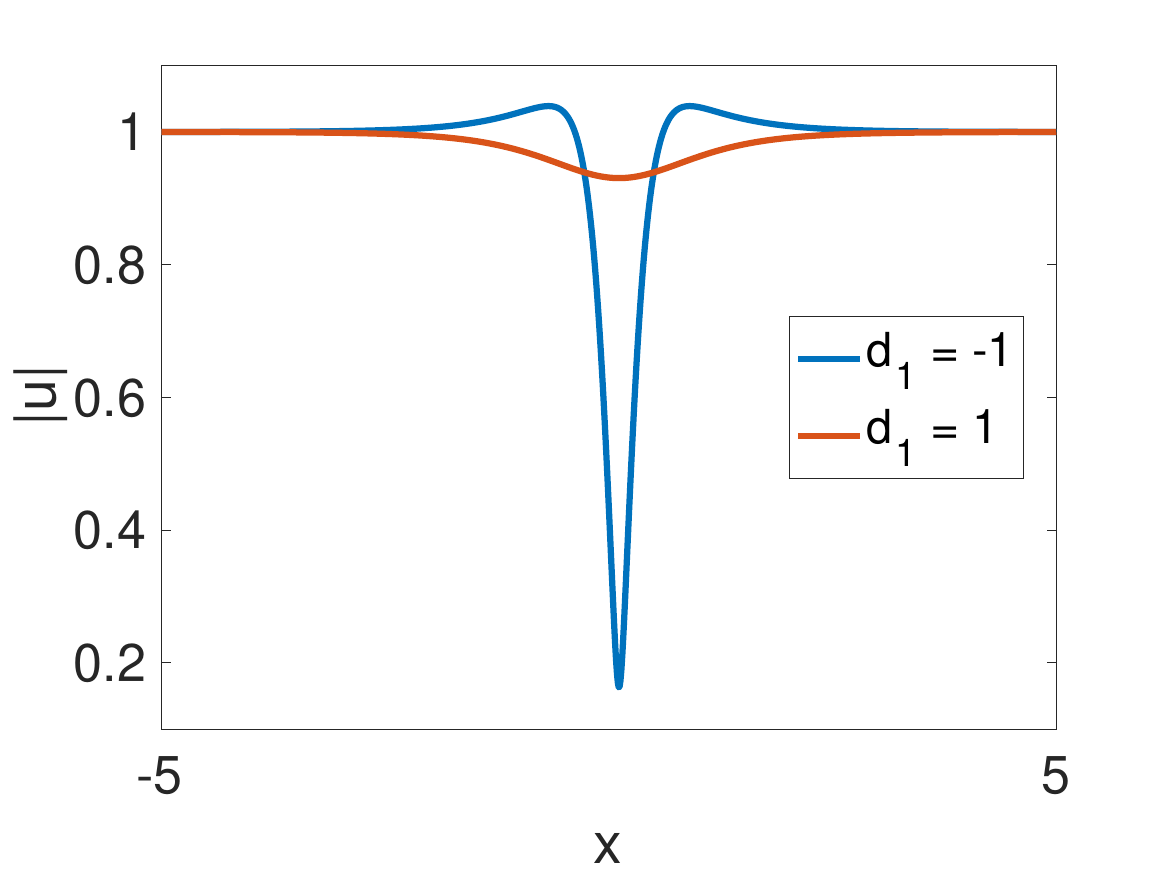}
    }
    \subfigure[]{
        \includegraphics[width=60mm]{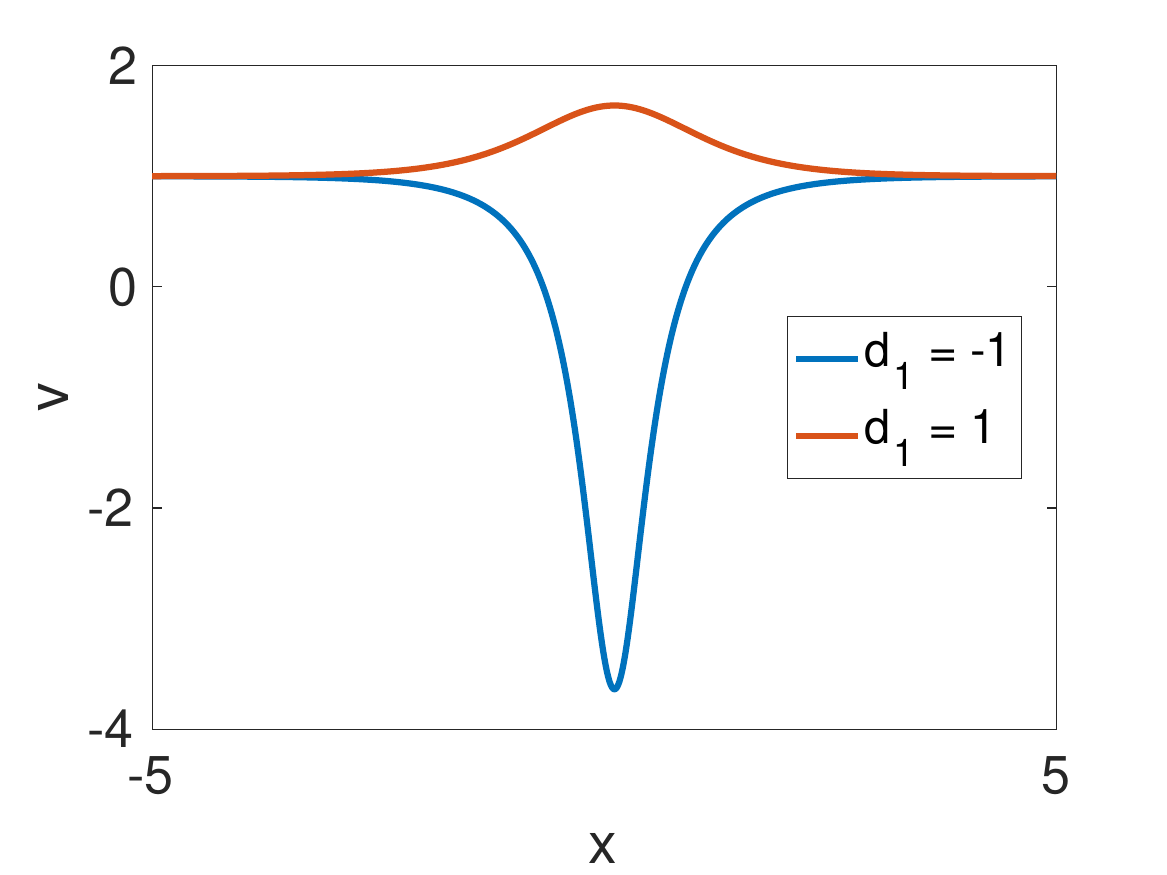}
    }
    \caption{Dark soliton solution to Eq.~\eqref{ss_mkdv_1}-\eqref{ss_mkdv_2} with parameters \(p_1= 2, p_2 \approx -0.42049, c_1 = c_2 = -1, \alpha = 2, \rho_1 = 1, \rho_2 = 1, \xi_{1,0} = 0\) and the value of \(d_1\) is shown on the legend.}
    \label{fig:SS_mKdV_dark_N=2_amh}
\end{figure}

\begin{figure}[!ht]
    \centering
    \subfigure[]{
        \includegraphics[width=60mm]{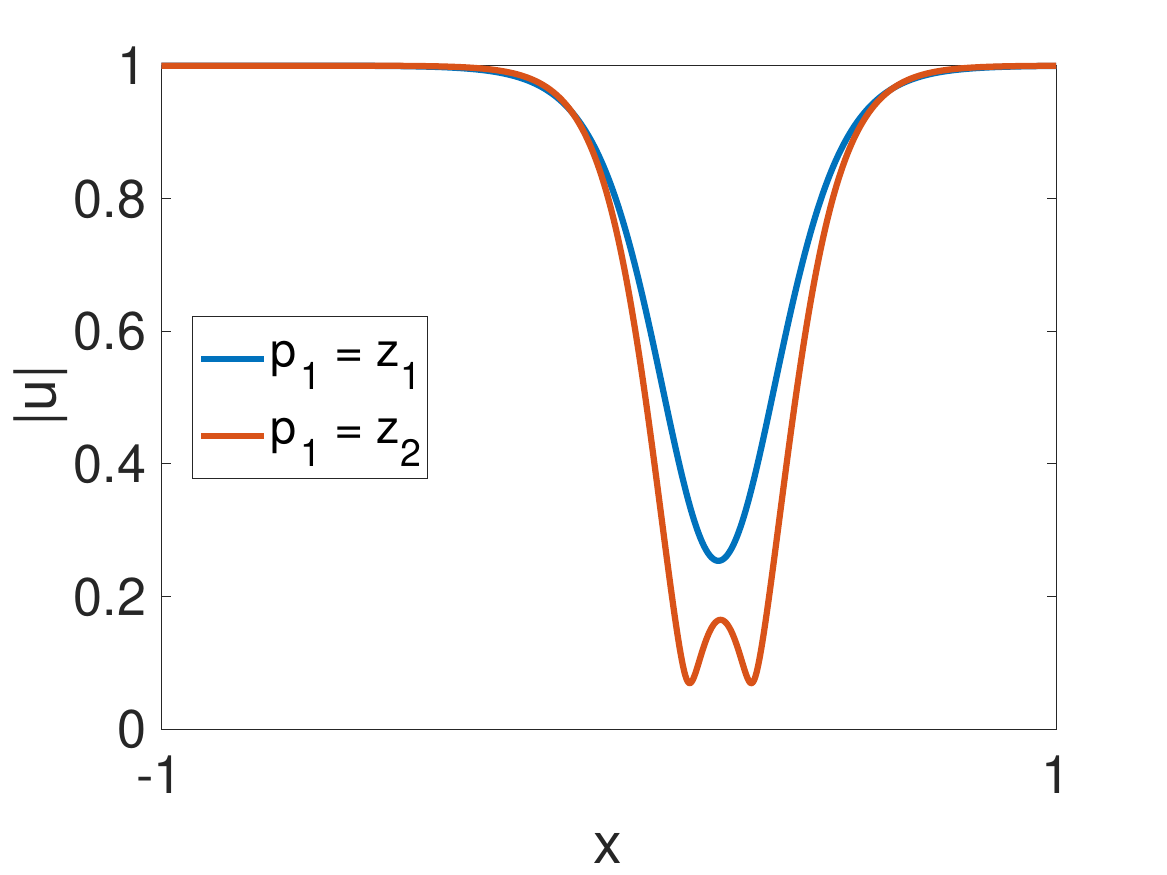}
    }
    \subfigure[]{
        \includegraphics[width=60mm]{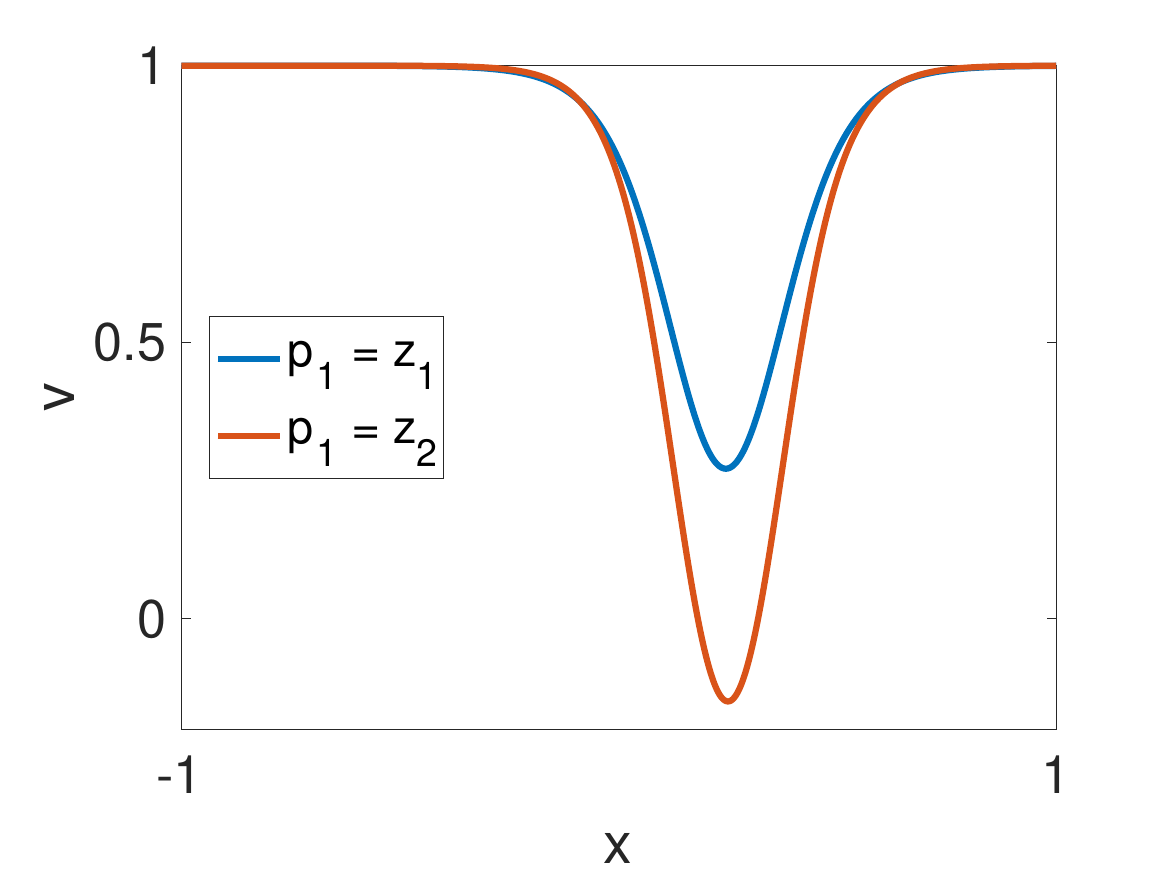}
    }
    \caption{Dark soliton solution to Eq.~\eqref{ss_mkdv_1}-\eqref{ss_mkdv_2} with parameters \(c_1 = c_2  = 20, \alpha = 1, \rho_1 = 1, \rho_2 = 1, d_1 = 1, \xi_{1,0} = 0\) and \(z_1 = 6+4.93905\i, z_2 = 7+3.28749\i\).}
    \label{fig:SS_mKdV_dark_N=2_hole}
\end{figure}

The second order soliton solution is obtained by taking \(N=2\) in \cref{thm:d-d}, which is expressed as \(u = \rho_1 \exp(\i \theta_1) h_1/f, \ v = \rho_2 h_2/f\) and
\begin{align*}
    h_1 &= \begin{vmatrix}
        d_1 \exp(- \xi_1 - \eta_1) + \dfrac{1}{p_1 + q_1}\left(-\dfrac{p_1 - \i \alpha}{q_1 + \i \alpha}\right) & \dfrac{1}{p_1 + q_2}\left(-\dfrac{p_1 - \i \alpha}{q_2 + \i \alpha}\right) \\
        \dfrac{1}{p_2 + q_1}\left(-\dfrac{p_2 - \i \alpha}{q_1 + \i \alpha}\right) & d_2 \exp(- \xi_2 - \eta_2) + \dfrac{1}{p_2 + q_2}\left(-\dfrac{p_2 - \i \alpha}{q_2 + \i \alpha}\right) \\
    \end{vmatrix},\\
    h_2 &= \begin{vmatrix}
        d_1 \exp(- \xi_1 - \eta_1) + \dfrac{1}{p_1 + q_1}\left(-\dfrac{p_1}{q_1}\right) & \dfrac{1}{p_1 + q_2}\left(-\dfrac{p_1}{q_2}\right) \\
        \dfrac{1}{p_2 + q_1}\left(-\dfrac{p_2}{q_1}\right) & d_2 \exp(- \xi_2 - \eta_2) + \dfrac{1}{p_2 + q_2}\left(-\dfrac{p_2}{q_2}\right) \\
    \end{vmatrix},\\
    f &= \begin{vmatrix}
        d_1 \exp(- \xi_1 - \eta_1) + \dfrac{1}{p_1 + q_1} & \dfrac{1}{p_1 + q_2} \\
        \dfrac{1}{p_2 + q_1} & d_2 \exp(- \xi_2 - \eta_2) + \dfrac{1}{p_2 + q_2} \\
    \end{vmatrix},
\end{align*}
where \(\xi_i = p_i (x - 3 c (2\rho_1^2 + \rho_2^2) t) + p_i^3 t + \xi_{i 0}, \eta_i = q_i (x - 3 c (2\rho_1^2 + \rho_2^2) t) + q_i^3 t + \xi_{i 0}, i = 1, 2\) and \(c_2 = c_1 \in \mathbb{R}, \xi_{2,0} = \xi_{1,0} \in \mathbb{R}\). Taking \(N=2\) gives us two possible choices for \(h\), which means parameters \(p_1, p_2, q_1, q_2\) need to satisfy one of the following conditions
\begin{align*}
    &h=0: \ q_2 = p_1\in \mathbb{R}, \ q_1 = p_2\in \mathbb{R},  &h=1: \ p_1 = q_1^* = p_2^* = q_2.
\end{align*}
For \(h=0\) case, solutions such as Mexican hat, Anti-Mexican hat, dark, and anti-dark solitons are identifiable within the component \(u\), while dark and anti-dark solitons manifest within the component \(v\) (refer to \cref{fig:SS_mKdV_dark_N=2_mh,fig:SS_mKdV_dark_N=2_amh}).
For \(h = 1\), we have double-hole and single-hole solitons for the component \(u\), while single hole solitons appear within the component \(v\) (refer to \cref{fig:SS_mKdV_dark_N=2_hole}). The condition for obtaining Mexican hat, Anti-Mexican hat and double-hole soliton solutions is similar to the case of coupled Hirota equation and Sasa-Satsuma equation, see Refs. \cite{shi2025general,zhang2025dark} for details.

Next, let us consider the dynamics for \(N=3\) case. Parameters are required to satisfy the following restrictions
\begin{align*}
    &\xi_{1,0} = \xi_{3,0} \in \mathbb R, \quad \xi_{2,0} \in \mathbb{R}, \quad d_1 = d_3 \in \mathbb{R}, \quad d_2 \in\mathbb{R}, \quad q_2 = p_2\in \mathbb R, \quad q_3 = p_1, \quad q_1 = p_3
\end{align*}
and above \(p_1, p_3\) need to satisfy one of the following condition
\begin{align}\label{dark_n3_hcase}
    &(h=0 \ \text{case})\ p_1, p_3\in \mathbb{R},  &(h=1 \ \text{case})\ p_3 = p_1^*.
\end{align}
Recall the definition of \(\xi_i\) and \(\eta_i\) in both \(h=0,h=1\) case, we have \(\eta_1 = \xi_3,\ \eta_2 = \xi_2,\ \eta_3 = \xi_1\).
We would like to discuss the general form of solution expression without apply the condition \eqref{dark_n3_hcase}.
We can denote soliton 1 as the one corresponding to \(\xi_1 + \xi_3 = 0\) and soliton 2 as \(\xi_2 = 0\). This setup allows us to analysis cases \(h=0\) and \(h=1\) together. We assume soliton 1 is on the left of soliton 2 when \(t \to -\infty\).
Denote
\begin{align*}
    A=\frac{\left(p_1-p_2\right) \left(p_2-p_3\right)}{\left(p_1+p_2\right) \left(p_2+p_3\right)},\quad
    B=\frac{\left(\alpha +\i p_1\right) \left(\alpha +\i p_3\right)}{\left(\alpha -\i p_1\right) \left(\alpha -\i p_3\right)},\quad
    C=\frac{\left(p_1-p_3\right)^2}{p_1 p_3 \left(p_1+p_3\right)^2},
\end{align*}

\begin{figure}[!ht]
    \centering
    \subfigure[]{
        \includegraphics[width=60mm]{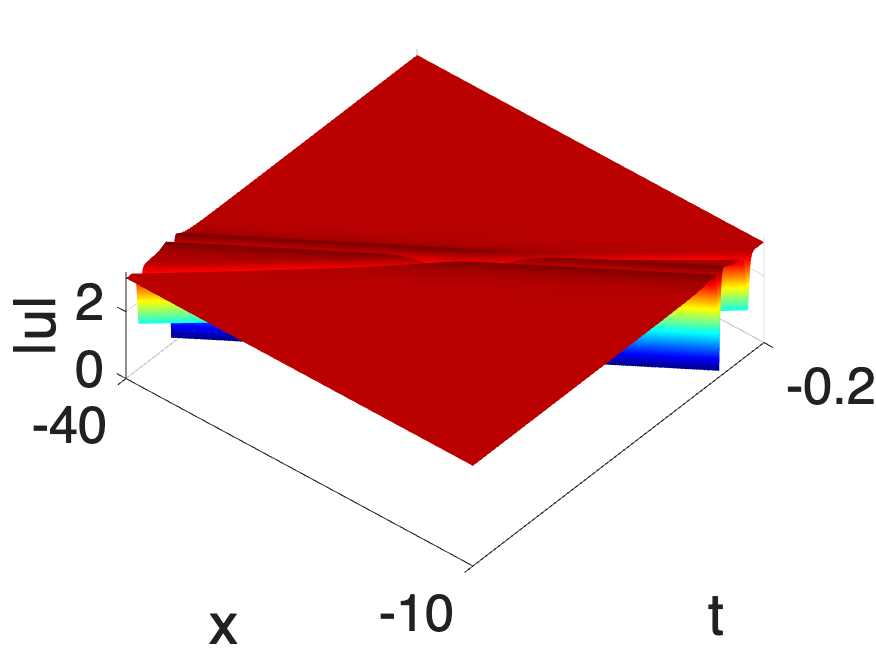}
    }
    \subfigure[]{
        \includegraphics[width=60mm]{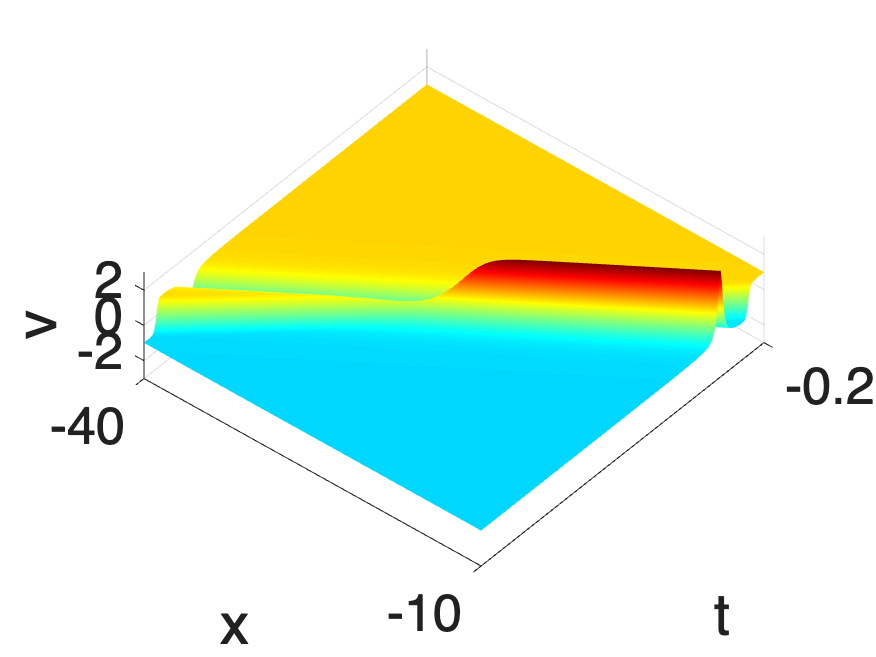}
    }
    \caption{Dark soliton solution to Eq.~\eqref{ss_mkdv_1}-\eqref{ss_mkdv_2} in \(h=0\) case with parameters \(c_1 = c_2 = 1, d_1 = d_2 = 1, \alpha = 2, p_1= 1, p_2 \approx 3.90667, p_3 \approx -1.22479, \rho_1 = 3, \rho_2 = 1, \xi_{1,0} = \xi_{2,0} = 0\).}
    \label{fig:SS_mKdV_dark_N=3_mh_kink}
\end{figure}

\begin{figure}[!ht]
    \centering
    \subfigure[]{
        \includegraphics[width=60mm]{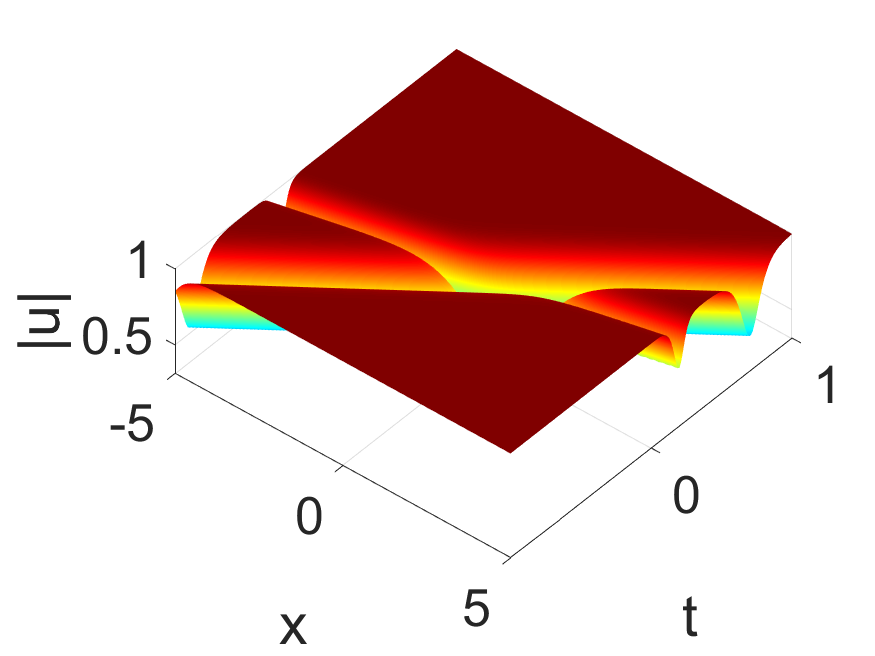}
    }
    \subfigure[]{
        \includegraphics[width=60mm]{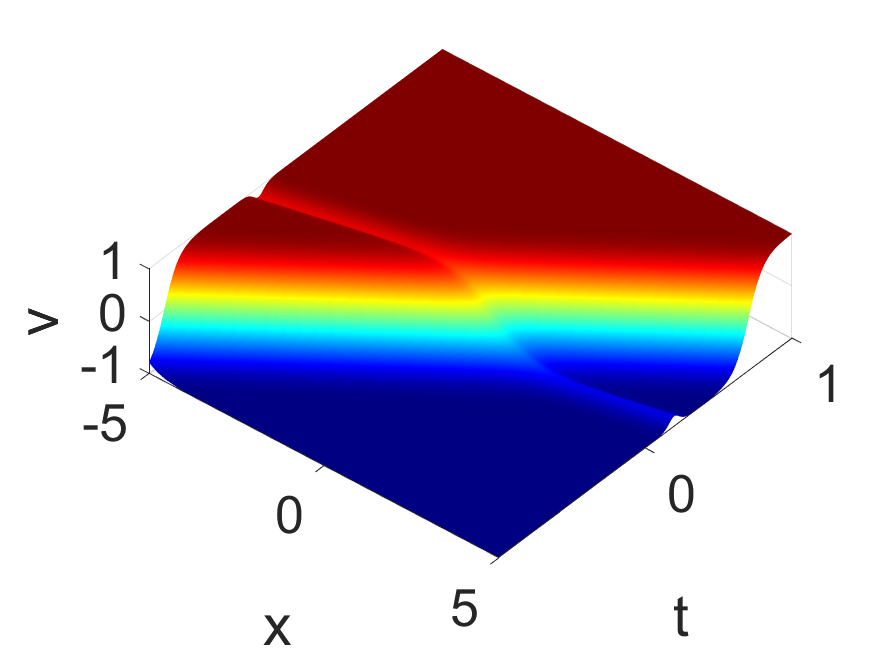}
    }
    \caption{Dark soliton solution to Eq.~\eqref{ss_mkdv_1}-\eqref{ss_mkdv_2} in \(h=1\) case with parameters \(c_1 = c_2 = 1, d_1 = d_2 = 1, \alpha = 1, p_1 \approx 1+ 2.1007\i, p_2 \approx 1.5538, \rho_1 = 2, \rho_2 = 1, \xi_{1,0} = \xi_{2,0} = 0\).}
    \label{fig:SS_mKdV_dark_N=3_hole_kink}
\end{figure}

\begin{figure}[!ht]
    \centering
    \subfigure[]{
        \includegraphics[width=60mm]{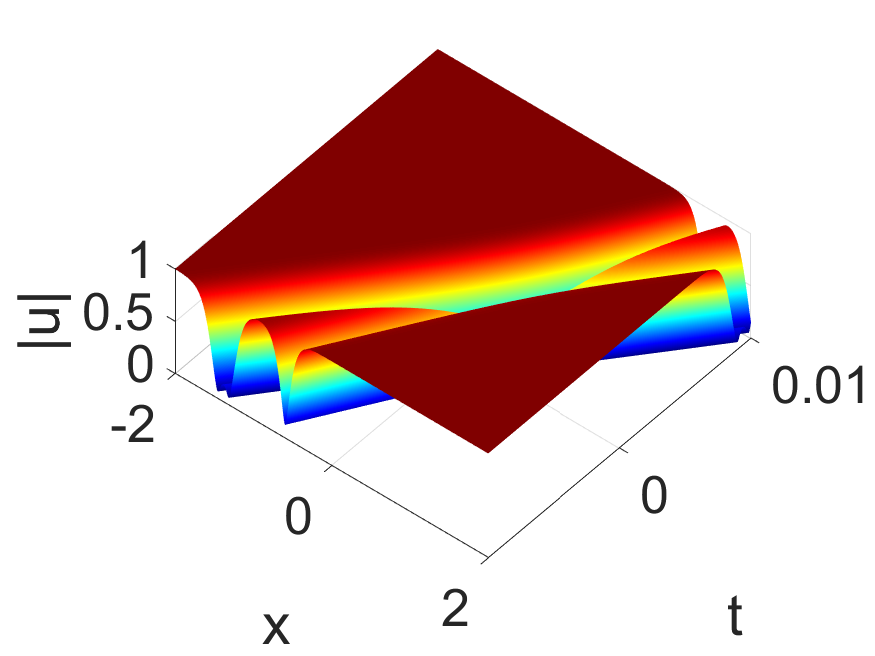}
    }
    \subfigure[]{
        \includegraphics[width=60mm]{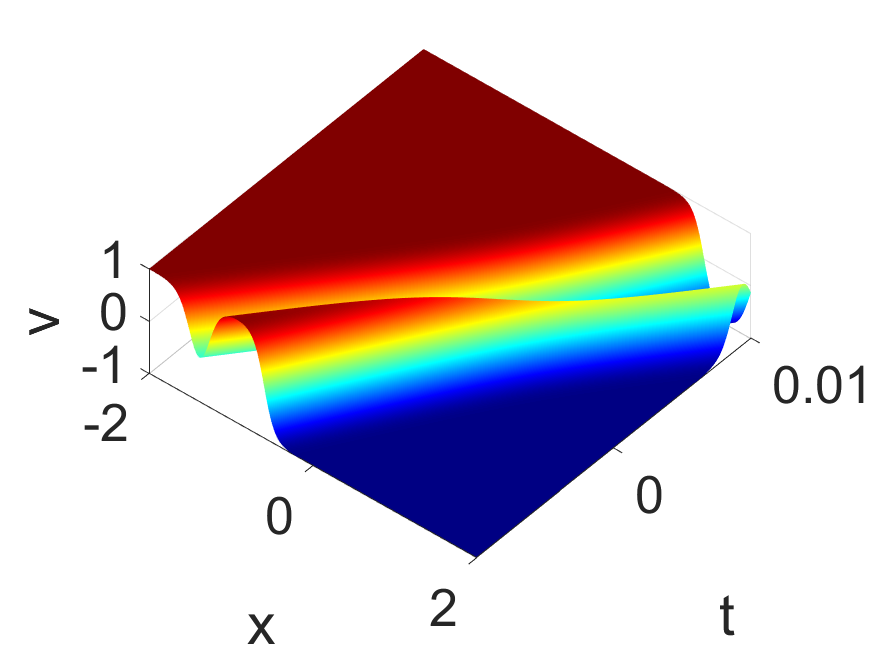}
    }
    \caption{Dark soliton solution to Eq.~\eqref{ss_mkdv_1}-\eqref{ss_mkdv_2} in \(h=1\) case with parameters \(c_1 = c_2 = 20, d_1 = 1, d_2 = 2, \alpha = 1, p_1 \approx 7 +3.2875\i, p_2 \approx 7.7031, \rho_1 = 2, \rho_2 = 1, \xi_{1,0} = \xi_{2,0} = 0\).}
    \label{fig:SS_mKdV_dark_N=3_doublehole_kink}
\end{figure}

\begin{enumerate}[(1)]
    \item Before collision, i.e., \(t \rightarrow - \infty\)

    Soliton 1 (\(\xi_1 + \xi_3 \approx 0,\ \xi_2\rightarrow - \infty\))
    \begin{align*}
        \frac{h_1}f &\simeq \frac{\left(d_1 e^{-\xi_1-\xi_3}+\frac{1}{p_1+p_3}-\frac{1}{p_1+\i \alpha }\right) \left(d_1 e^{-\xi_1-\xi_3}+\frac{1}{p_1+p_3}-\frac{1}{p_3+\i \alpha }\right)-\frac{B}{4 p_1 p_3 }}{\left(d_1 e^{-\xi_1-\xi_3}+\frac{1}{p_1+p_3}\right)^2-\frac{1}{4 p_1 p_3}},\\
        \frac{h_2}f &\simeq \frac{4 d_1 e^{-2 \left(\xi_1+\xi_3\right)} \left(d_1-\frac{e^{\xi_1+\xi_3} \left(p_1^2+p_3^2\right)}{p_1 p_3 \left(p_1+p_3\right)}\right)-C}{4\left(d_1 e^{-\xi_1-\xi_3}+\frac{1}{p_1+p_3}\right)^2-\frac{1}{p_1 p_3}}.
    \end{align*}

    Soliton 2 (\(\xi_2\approx 0,\ \xi_1 + \xi_3 \rightarrow + \infty\))
    \begin{align*}
        \frac{h_1}f &\simeq \frac{B}{p_2+\i \alpha } \left(\frac{4 d_2 p_2^2}{A^2 e^{2 \xi_2}+2 d_2 p_2}-p_2 +\i \alpha\right),\\
        \frac{h_2}f &\simeq \frac{4 d_2 p_2}{A^2 e^{2 \xi_2}+2 d_2 p_2}-1 = \tanh \left(-2\xi_2 + \frac 1 2 \log \left(2 d_2 p_2\right) - \frac 1 2 \log (A^2)\right).
    \end{align*}

    \item After collision, i.e., \(t \rightarrow + \infty\)

    Soliton 1 (\(\xi_1 + \xi_3 \approx 0,\ \xi_2\rightarrow + \infty\))
    \begin{align*}
        \frac{h_1}f &\simeq \frac{\left(\alpha +\i p_2\right) \left(-A^2 B C+4 d_1 e^{-2 \left(\xi_1+\xi_3\right)} \left(d_1+\frac{A e^{\xi_1+\xi_3} \left(2 \alpha ^2+p_1^2+p_3^2\right)}{\left(p_1+\i \alpha \right) \left(p_1+p_3\right) \left(p_3+\i \alpha \right)}\right)\right)}{\left(\alpha -\i p_2\right) \left(4 d_1 e^{-2 \left(\xi_1+\xi_3\right)} \left(d_1-\frac{2 A e^{\xi_1+\xi_3}}{p_1+p_3}\right)-A^2 C\right)},\\
        \frac{h_2}f &\simeq \frac{4 A d_1 e^{\xi_1+\xi_3} \left(p_1+p_3\right)^2}{p_1 p_3 \left(A^2 C e^{2 \left(\xi_1+\xi_3\right)} \left(p_1+p_3\right)+8 A d_1 e^{\xi_1+\xi_3}-4 d_1^2 \left(p_1+p_3\right)\right)}-1.
    \end{align*}

    Soliton 2 (\(\xi_2\approx 0,\ \xi_1 + \xi_3 \rightarrow - \infty\))
    \begin{align*}
        \frac{h_1}f &\simeq \frac{1}{\left(p_2+\i \alpha \right)} \left(\frac{4 d_2 p_2^2}{ 2 d_2 p_2+e^{2 \xi_2}}-p_2+\i \alpha \right),\\
        \frac{h_2}f &\simeq \frac{4 d_2 p_2}{2 d_2 p_2+e^{2 \xi_2}}-1 = \tanh \left(-2\xi_2 + \frac 1 2 \log \left(2 d_2 p_2\right)\right).
    \end{align*}
\end{enumerate}
It is particularly noted that, the asymptotic expression of soliton 2 in component \(v\) before and after collision is expressed by a hyperbolic tangent function. Since \(v\) is a real-valued function, we can observe the collision behavior between the first order kink soliton and second order dark solitons: see examples for anti-Mexican-kink interaction \cref{fig:SS_mKdV_dark_N=3_mh_kink}, single-hole-kink interaction \cref{fig:SS_mKdV_dark_N=3_hole_kink}, and double-hole-kink interaction \cref{fig:SS_mKdV_dark_N=3_doublehole_kink}.

\begin{figure}[!ht]
    \centering
    \subfigure[]{
        \includegraphics[width=60mm]{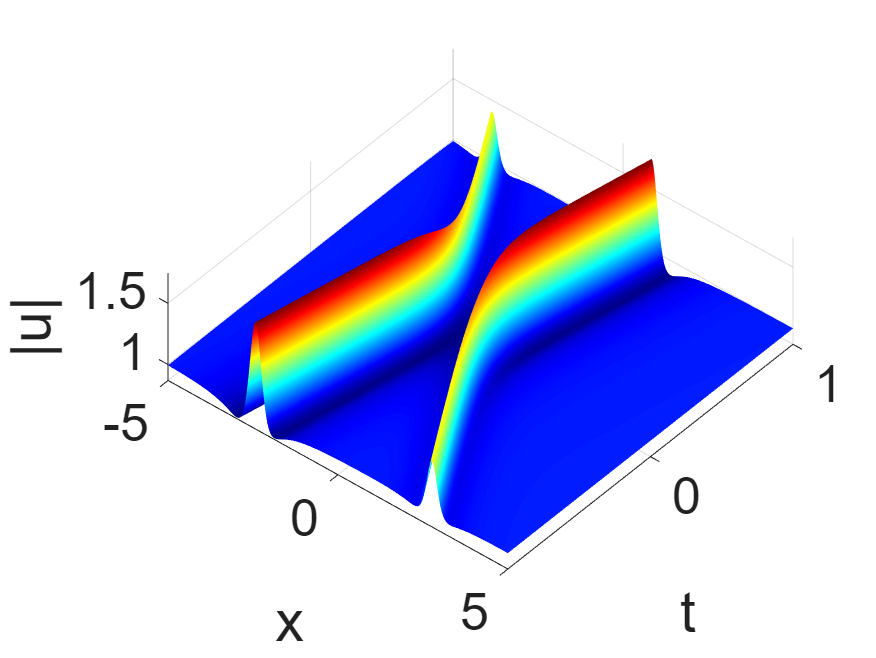}
    }
    \subfigure[]{
        \includegraphics[width=60mm]{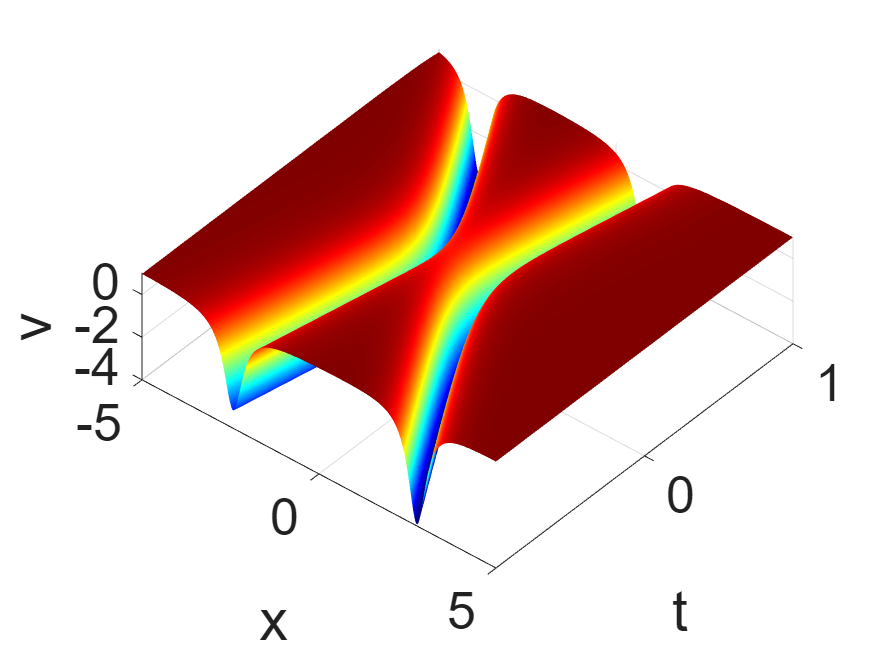}
    }
    \caption{Dark soliton solution to Eq.~\eqref{ss_mkdv_1}-\eqref{ss_mkdv_2} in \(h=0\) case with parameters \(c_1 = c_2 = 1, d_1 = d_2 = 1, \alpha = 1, p_1 = 1, p_2 = 0.5, p_3 \approx -2.20557, p_4 = -3.18546, \rho_1 = 1, \rho_2 = 1, \xi_{1,0} = \xi_{2,0} = 0\).}
    \label{fig:SS_mKdV_dark_N=4_MHs}
\end{figure}

\begin{figure}[!ht]
    \centering
    \subfigure[]{
        \includegraphics[width=60mm]{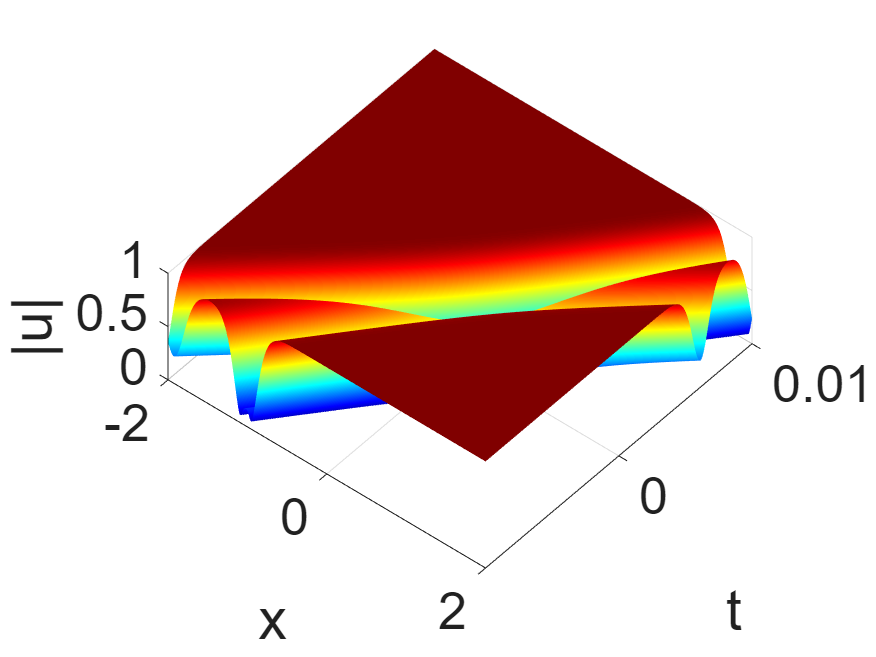}
    }
    \subfigure[]{
        \includegraphics[width=60mm]{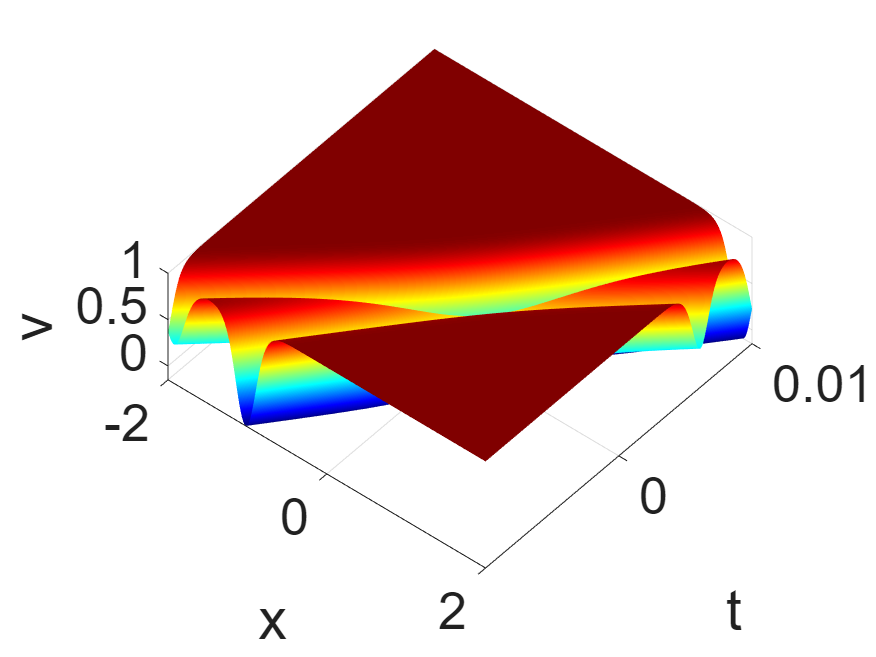}
    }
    \caption{Dark soliton solution to Eq.~\eqref{ss_mkdv_1}-\eqref{ss_mkdv_2} in \(h=2\) case with parameters \(c_1 = c_2 = 20, d_1 = d_2 = 1, \alpha = 1, p_1 \approx 6+4.9390\i, p_2 \approx 7 +3.2875\i, \rho_1 = 2, \rho_1 = 1, \xi_{1,0} = \xi_{2,0} = 0\).}
    \label{fig:SS_mKdV_dark_N=4_hole_doublehole}
\end{figure}

Furthermore, illustrations for \(n=4\) is also obtained (see \cref{fig:SS_mKdV_dark_N=4_MHs,fig:SS_mKdV_dark_N=4_hole_doublehole}).

\section{Dynamics of bright-dark solitons}\label{section:dynmics_bd}
\begin{figure}[!ht]
    \centering
    \subfigure[]{
        \includegraphics[width=60mm]{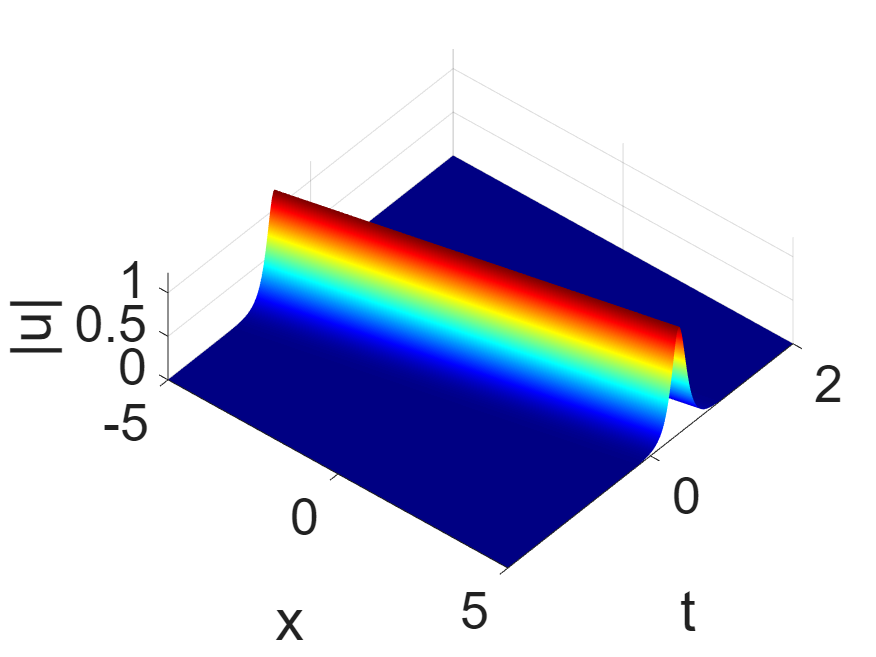}
    }
    \subfigure[]{
        \includegraphics[width=60mm]{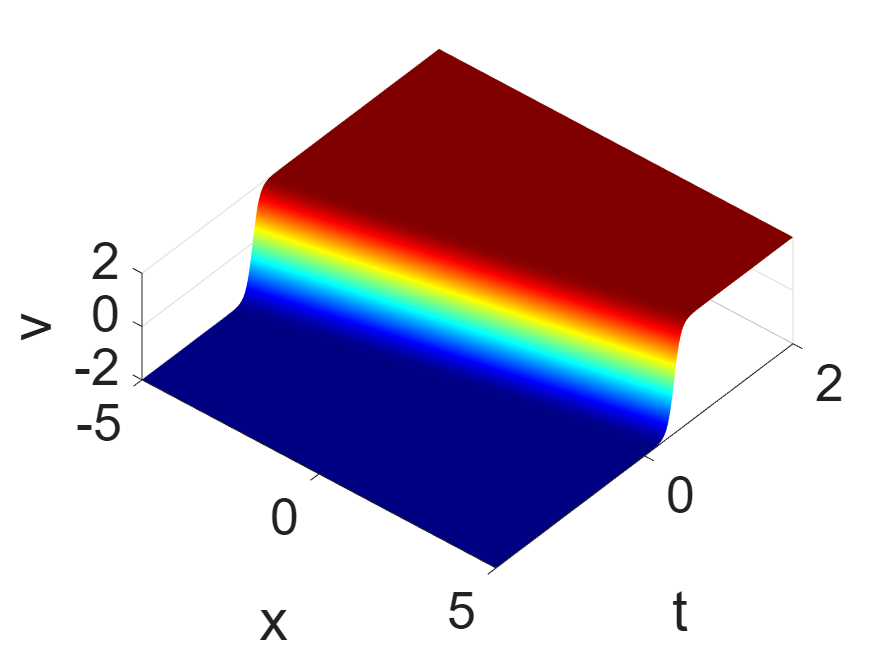}
    }
    \caption{One-bright-dark soliton solution to Eq.~\eqref{ss_mkdv_1}-\eqref{ss_mkdv_2} with parameters \(c_1 = c_2 = 1, p_1 = 1, \rho_2 = 2, C_1 = 1+2\i\).}
    \label{fig:SS_mKdV_bd_N=1}
\end{figure}

With bright-dark soliton solution given by \cref{thm:b-d}. 
  One-soliton solution (\(N=1\)) can be expressed by the following formula
\begin{align*}
    u &= \frac{C_1}{|C_1|} \sqrt{2\frac{c_2 \rho_2^2 - p_1^2}{c_1}} \sech \left(p_1 \left(x - 3c_2\rho_2^2 t\right) + p_1^3 t +\xi_{1,0} - \log \sqrt{\frac{2c_1 |C_1|^2}{(c_2 \rho_2^2)/p_1^2 - 1}}\right),\\
    v &= -\rho_2 \tanh \left(p_1 \left(x - 3c_2\rho_2^2 t\right) + p_1^3 t +\xi_{1,0} - \log \sqrt{\frac{2c_1 |C_1|^2}{(c_2 \rho_2^2)/p_1^2 - 1}}\right),
\end{align*}
where \(p_1,\xi_{1,0}\in \mathbb{R}\). To avoid singularity, we require \(\frac{2c_1 |C_1|^2}{(c_2 \rho_2^2)/p_1^2 - 1} > 0\). The energy intensity of above solution is
\begin{align*}
    N(u) = \int_{-\infty}^\infty |u|^2 \,dx = \frac{c_2 \rho_2^2 - p_1^2}{c_1 p_1}, \quad N(v) = \int_{-\infty}^\infty v^2 - \rho^2 \,dx = -\frac{2 \rho_2^2}{p_1}.
\end{align*}
An example  is illustrated in \cref{fig:SS_mKdV_bd_N=1}.

From \eqref{bd_cij}, we have \(c_{i,j}^* = c_{j,i}\) and hence when \(N=2\)
\begin{align*}
    c_{1,2} = c_{2,1}^* = \frac{c_1 C_1^* C_2+c_1 C_2 C_1^* }{(c_2 \rho_2^2)/(p_1 p_2^*) - 1}, \quad c_{1,1} = c_{2,2} = \frac{c_1 |C_1|^2 + c_1 |C_2|^2}{(c_2 \rho_2^2)/|p_1|^2 - 1} \in \mathbb{R}.
\end{align*}
Thus, the second order bright-dark soliton can be simplified as 
\begin{align*}
    f ={}& \left(\dfrac{1}{p_1 + p_1^*} \left(e^{\xi_1 + \xi_1^*} + c_{1,1}\right)\right)^2 - \frac{1}{|2p_1|^2} \left(e^{2\xi_1+2\xi_1^*} + c_{1,2}^* e^{2\xi_1} + c_{1,2} e^{2\xi_1^*} + |c_{1,2}|^2\right),\\
    g_1 ={}& \frac{C_1}{2 p_1 (p_1 + p_1^*)} \left(2 p_1 c_{1,1} \exp(\xi_1) - c_{1,2} (p_1 + p_1^*) \exp(\xi_1^*) + (p_1-p_1^*) \exp(2\xi_1 + \xi_1^*) \right)\\
    & + \frac{C_2}{2 p_1 (p_1 + p_1^*)} \left(2 p_1^* c_{1,1} \exp(\xi_1^*) - c_{1,2}^* (p_1 + p_1^*) \exp(\xi_1) + (p_1^*-p_1) \exp(\xi_1 + 2\xi_1^*) \right),\\
    h_2 ={}& \left|\dfrac{1}{p_1 + p_1^*} \left(-\dfrac{p_1}{p_1^*}e^{\xi_1 + \xi_1^*} + c_{1,1}\right)\right|^2 -  \frac{1}{|2p_1|^2} \left(e^{2\xi_1+2\xi_1^*} - c_{1,2}^* e^{2\xi_1} - c_{1,2} e^{2\xi_1^*} + |c_{1,2}|^2\right),
\end{align*}
where \(\xi_1 = p_1 \left(x-3 c_2 \rho_2^2 t\right)+p_1^3 t + \xi_{1,0}, \xi_2 = p_2 \left(x-3 c_2 \rho_2^2 t\right)+p_2^3 t + \xi_{1,0}\).

\begin{figure}[!ht]
    \centering
    \subfigure[]{
        \includegraphics[width=60mm]{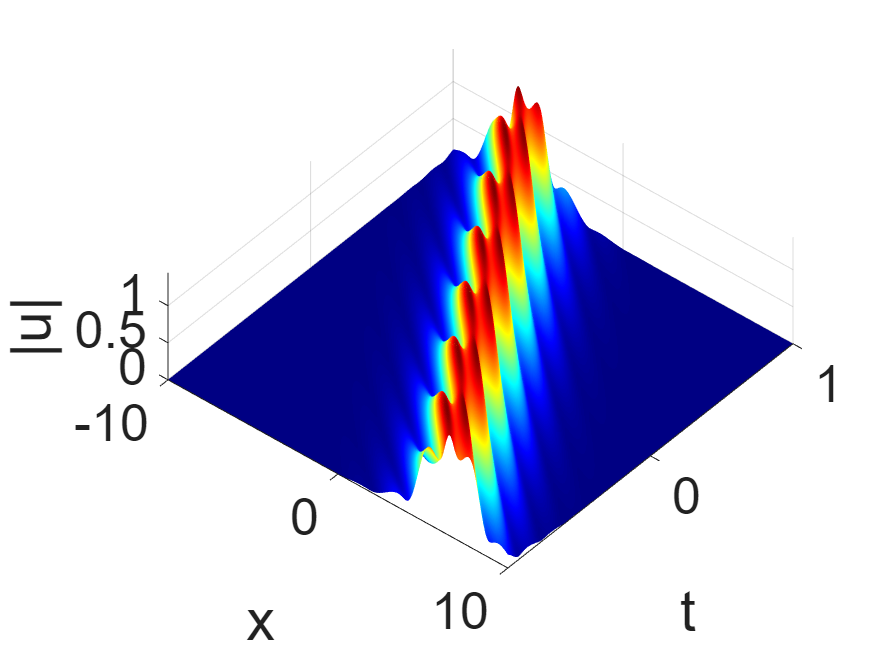}
    }
    \subfigure[]{
        \includegraphics[width=60mm]{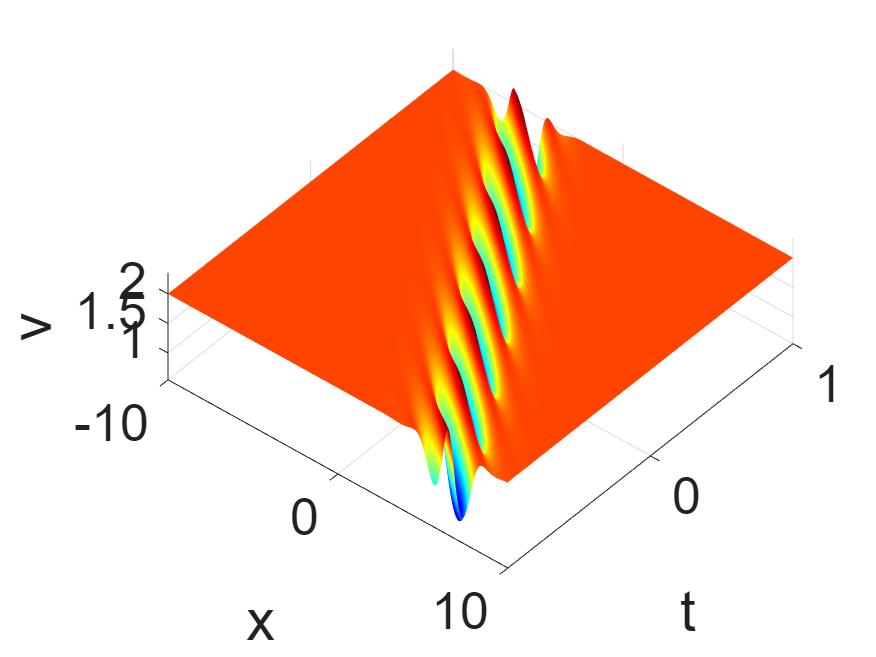}
    }
    \caption{Oscillated soliton and breather soliton solution to Eq.~\eqref{ss_mkdv_1}-\eqref{ss_mkdv_2} with parameters \(c_1 = c_2 = -1, p_1 = 1+1.5\i, \rho_2 = 2, C_1 = 1+2\i, C_2 = 1\).}
    \label{fig:SS_mKdV_bd_N=2}
\end{figure}

\begin{figure}[!ht]
    \centering
    \subfigure[]{
        \includegraphics[width=60mm]{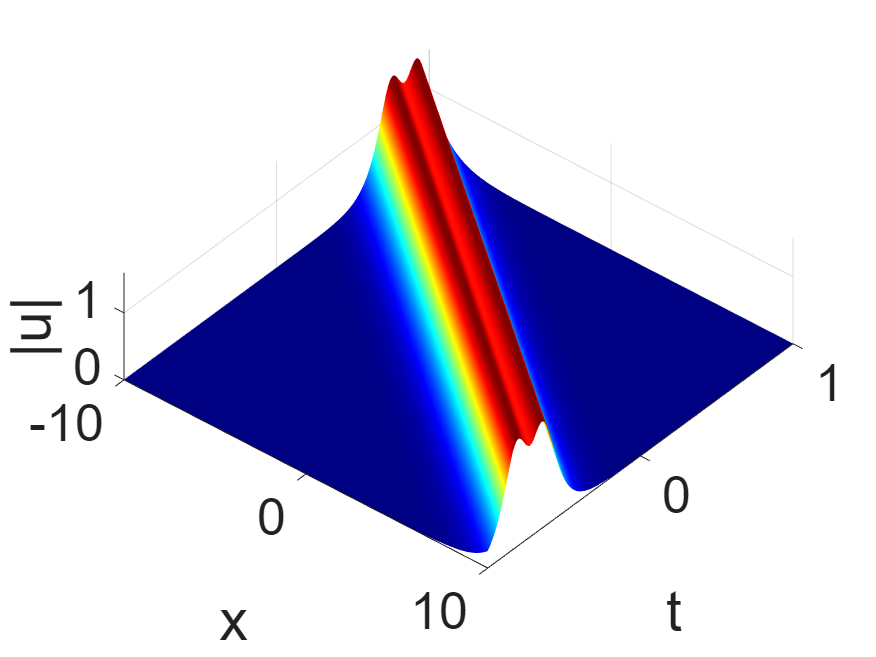}
    }
    \subfigure[]{
        \includegraphics[width=60mm]{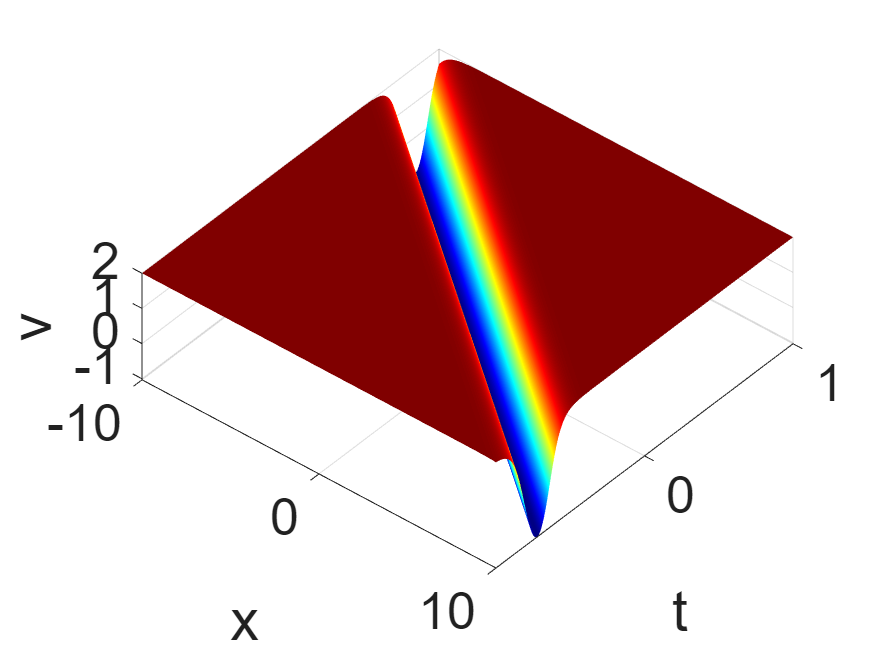}
    }
    \caption{Double hump and dark soliton solution to Eq.~\eqref{ss_mkdv_1}-\eqref{ss_mkdv_2} with parameters \(c_1 = c_2 = -1, p_1 = 1+0.25\i, \rho_2 = 2, C_1 = 1+2\i, C_2 = 0\).}
    \label{fig:SS_mKdV_bd_N=2_doublehump}
\end{figure}

Note that if \(p_1\) is not real, we have oscillated soliton and breather in each of the components \(u,v\) (see \cref{fig:SS_mKdV_bd_N=2}). \(C_1 = 0\) or \(C_2 = 0\) can lead to double-hump soliton in component \(u\) ( see \cref{fig:SS_mKdV_bd_N=2_doublehump}). Similar to the bright-bright case,  \(p_1\in \mathbb R\)  reduces above solution to the first order one.

\begin{figure}[!ht]
    \centering
    \subfigure[]{
        \includegraphics[width=60mm]{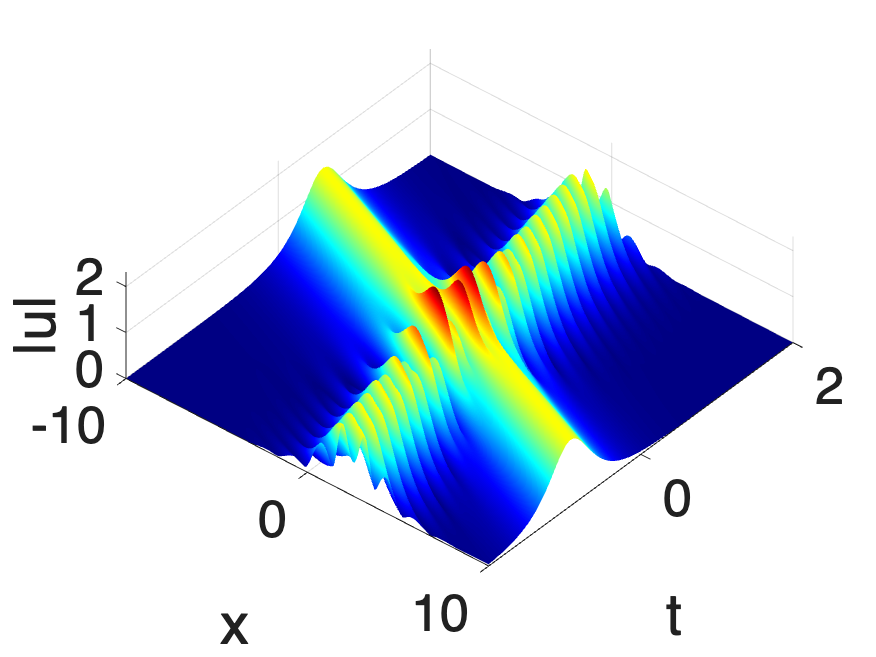}
    }
    \subfigure[]{
        \includegraphics[width=60mm]{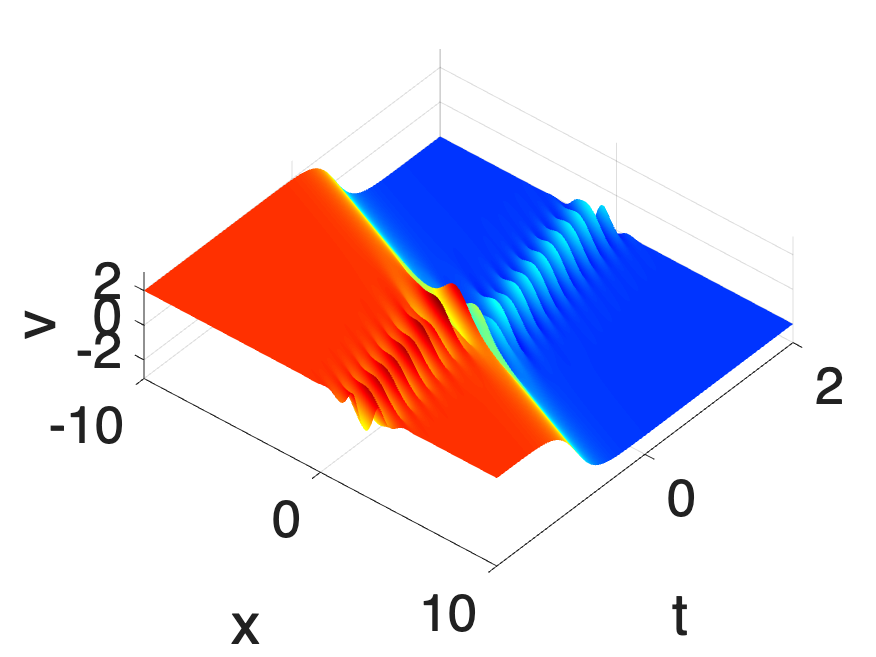}
    }
    \caption{Bright-dark soliton solution to Eq.~\eqref{ss_mkdv_1}-\eqref{ss_mkdv_2} with parameters \(c_1 = c_2 = -1, p_1 = 1+2\i, p_2 = \frac 1 3, \rho_2 = 2, C_1 = 1+\i, C_2 = 2-\i, C_3 = 1+2\i\).}
    \label{fig:SS_mKdV_bd_N=3}
\end{figure}

When \(N=3\), the parameter restrictions \eqref{cond_bd} gives \(p_1 = p_3^*, p_2 \in \mathbb R\), and collision between solitons can be observed (see \cref{fig:SS_mKdV_bd_N=3}). To discuss the asymptotic behavior in this case, we denote soliton 1 and soliton 2 corresponding to solitons which determined by \(\xi_1\) and \(\xi_2\) respectively. Assume soliton 1 is on the left of soliton 2 when \(t \to - \infty\). 

\begin{figure}[!ht]
    \centering
    \subfigure[]{
        \includegraphics[width=60mm]{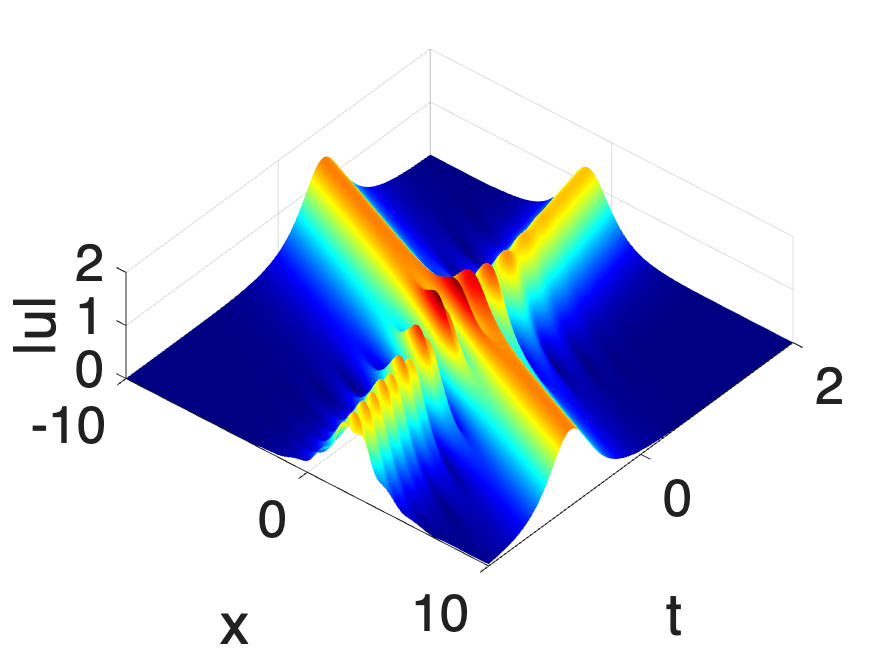}
    }
    \subfigure[]{
        \includegraphics[width=60mm]{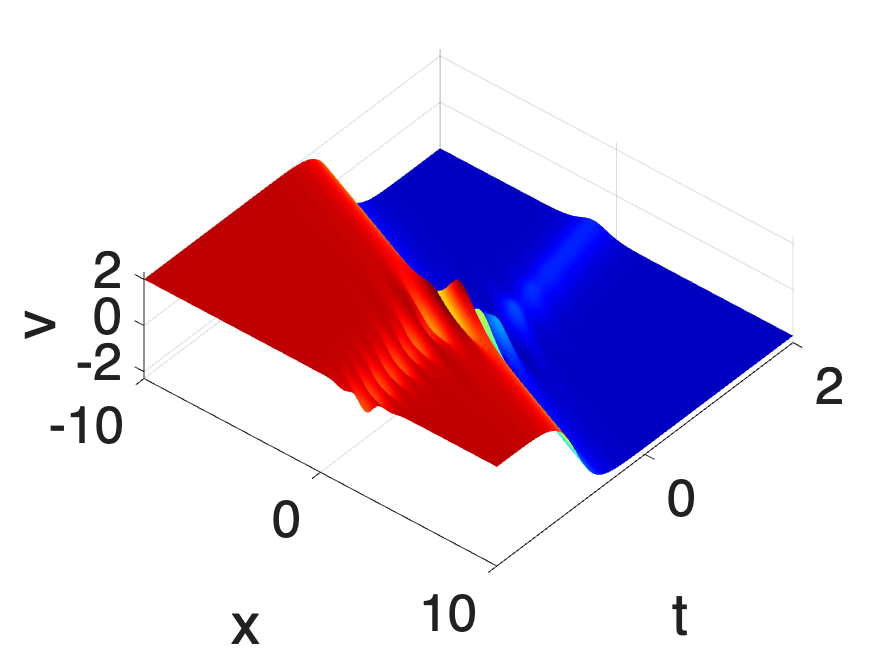}
    }
    \caption{Bright-dark soliton solution to Eq.~\eqref{ss_mkdv_1}-\eqref{ss_mkdv_2} with parameters \(c_1 = c_2 = -1, p_1 = 1+2\i, p_2 = \frac 1 3, \rho_2 = 2, C_1 = 0, C_2 = 2-\i, C_3 = 1+2\i\).}
    \label{fig:SS_mKdV_bd_N=3_shapechange}
\end{figure}

\begin{enumerate}[(1)]
    \item Before collision, i.e., \(t \rightarrow - \infty\)

    Soliton 1 (\(\xi_1 + \xi_1^* \approx 0,\ \xi_2\rightarrow - \infty\))
    {\allowdisplaybreaks
    \begin{align*}
        f &\simeq \begin{vmatrix}
        \dfrac{1}{p_1 + p_1^*} \left(e^{\xi_1+\xi_1^*} + c_{1, 1}\right) &
        \dfrac{c_{1, 2}}{p_1 + p_2} &
        \dfrac{1}{2p_1} \left(e^{2\xi_1} + c_{1, 3}\right)  \\
        \dfrac{c_{1, 2}^*}{p_1^* + p_2} &
        \dfrac{c_{2, 2}}{2p_2}  &
        \dfrac{c_{2, 3}}{p_1+p_2} \\
        \dfrac{1}{2p_1^*} \left(e^{2\xi_1^*} + c_{1, 3}^*\right)&
        \dfrac{c_{2, 3}^*}{p_1^*+p_2}  &
        \dfrac{1}{p_1 + p_1^*} \left(e^{\xi_1+\xi_1^*} + c_{3, 3}\right)
        \end{vmatrix},\\
        g_1 & \simeq \begin{vmatrix}
        \dfrac{1}{p_1 + p_1^*} \left(e^{\xi_1+\xi_1^*} + c_{1, 1}\right) &
        \dfrac{c_{1, 2}}{p_1 + p_2} &
        \dfrac{1}{2p_1} \left(e^{2\xi_1} + c_{1, 3}\right) & \exp(\xi_1) \\
        \dfrac{c_{1, 2}^*}{p_1^* + p_2} &
        \dfrac{c_{2, 2}}{2p_2}  &
        \dfrac{c_{2, 3}}{p_1+p_2} & 0 \\
        \dfrac{1}{2p_1^*} \left(e^{2\xi_1^*} + c_{1, 3}^*\right)&
        \dfrac{c_{2, 3}^*}{p_1^*+p_2} &
        \dfrac{1}{p_1 + p_1^*} \left(e^{\xi_1+\xi_1^*} + c_{3, 3}\right) & \exp(\xi_1^*) \\
         -C_1 & -C_2 & -C_3 &  0
        \end{vmatrix},\\
        h_2 &\simeq \begin{vmatrix}
        \dfrac{1}{p_1 + p_1^*} \left(-\dfrac{p_1}{p_1^*}e^{\xi_1+\xi_1^*} + c_{1, 1}\right) &
        \dfrac{c_{1, 2}}{p_1 + p_2} &
        \dfrac{1}{2p_1} \left(-e^{2\xi_1} + c_{1, 3}\right)  \\
        \dfrac{c_{1, 2}^*}{p_1^* + p_2} &
        \dfrac{c_{2, 2}}{2p_2}  &
        \dfrac{c_{2, 3}}{p_1+p_2} \\
        \dfrac{1}{2p_1^*} \left(-e^{2\xi_1^*} + c_{1, 3}^*\right)&
        \dfrac{c_{2, 3}^*}{p_1^*+p_2}  &
        \dfrac{1}{p_1 + p_1^*} \left(-\dfrac{p_1^*}{p_1}e^{\xi_1+\xi_1^*} + c_{3, 3}\right)
        \end{vmatrix}.
    \end{align*}
    }

    Soliton 2 (\(\xi_2\approx 0,\ \xi_1 + \xi_1^* \rightarrow + \infty\))
    \begin{align}
    \begin{split}
        u &= \frac {g_1} f \simeq \frac{ e^{\xi_2} C_1 }{e^{2 \xi_2} A_1  + B_1} = C_1 \sqrt{\frac {B_1} {A_1}} \sech \left(\xi_2 + \log \sqrt {\frac {A_1} {B_1}}\right),\\
        v &= \rho_2 \frac {h_2} f \simeq \rho_2 \frac{-e^{2 \xi_2} A_1 + B_1}{e^{2 \xi_2} A_1 + B_1} = -\rho_2 \tanh \left(\xi_2 + \log \sqrt {\frac {A_1} {B_1}}\right),
    \end{split}
    \end{align}
    where
    \begin{align*}
        A_1 &= (p_1-p_2)^2 (p_1^*-p_2)^2 ,\\
        B_1 &= c_{2,2} (p_1+p_2)^2 (p_1^*+p_2)^2,\\
        C_1 &= 2 C_2 p_2 (p_2-p_1) (p_1+p_2) (p_2-p_1^*) (p_1^*+p_2).
    \end{align*}

    \item After collision, i.e., \(t \rightarrow + \infty\)

    Soliton 1 (\(\xi_1 + \xi_1^* \approx 0,\ \xi_2\rightarrow + \infty\))
    \begin{align*}
        f &\simeq \begin{vmatrix}
        \dfrac{1}{p_1 + p_1^*} \left(e^{\xi_1 + \xi_1^*} + c_{1,1}\right) &
        \dfrac{1}{p_1 + p_2} e^{\xi_1} &
        \dfrac{1}{2p_1} \left(e^{2\xi_1} + c_{1,3}\right) \\
        \dfrac{1}{p_1^* + p_2} e^{\xi_1^*} &
        \dfrac{1}{2p_2} &
        \dfrac{1}{p_1+p_2} e^{\xi_1} \\
        \dfrac{1}{2p_1^*} \left(e^{2\xi_1^*} + c_{1,3}^*\right) &
        \dfrac{1}{p_1^*+p_2} e^{\xi_1^*} &
        \dfrac{1}{p_1 + p_1^*} \left(e^{\xi_1 + \xi_1^*} + c_{1,1}\right)
        \end{vmatrix},\\
        g_1 &\simeq \begin{vmatrix}
        \dfrac{1}{p_1 + p_1^*} \left(e^{\xi_1 + \xi_1^*} + c_{1,1}\right) &
        \dfrac{1}{p_1 + p_2} e^{\xi_1} &
        \dfrac{1}{2p_1} \left(e^{2\xi_1} + c_{1,3}\right) & \exp(\xi_1) \\
        \dfrac{1}{p_1^* + p_2} e^{\xi_1^*} &
        \dfrac{1}{2p_2} &
        \dfrac{1}{p_1+p_2} e^{\xi_1} & 1 \\
        \dfrac{1}{2p_1^*} \left(e^{2\xi_1^*} + c_{1,3}^*\right) &
        \dfrac{1}{p_1^*+p_2} e^{\xi_1^*} &
        \dfrac{1}{p_1 + p_1^*} \left(e^{\xi_1 + \xi_1^*} + c_{3,3}\right) & \exp(\xi_1^*) \\
        -C_1 & 0 & -C_3 &  0
        \end{vmatrix},\\
        h_2 &\simeq \begin{vmatrix}
        \dfrac{1}{p_1 + p_1^*} \left(-\dfrac{p_1}{p_1^*}e^{\xi_1 + \xi_1^*} + c_{1,1}\right) &
        -\dfrac{p_1} {p_2}\dfrac{1}{p_1 + p_2} e^{\xi_1}  &
        \dfrac{1}{2p_1} \left(-e^{2\xi_1} + c_{1,3}\right) \\
        -\dfrac{p_2} {p_1^*}\dfrac{1}{p_1^* + p_2} e^{\xi_1^*}  &
        -\dfrac{1}{2p_2} &
        -\dfrac{p_2} {p_1}\dfrac{1}{p_1+p_2} e^{\xi_1} \\
        \dfrac{1}{2p_1^*} \left(-e^{2\xi_1^*} + c_{1,3}^*\right) &
        -\dfrac{p_1^*} {p_2}\dfrac{1}{p_1^*+p_2} e^{\xi_1^*} &
        \dfrac{1}{p_1 + p_1^*} \left(-\dfrac{p_1^*}{p_1}e^{\xi_1 + \xi_1^*} + c_{3,3}\right)
        \end{vmatrix}.
    \end{align*}

    Soliton 2 (\(\xi_2\approx 0,\ \xi_1 + \xi_1^* \rightarrow - \infty\))
    \begin{align*}
        u &= \frac {g_1} f \simeq \frac{e^{\xi_2} C_2}{e^{2\xi_2} A_2 + B_2} = C_2 \sqrt{\frac {B_2} {A_2}} \sech \left(\xi_2 + \log \sqrt{\frac {A_2} {B_2}}\right),\\
        v &= \rho_2 \frac {h_2} f \simeq \rho_2\frac{-e^{2\xi_2} A_2+ B_2}{e^{2\xi_2} A_2 + B_2} = -2\rho_2 \tanh \left(\xi_2 + \log \sqrt {\frac {A_2} {B_2}}\right),
    \end{align*}
    where
    \begin{align*}
        A_2 ={}& \frac 1 {2p_2} \left(\frac{c_{1,1} c_{3,3}}{(p_1+p_1^*)^2}-\frac{c_{1,3} \text{c}_{3,1}}{4 p_1 p_1^*}\right),\\
        B_2 ={}& c_{2,2} A - \frac{c_{1,2}^*}{p_1^*+p_2} \left(\frac{c_{1,2} c_{3,3}}{(p_1+p_1^*) (p_1+p_2)}-\frac{c_{1,3} c_{2,3}^*}{2 p_1 (p_1^*+p_2)}\right) \\
        &- \frac{c_{2,3}}{p_1+p_2} \left(\frac{c_{1,1} c_{2,3}^*}{(p_1+p_1^*) (p_1^*+p_2)}-\frac{c_{1,2} \text{c}_{3,1}}{2 p_1^* (p_1+p_2)}\right),\\
        C_2 ={}& \frac{C_1 c_{1,3} c_{2,3}^*}{2 p_1 (p_1^*+p_2)}-\frac{C_1 c_{1,2} c_{3,3}}{(p_1+p_1^*) (p_1+p_2)}-\frac{C_2 c_{1,3} \text{c}_{3,1}}{4 p_1 p_1^*}\\
        &+\frac{C_2 c_{1,1} c_{3,3}}{(p_1+p_1^*)^2}+\frac{C_3 c_{1,2} \text{c}_{3,1}}{2 p_1^* (p_1+p_2)}-\frac{C_3 c_{1,1} c_{2,3}^*}{(p_1+p_1^*) (p_1^*+p_2)}.
    \end{align*}
\end{enumerate}
Unlike the case of bright-bright soliton solution, we cannot have the Y-shaped solutions by taking \(C_1 = 0\) or \(C_3 = 0\) in above bright-dark soliton solution. Instead, the breather can change into bright soliton by interacting with kink (see \cref{fig:SS_mKdV_bd_N=3_shapechange}). For example, taking \(C_1 = 0\), soliton 2 after collision becomes
\begin{align*}
    u &\simeq \frac{-2 C_3 e^{\xi_1^*} p_1 (p_1+p_1^*) (p_1+p_2) (p_1^*-p_2) \left(e^{\xi_1 + \xi_1^*} (p_1-p_1^*) |p_1-p_2|^2-2 p_1^* c_{1,1} |p_1+p_2|^2\right)}
    {4 |p_1|^2 |p_1+p_2|^2 \left(\left(c_{1,1}+c_{3,3}\right) e^{\xi_1 + \xi_1^*} |p_1-p_2|^2+c_{1,1} c_{3,3} |p_1+p_2|^2\right)-e^{2 (\xi_1 + \xi_1^*)} (p_1-p_1^*)^2 |p_1-p_2|^4},\\
    v &\simeq \rho_2 \frac{4 |p_1+p_2|^2 \left(e^{\xi_1 + \xi_1^*} |p_1-p_2|^2 \left(p_1^2 c_{3,3}+(p_1^*)^2 c_{1,1}\right)-|p_1|^2 c_{1,1} c_{3,3} |p_1+p_2|^2\right)+e^{2 (\xi_1 + \xi_1^*)} (p_1-p_1^*)^2 |p_1-p_2|^4}{4 |p_1|^2 |p_1+p_2|^2 \left(\left(c_{1,1}+c_{3,3}\right) e^{\xi_1 + \xi_1^*} |p_1-p_2|^2+c_{1,1} c_{3,3} |p_1+p_2|^2\right)-e^{2 (\xi_1 + \xi_1^*)} (p_1-p_1^*)^2 |p_1-p_2|^4},
\end{align*}
note that above \(u\) is in a similar form to the single/double hump soliton solution in \cite{sasa1991new,shi2025general}.

\begin{figure}[!ht]
    \centering
    \subfigure[]{
        \includegraphics[width=60mm]{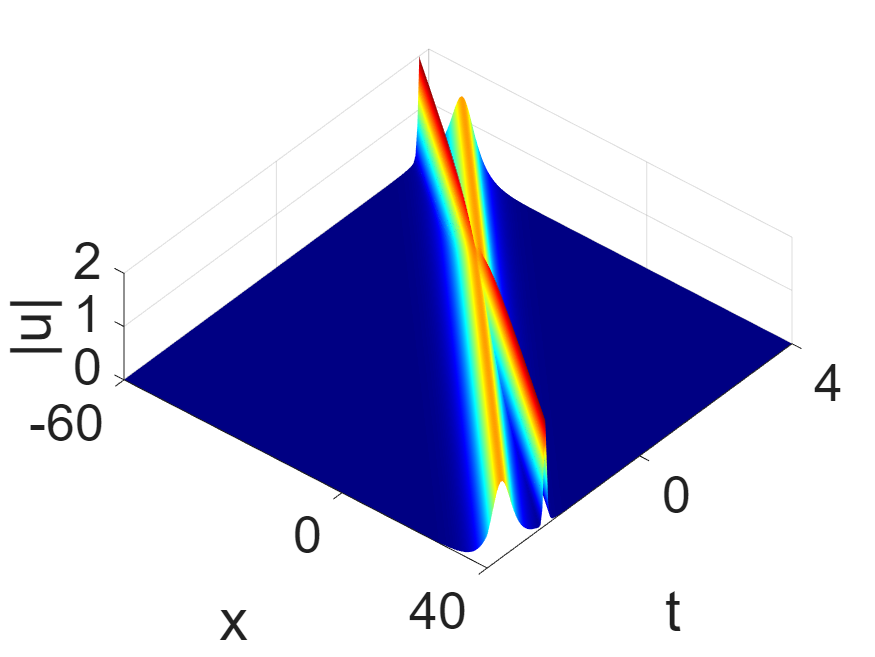}
    }
    \subfigure[]{
        \includegraphics[width=60mm]{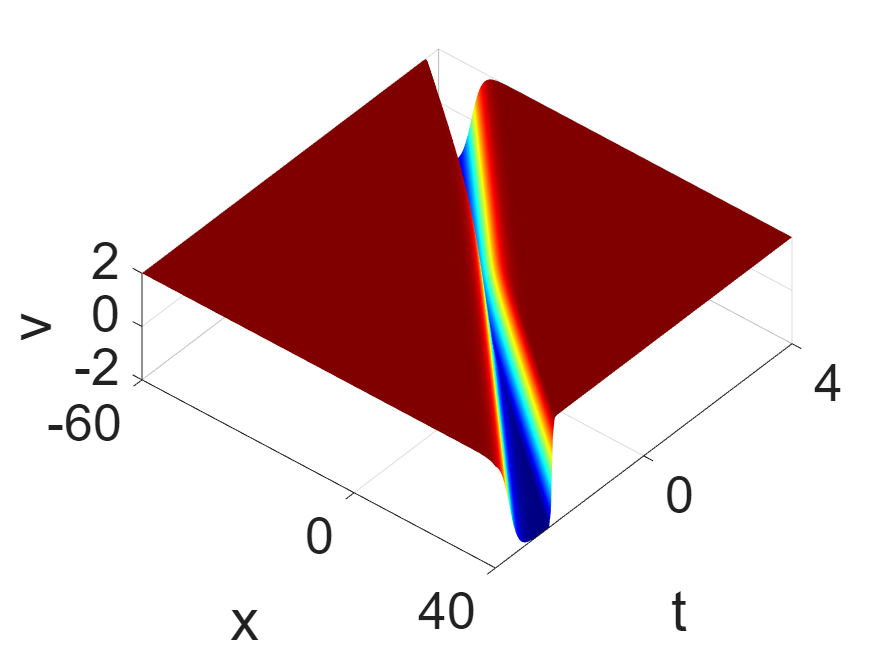}
    }
    \caption{Bright-dark soliton solution to Eq.~\eqref{ss_mkdv_1}-\eqref{ss_mkdv_2} with parameters \(c_1 = c_2 = -1, p_1 = 2, p_2 = \frac 1 3, \rho_2 = 2, C_1 = 1+\i, C_2 = 2-\i, C_3 = 1+2\i\).}
    \label{fig:SS_mKdV_bd_N=3_realcase}
\end{figure}

Now, let us discuss the case for \(p_1 \in \mathbb R\). It is particularly noted that, soliton 1 of component \(v\) reduced to the first order solution, i.e., a hyperbolic tangent shaped soliton. We have an interaction between two kinks (see \cref{fig:SS_mKdV_bd_N=3_realcase}). Let us take the following parameters for example,
\begin{align}\label{bd_n3_preal_conds}
    p_1 = 2, \quad p_2 = \frac 1 3, \quad c_1 = c_2 = -1, \quad \rho_2 = 2, \quad C_1 = 1+\i, \quad C_2 = 2-\i, \quad C_3 = 1+2\i,
\end{align}
and the asymptotic expression of soliton 1 and soliton 2 being
\begin{enumerate}[(1)]
    \item Before collision, i.e., \(t \rightarrow - \infty\)

    Soliton 1 (\(\xi_1 + \xi_1^* \approx 0,\ \xi_2\rightarrow - \infty\))
    \begin{align*}
        u &\simeq \frac{(88800+30600\i) e^{\xi_1}}{19604 e^{2 \xi_1}+28125} = (88800+30600\i) \sqrt{\frac{19604}{28125}}\sech \left(\xi_1 + \log \sqrt{\frac{19604}{28125}}\right),\\
        v &\simeq \rho_2\frac{28125-19604 e^{2 \xi_1}}{19604 e^{2 \xi_1}+28125} = -\rho_2 \tanh \left(\xi_1 + \log \sqrt{\frac{19604}{28125}}\right).
    \end{align*}

    Soliton 2 (\(\xi_2\approx 0,\ \xi_1 + \xi_1^* \rightarrow + \infty\))
    \begin{align*}
        u &\simeq -\frac{(129500-126910\i) e^{\xi_2}}{135975 e^{2 \xi_2}+29406} = -\frac{(129500-126910\i)}{3} \sqrt{\frac{45325}{9802}}\sech \left(\xi_2 + \log \sqrt{\frac{45325}{9802}}\right),\\
        v &\simeq \rho_2 \frac{45325 e^{2 \xi_2}-9802}{45325 e^{2 \xi_2}+9802} = \rho_2 \tanh \left(\xi_2 + \log \sqrt{\frac{45325}{9802}}\right).
    \end{align*}

    \item After collision, i.e., \(t \rightarrow + \infty\)

    Soliton 1 (\(\xi_1 + \xi_1^* \approx 0,\ \xi_2\rightarrow + \infty\))
    \begin{align*}
        u &\simeq \frac{840\i e^{\xi_1}}{100 e^{2 \xi_1}+441} = 840\i \sqrt{\frac{100}{441}} \sech \left(\xi_1 + \log \sqrt{\frac{100}{441}}\right),\\
        v &\simeq \rho_2\frac{100 e^{2 \xi_1}-441}{100 e^{2 \xi_1}+441} = \rho_2 \tanh \left(\xi_1 + \log \sqrt{\frac{100}{441}}\right).
    \end{align*}

    Soliton 2 (\(\xi_2\approx 0,\ \xi_1 + \xi_1^* \rightarrow - \infty\))
    \begin{align*}
        u &\simeq \frac{(88800+30600\i) e^{\xi_1}}{19604 e^{2 \xi_1}+28125} = (88800+30600\i) \sqrt{\frac{19604}{30600}} \sech \left(\xi_2 + \log \sqrt{\frac{19604}{30600}}\right),\\
        v &\simeq \rho_2\frac{28125-19604 e^{2 \xi_1}}{19604 e^{2 \xi_1}+28125} = -\rho_2 \tanh \left(\xi_2 + \log \sqrt{\frac{19604}{30600}}\right).
    \end{align*}
\end{enumerate}

\section{Dynamics of dark-bright solitons}\label{section:dynmics_db}
The following first order dark-bright soliton solution can be obtained from 
\cref{thm:d-b} for \(N=1\) 
\begin{align*}
    u &= \frac{\rho_1 \exp(\i \theta_1)}{p_1 + \i \alpha} \left(\i \alpha - p_1\tanh\left(p_1 (x - 3(2c_1\rho_1^2 + c_2\rho_2^2) t) + p_1^3 t + \xi_{1,0} - \log\sqrt{\frac{c_2 D_1^2 (p_1^2 + \alpha^2)}{2c_1 \rho^2 - p_1^2 - \alpha^2}}\right) \right),\\
    v &= -\frac{D_1}{|D_1|}\frac{2p_1}{\sqrt{c_2}} \sqrt{\dfrac{2c_1 \rho^2}{p_1^2 + \alpha^2} - 1} \sech \left(p_1 (x - 3(2c_1\rho_1^2 + c_2\rho_2^2) t) + p_1^3 t + \xi_{1,0} - \log\sqrt{\frac{c_2 D_1^2 (p_1^2 + \alpha^2)}{2c_1 \rho^2 - p_1^2 - \alpha^2}}\right).
\end{align*}
Where \(p_1,\xi_{0,1}, D_1 \in \mathbb R\) and \(\theta = \alpha x - \left(\alpha^3 + 6c_1\alpha\rho_1^2\right) t\). The background intensity of components \(u\), \(v\) are
\begin{align*}
    N(u) = \int_{-\infty}^{\infty} \left(\left|u\right|^2 - \rho_1^2\right)\, dx = -\frac{2\rho_1^2 p_1}{p_1^2 + \alpha^2}, \quad N(v) = \int_{-\infty}^{\infty} v^2\,dx = \frac{2 p_1}{c_2}\left(\frac{2 c_1 \rho_1^2}{p_1^2+\alpha^2}-1\right).
\end{align*}

\begin{figure}[!ht]
    \centering
    \subfigure[]{
        \includegraphics[width=60mm]{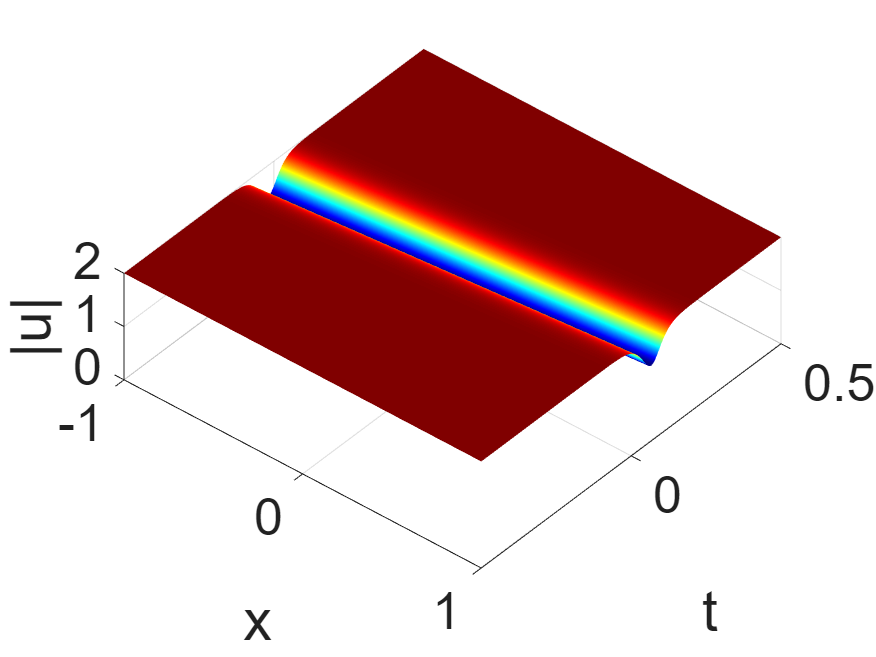}
    }
    \subfigure[]{
        \includegraphics[width=60mm]{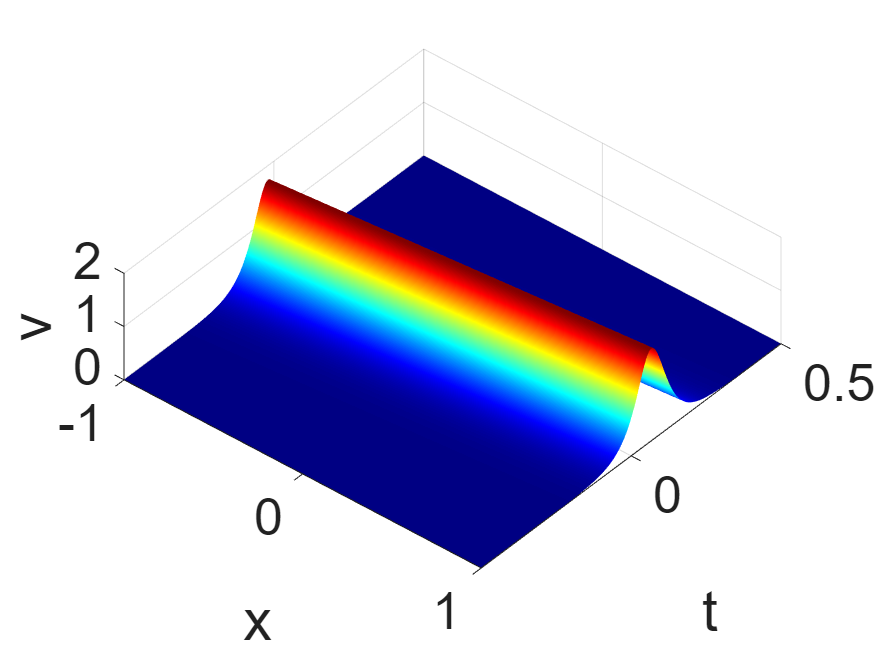}
    }
    \caption{One-dark-one-bright soliton solution to Eq.~\eqref{ss_mkdv_1}-\eqref{ss_mkdv_2} with parameters \(c_1 = c_2 = 1, p_1 = 1, \rho_1 = 1, \alpha=1, D_1 = 1\).}
    \label{fig:SS_mKdV_db_N=1}
\end{figure}
\begin{figure}[!ht]
    \centering
    \subfigure[]{
        \includegraphics[width=60mm]{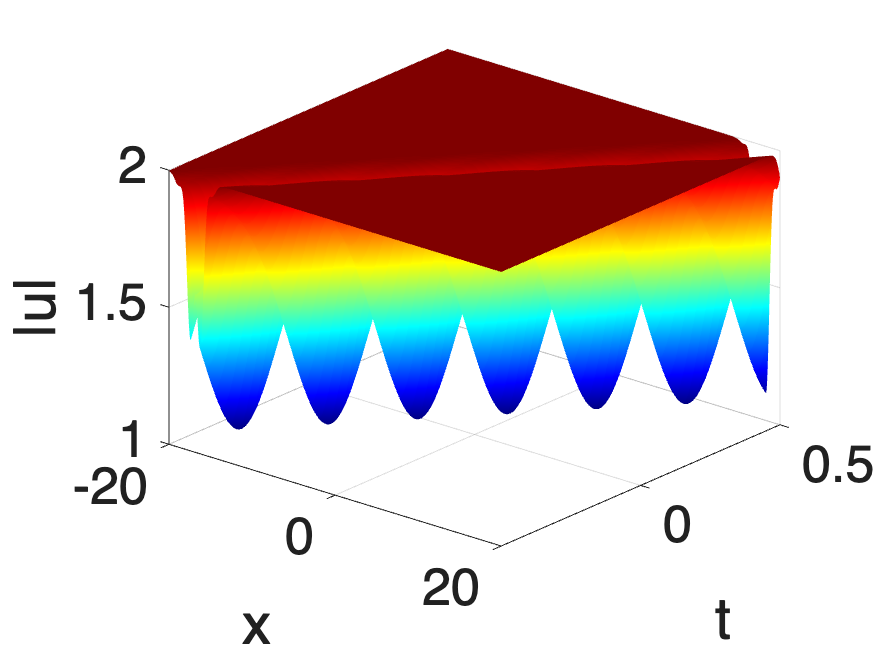}
    }
    \subfigure[]{
        \includegraphics[width=60mm]{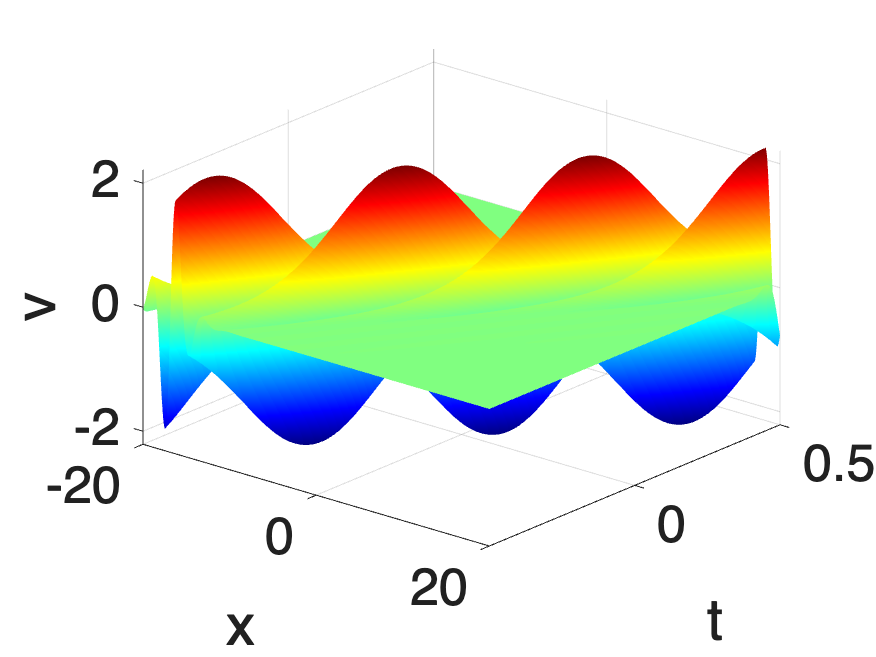}
    }
    \caption{Breather and oscillated soliton solution to Eq.~\eqref{ss_mkdv_1}-\eqref{ss_mkdv_2} with parameters \(c_1 = c_2 = 1, p_1 = 1+2\i, \rho_1 = 2, \alpha=1, D_1 = 2+1\i\).}
    \label{fig:SS_mKdV_db_N=2}
\end{figure}

An example is illustrated in  \cref{fig:SS_mKdV_db_N=1}.
For \(N=2\) case, we have breather \(u\)-component and oscillated soliton in  \(v\)-component (see \cref{fig:SS_mKdV_db_N=2}). The  solution is expressed as
\begin{align}
    f ={}& \left(\dfrac{1}{p_1 + p_1^*} \left(e^{\xi_1 + \xi_1^*} + c_{1,1}\right)\right)^2 - \frac{1}{|2p_1|^2} \left(e^{2\xi_1+2\xi_1^*} + c_{1,2}^* e^{2\xi_1} + c_{1,2} e^{2\xi_1^*} + |c_{1,2}|^2\right),\\
    h_1 ={}& \dfrac{1}{(p_1 + p_1^*)^2} \left|\left(-\frac{p_1 - \i \alpha}{p_1^*+\i \alpha}\right)e^{\xi_1 + \xi_1^*}
    + c_{1,1}\right|^2 - \frac{1}{|2p_1|^2}\left(\left|\frac{p_1 - \i \alpha}{p_1^*+\i \alpha}\right|^2e^{2\xi_1+2\xi_1^*} \right. \nonumber \\
    & \left. -c_{1,2}^*\frac{p_1 - \i \alpha}{p_1+\i \alpha} e^{2\xi_1} - c_{1,2}\frac{p_1^* - \i \alpha}{p_1^*+\i \alpha} e^{2\xi_1^*} + |c_{1,2}|^2\right),\\
    g_2 = {}& \frac{D_1}{2 p_1 (p_1 + p_1^*)} \left(2 p_1 c_{1,1} \exp(\xi_1) - c_{1,2} (p_1 + p_1^*) \exp(\xi_1^*) + (p_1-p_1^*) \exp(2\xi_1 + \xi_1^*) \right)\\
    & + \frac{D_1^*}{2 p_1 (p_1 + p_1^*)} \left(2 p_1^* c_{1,1} \exp(\xi_1^*) - c_{1,2}^* (p_1 + p_1^*) \exp(\xi_1) + (p_1^*-p_1) \exp(\xi_1 + 2\xi_1^*) \right)
\end{align}
where
\begin{align*}
    c_{1,1} = \frac{c_2 |D_1|^2}{\frac{2c_1 (|p_1|^2 + \alpha^2) \rho_1^2}{|p_1^2 + \alpha^2|^2} - 1} \in \mathbb{R}, \quad c_{1,2} = \frac{c_2 (D_1^*)^2}{\frac{2c_1 \rho_1^2}{p_1^2 + \alpha^2} - 1}.
\end{align*}
Note that if we take \(p_1 \in \mathbb R\), above solution does not degenerate to \(N=1\) form, instead, we have trivial solution
\[
    u = -\frac{p_1 - \i \alpha}{p_1 + \i \alpha}\rho_1 e^{\i \theta_1}, \quad v = 0.
\]

\begin{figure}[!ht]
    \centering
    \subfigure[]{
        \includegraphics[width=60mm]{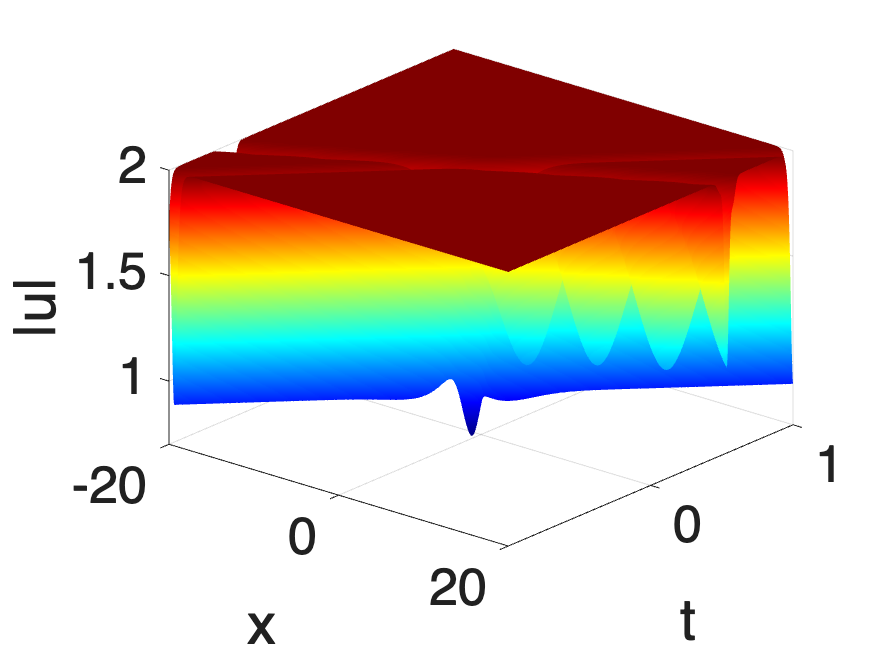}
    }
    \subfigure[]{
        \includegraphics[width=60mm]{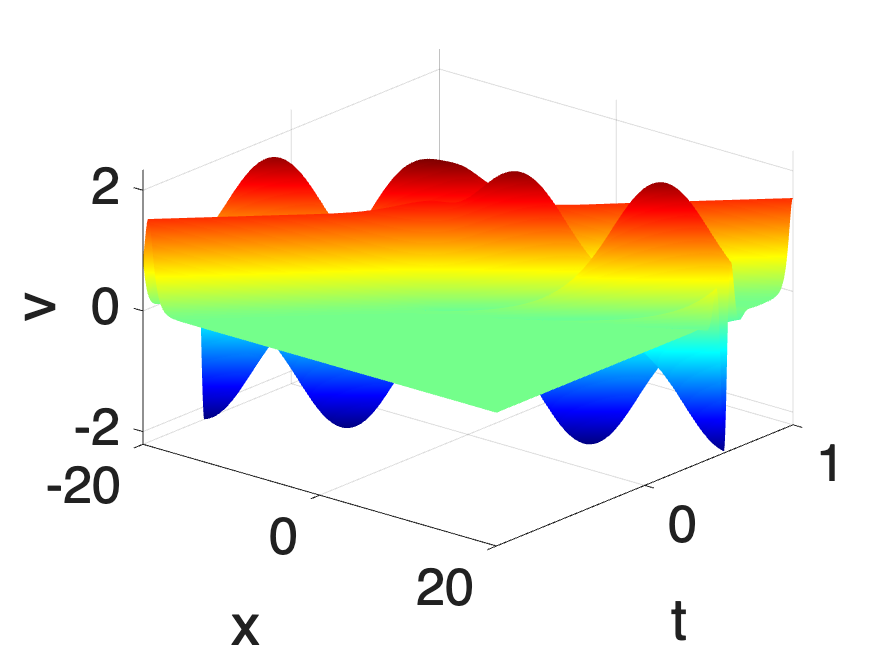}
    }
    \caption{Dark-bright soliton solution to Eq.~\eqref{ss_mkdv_1}-\eqref{ss_mkdv_2} with parameters \(c_1 = c_2 = 1, p_1 = 1+2\i, p_2 = 2, \rho_2 = 2, \alpha=1, D_1 = 2+1i, D_2 = 1\).}
    \label{fig:SS_mKdV_db_N=3}
\end{figure}

Next, let us consider the \(N=3\) case, where the collision between breather/oscillated soliton and regular soliton is observed (see \cref{fig:SS_mKdV_db_N=3}). We would like to perform the asymptotic analysis to investigate this solution. Denote soliton 1 and soliton 2 corresponding to solitons which determined by \(\xi_1\) and \(\xi_2\) respectively, and assume soliton 1 is on the left of soliton 2 when \(t \to - \infty\).
\begin{enumerate}[(1)]
    \item Before collision, i.e., \(t \rightarrow - \infty\)

    Soliton 1 (\(\xi_1 + \xi_1^* \approx 0,\ \xi_2\rightarrow - \infty\))
    {\allowdisplaybreaks
    \begin{align*}
        f &\simeq \begin{vmatrix}
        \dfrac{1}{p_1 + p_1^*} \left(e^{\xi_1+\xi_1^*} + c_{1, 1}\right) &
        \dfrac{c_{1, 2}}{p_1 + p_2} &
        \dfrac{1}{2p_1} \left(e^{2\xi_1} + c_{1, 3}\right)  \\
        \dfrac{c_{1, 2}^*}{p_1^* + p_2} &
        \dfrac{c_{2, 2}}{2p_2}  &
        \dfrac{c_{2, 3}}{p_1+p_2} \\
        \dfrac{1}{2p_1^*} \left(e^{2\xi_1^*} + c_{1, 3}^*\right)&
        \dfrac{c_{2, 3}^*}{p_1^*+p_2}  &
        \dfrac{1}{p_1 + p_1^*} \left(e^{\xi_1+\xi_1^*} + c_{3, 3}\right)
        \end{vmatrix}\\
        g_1 & \simeq \begin{vmatrix}
        \dfrac{1}{p_1 + p_1^*} \left(-\dfrac{p_1-\i \alpha}{p_1^* + \i \alpha}e^{\xi_1+\xi_1^*} + c_{1, 1}\right) &
        \dfrac{c_{1, 2}}{p_1 + p_2} &
        \dfrac{1}{2p_1} \left(-\dfrac{p_1-\i \alpha}{p_1 + \i \alpha}e^{2\xi_1} + c_{1, 3}\right)  \\
        \dfrac{c_{1, 2}^*}{p_1^* + p_2} &
        \dfrac{c_{2, 2}}{2p_2}  &
        \dfrac{c_{2, 3}}{p_1+p_2} \\
        \dfrac{1}{2p_1^*} \left(-\dfrac{p_1^* -\i \alpha}{p_1^* +\i \alpha}e^{2\xi_1^*} + c_{1, 3}^*\right)&
        \dfrac{c_{2, 3}^*}{p_1^*+p_2}  &
        \dfrac{1}{p_1 + p_1^*} \left(-\dfrac{p_1^* -\i \alpha}{p_1 + \i \alpha}e^{\xi_1+\xi_1^*} + c_{3, 3}\right)
        \end{vmatrix},\\
        h_2 &\simeq \begin{vmatrix}
        \dfrac{1}{p_1 + p_1^*} \left(e^{\xi_1+\xi_1^*} + c_{1, 1}\right) &
        \dfrac{c_{1, 2}}{p_1 + p_2} &
        \dfrac{1}{2p_1} \left(e^{2\xi_1} + c_{1, 3}\right) & \exp(\xi_1) \\
        \dfrac{c_{1, 2}^*}{p_1^* + p_2} &
        \dfrac{c_{2, 2}}{2p_2}  &
        \dfrac{c_{2, 3}}{p_1+p_2} & 0 \\
        \dfrac{1}{2p_1^*} \left(e^{2\xi_1^*} + c_{1, 3}^*\right)&
        \dfrac{c_{2, 3}^*}{p_1^*+p_2} &
        \dfrac{1}{p_1 + p_1^*} \left(e^{\xi_1+\xi_1^*} + c_{3, 3}\right) & \exp(\xi_1^*) \\
         -D_1 & -D_2 & -D_1^* &  0
        \end{vmatrix}.
    \end{align*}
    }

    Soliton 2 (\(\xi_2\approx 0,\ \xi_1 + \xi_1^* \rightarrow + \infty\))
    \begin{align*}
        f &\simeq \begin{vmatrix}
        \dfrac{1}{p_1 + p_1^*} &
        \dfrac{1}{p_1 + p_2} e^{\xi_2} &
        \dfrac{1}{2p_1}  \\
        \dfrac{1}{p_1^* + p_2}  e^{\xi_2} &
        \dfrac{1}{2p_2} \left( e^{2\xi_2} + c_{2, 2} \right) &
        \dfrac{1}{p_1+p_2}  e^{\xi_2}\\
        \dfrac{1}{2p_1^*} &
        \dfrac{1}{p_1^*+p_2} e^{\xi_2} &
        \dfrac{1}{p_1 + p_1^*}
        \end{vmatrix}\\
        g_1 & \simeq \begin{vmatrix}
        -\dfrac{p_1-\i \alpha}{p_1^* + \i \alpha}\dfrac{1}{p_1 + p_1^*} &
        -\dfrac{p_1-\i \alpha}{p_2 + \i \alpha}\dfrac{1}{p_1 + p_2} e^{\xi_2} &
        -\dfrac{p_1-\i \alpha}{p_1 + \i \alpha}\dfrac{1}{2p_1}  \\
        -\dfrac{p_2-\i \alpha}{p_1^* + \i \alpha}\dfrac{1}{p_1^* + p_2}  e^{\xi_2} &
        \dfrac{1}{2p_2} \left(-\dfrac{p_2-\i \alpha}{p_2 + \i \alpha}e^{2\xi_2} + c_{2, 2} \right) &
        -\dfrac{p_2-\i \alpha}{p_1 + \i \alpha}\dfrac{1}{p_1+p_2}  e^{\xi_2}\\
        -\dfrac{p_1^*-\i \alpha}{p_1^* + \i \alpha}\dfrac{1}{2p_1^*} &
        -\dfrac{p_1^*-\i \alpha}{p_2 + \i \alpha}\dfrac{1}{p_1^*+p_2} e^{\xi_2} &
        -\dfrac{p_1^*-\i \alpha}{p_1 + \i \alpha}\dfrac{1}{p_1 + p_1^*}
        \end{vmatrix},\\
        h_2 &\simeq \begin{vmatrix}
        \dfrac{1}{p_1 + p_1^*} &
        \dfrac{1}{p_1 + p_2} e^{\xi_2} &
        \dfrac{1}{2p_1}  & 1 \\
        \dfrac{1}{p_1^* + p_2}  e^{\xi_2} &
        \dfrac{1}{2p_2} \left( e^{2\xi_2} + c_{2, 2} \right) &
        \dfrac{1}{p_1+p_2}  e^{\xi_2}& \exp(\xi_2) \\
        \dfrac{1}{2p_1^*} &
        \dfrac{1}{p_1^*+p_2} e^{\xi_2} &
        \dfrac{1}{p_1 + p_1^*}& 1 \\
        0 & -D_2 & 0 &  0
        \end{vmatrix}.
    \end{align*}

    \item After collision, i.e., \(t \rightarrow + \infty\)

    Soliton 1 (\(\xi_1 + \xi_1^* \approx 0,\ \xi_2\rightarrow + \infty\))
    \begin{align*}
        f &\simeq \begin{vmatrix}
        \dfrac{1}{p_1 + p_1^*} \left(e^{\xi_1 + \xi_1^*} + c_{1,1}\right) &
        \dfrac{1}{p_1 + p_2} e^{\xi_1} &
        \dfrac{1}{2p_1} \left(e^{2\xi_1} + c_{1,3}\right) \\
        \dfrac{1}{p_1^* + p_2} e^{\xi_1^*} &
        \dfrac{1}{2p_2} &
        \dfrac{1}{p_1+p_2} e^{\xi_1} \\
        \dfrac{1}{2p_1^*} \left(e^{2\xi_1^*} + c_{1,3}^*\right) &
        \dfrac{1}{p_1^*+p_2} e^{\xi_1^*} &
        \dfrac{1}{p_1 + p_1^*} \left(e^{\xi_1 + \xi_1^*} + c_{1,1}\right)
        \end{vmatrix},\\
        g_1 &\simeq \begin{vmatrix}
        \dfrac{1}{p_1 + p_1^*} \left(-\dfrac{p_1 - \i \alpha}{p_1^* + \i \alpha}e^{\xi_1 + \xi_1^*} + c_{1,1}\right) &
        -\dfrac{p_1 - \i \alpha} {p_2 + \i \alpha}\dfrac{1}{p_1 + p_2} e^{\xi_1}  &
        \dfrac{1}{2p_1} \left(-\dfrac{p_1 - \i \alpha} {p_1 + \i \alpha} e^{2\xi_1} + c_{1,3}\right) \\
        -\dfrac{p_2 - \i \alpha}{p_1^* + \i \alpha}\dfrac{1}{p_1^* + p_2} e^{\xi_1^*}  &
        -\dfrac{p_2 - \i \alpha}{p_2 + \i \alpha}\dfrac{1}{2p_2} &
        -\dfrac{p_2 - \i \alpha}{p_1 + \i \alpha}\dfrac{1}{p_1+p_2} e^{\xi_1} \\
        \dfrac{1}{2p_1^*} \left(-\dfrac{p_1^* - \i \alpha}{p_1^* + \i \alpha}e^{2\xi_1^*} + c_{1,3}^*\right) &
        -\dfrac{p_1^* - \i \alpha} {p_2 + \i \alpha}\dfrac{1}{p_1^*+p_2} e^{\xi_1^*} &
        \dfrac{1}{p_1 + p_1^*} \left(-\dfrac{p_1^* - \i \alpha}{p_1 + \i \alpha}e^{\xi_1 + \xi_1^*} + c_{3,3}\right)
        \end{vmatrix},\\
        h_2 &\simeq \begin{vmatrix}
        \dfrac{1}{p_1 + p_1^*} \left(e^{\xi_1 + \xi_1^*} + c_{1,1}\right) &
        \dfrac{1}{p_1 + p_2} e^{\xi_1} &
        \dfrac{1}{2p_1} \left(e^{2\xi_1} + c_{1,3}\right) & \exp(\xi_1) \\
        \dfrac{1}{p_1^* + p_2} e^{\xi_1^*} &
        \dfrac{1}{2p_2} &
        \dfrac{1}{p_1+p_2} e^{\xi_1} & 1 \\
        \dfrac{1}{2p_1^*} \left(e^{2\xi_1^*} + c_{1,3}^*\right) &
        \dfrac{1}{p_1^*+p_2} e^{\xi_1^*} &
        \dfrac{1}{p_1 + p_1^*} \left(e^{\xi_1 + \xi_1^*} + c_{3,3}\right) & \exp(\xi_1^*) \\
        -D_1 & 0 & -D_1^* &  0
        \end{vmatrix}
    \end{align*}

    Soliton 2 (\(\xi_2\approx 0,\ \xi_1 + \xi_1^* \rightarrow - \infty\))
    \begin{align*}
        f &\simeq \begin{vmatrix}
        \dfrac{c_{1, 1}}{p_1 + p_1^*} &
        \dfrac{c_{1, 2}}{p_1 + p_2} &
        \dfrac{c_{1, 3}}{2p_1}  \\
        \dfrac{c_{1, 2}^*}{p_1^* + p_2} &
        \dfrac{1}{2p_2} \left(e^{2\xi_2} + c_{2, 2}\right) &
        \dfrac{c_{2, 3}}{p_1+p_2} \\
        \dfrac{c_{1, 3}^*}{2p_1^*} &
        \dfrac{c_{2, 3}^*}{p_1^*+p_2}  &
        \dfrac{c_{3, 3}}{p_1 + p_1^*}
        \end{vmatrix},\\
        g_1 &\simeq \begin{vmatrix}
        \dfrac{1}{p_1 + p_1^*} \left(-\dfrac{p_1 - \i \alpha}{p_1^* + \i \alpha}e^{\xi_1 + \xi_1^*} + c_{1,1}\right) &
        -\dfrac{p_1 - \i \alpha}{p_2 + \i \alpha}\dfrac{1}{p_1 + p_2} e^{\xi_1}  &
        \dfrac{1}{2p_1} \left(-\dfrac{p_1 - \i \alpha}{p_1 + \i \alpha} e^{2\xi_1} + c_{1,3}\right) \\
        -\dfrac{p_2 - \i \alpha}{p_1^* + \i \alpha}\dfrac{1}{p_1^* + p_2} e^{\xi_1^*}  &
        -\dfrac{p_2 - \i \alpha}{p_2 + \i \alpha}\dfrac{1}{2p_2} &
        -\dfrac{p_2 - \i \alpha}{p_1 + \i \alpha}\dfrac{1}{p_1+p_2} e^{\xi_1} \\
        \dfrac{1}{2p_1^*} \left(-\dfrac{p_1^* - \i \alpha}{p_1^* + \i \alpha}e^{2\xi_1^*} + c_{1,3}^*\right) &
        -\dfrac{p_1^* - \i \alpha}{p_2 + \i \alpha}\dfrac{1}{p_1^*+p_2} e^{\xi_1^*} &
        \dfrac{1}{p_1 + p_1^*} \left(-\dfrac{p_1^* - \i \alpha}{p_1 + \i \alpha}e^{\xi_1 + \xi_1^*} + c_{3,3}\right)
        \end{vmatrix},\\
        h_2 &\simeq \begin{vmatrix}
        \dfrac{c_{1, 1}}{p_1 + p_1^*} &
        \dfrac{c_{1, 2}}{p_1 + p_2} &
        \dfrac{c_{1, 3}}{2p_1} & 0 \\
        \dfrac{c_{1, 2}^*}{p_1^* + p_2} &
        \dfrac{1}{2p_2} \left(e^{2\xi_2} + c_{2, 2}\right) &
        \dfrac{c_{2, 3}}{p_1+p_2} & \exp(\xi_2) \\
        \dfrac{c_{1, 3}^*}{2p_1^*} &
        \dfrac{c_{2, 3}^*}{p_1^*+p_2}  &
        \dfrac{c_{3, 3}}{p_1 + p_1^*} & 0 \\
        -D_1 & -D_2 & -D_1^* &  0
        \end{vmatrix}
    \end{align*}
\end{enumerate}

\section*{Acknowledgements}
B.F. Feng's work is partially supported by the U.S. Department of Defense (DoD), Air Force for Scientific Research (AFOSR) under grant No. W911NF2010276.

\bibliography{cite}

@article{hasegawa1973transmission,
	author = {Hasegawa, Akira and Tappert, Frederick},
	journal = {Appl. Phys. Lett.},
	pages = {171--172},
	publisher = {American Institute of Physics},
	title = {{Transmission of stationary nonlinear optical pulses in dispersive dielectric fibers. II. Normal dispersion}},
	volume = {23},
	year = {1973}
}

@book{fibich2015nonlinear,
	author = {Fibich, Gadi},
	publisher = {Springer},
	title = {{The nonlinear Schr{\"o}dinger equation}},
	volume = {192},
	year = {2015}
}

@article{dalfovo1999theory,
	author = {Dalfovo, Franco and Giorgini, Stefano and Pitaevskii, Lev P and Stringari, Sandro},
	journal = {Rev. Mod. Phys.},
	pages = {463},
	publisher = {APS},
	title = {{Theory of Bose-Einstein condensation in trapped gases}},
	volume = {71},
	year = {1999}
}

@book{pitaevskii2003bose,
	author = {Pitaevskii, L and Stringari, S},
	publisher = {USA},
	title = {{Bose-Einstein Condensation Oxford University Press}},
	year = {2003}
}

@article{zakharov1972collapse,
	author = {Zakharov, Vladimir E and others},
	journal = {Sov. Phys. JETP},
	pages = {908--914},
	title = {{Collapse of Langmuir waves}},
	volume = {35},
	year = {1972}
}

@incollection{kato2005nonlinear,
	author = {Kato, Tosio},
	booktitle = {{Schr{\"o}dinger Operators: Proceedings of the Nordic Summer School in Mathematics Held at Sandbjerg Slot, S{\o}nderborg, Denmark, August 1--12, 1988}},
	pages = {218--263},
	publisher = {Springer},
	title = {{Nonlinear schr{\"o}dinger equations}},
	year = {2005}
}

@article{benney1967propagation,
	author = {Benney, DJ and Newell, Alan C},
	journal = {J. Math. Phys.},
	pages = {133--139},
	publisher = {Wiley Online Library},
	title = {{The propagation of nonlinear wave envelopes}},
	volume = {46},
	year = {1967}
}

@incollection{agrawal2000nonlinear,
	author = {Agrawal, Govind P},
	booktitle = {Nonlinear Science at the Dawn of the 21st Century},
	pages = {195--211},
	publisher = {Springer},
	title = {Nonlinear fiber optics},
	year = {2000}
}

@article{nakkeeran2000exact,
	author = {Nakkeeran, K},
	journal = {Phys. Rev. E},
	pages = {1313},
	publisher = {APS},
	title = {{Exact soliton solutions for a family of N coupled nonlinear Schr{\"o}dinger equations in optical fiber media}},
	volume = {62},
	year = {2000}
}

@article{bindu2001dark,
	author = {Bindu, SG and Mahalingam, A and Porsezian, K},
	journal = {Phys. Lett. A},
	pages = {321--331},
	publisher = {Elsevier},
	title = {{Dark soliton solutions of the coupled Hirota equation in nonlinear fiber}},
	volume = {286},
	year = {2001}
}

@article{gilson2003sasa,
	author = {Gilson, C. and Hietarinta, J. and Nimmo, J. and Ohta, Y.},
	journal = {Phys. Rev. E},
	pages = {016614},
	publisher = {APS},
	title = {{Sasa-Satsuma higher-order nonlinear Schr{\"o}dinger equation and its bilinearization and multisoliton solutions}},
	volume = {68},
	year = {2003}
}

@article{porsezian1994coupled,
	author = {Porsezian, K. and {Sundaram Shanmugha}, P. and Mahalingam, A.},
	journal = {Phys. Rev. E},
	pages = {1543},
	publisher = {APS},
	title = {{Coupled higher-order nonlinear Schr{\"o}dinger equations in nonlinear optics: Painlev{\'e} analysis and integrability}},
	volume = {50},
	year = {1994}
}

@article{sakovich2000symmetrically,
	author = {Sakovich, S. and Tsuchida, T.},
	journal = {J. Phys. A},
	pages = {7217},
	publisher = {IOP Publishing},
	title = {{Symmetrically coupled higher-order nonlinear Schr{\"o}dinger equations: singularity analysis and integrability}},
	volume = {33},
	year = {2000}
}

@article{tasgal1992soliton,
	author = {Tasgal, R. and Potasek, M.},
	journal = {J. Math. Phys.},
	pages = {1208--1215},
	publisher = {American Institute of Physics},
	title = {{Soliton solutions to coupled higher-order nonlinear Schr{\"o}dinger equations}},
	volume = {33},
	year = {1992}
}

@article{chai2019localized,
	author = {Chai, Han-Peng and Tian, Bo and Du, Zhong},
	journal = {Commun. Nonlinear Sci. Numer. Simul.},
	pages = {181--192},
	publisher = {Elsevier},
	title = {{Localized waves for the mixed coupled Hirota equations in an optical fiber}},
	volume = {70},
	year = {2019}
}

@article{chan2017rogue,
	author = {Chan, HN and Chow, KW},
	journal = {Stud. Appl. Math.},
	pages = {78--103},
	publisher = {Wiley Online Library},
	title = {{Rogue waves for an alternative system of coupled Hirota equations: Structural robustness and modulation instabilities}},
	volume = {139},
	year = {2017}
}

@article{chen2013rogue,
	author = {Chen, Shihua and Song, Lian-Yan},
	journal = {Phys. Rev. E},
	pages = {032910},
	publisher = {APS},
	title = {{Rogue waves in coupled Hirota systems}},
	volume = {87},
	year = {2013}
}

@article{chen2014dark,
	author = {Chen, Shihua},
	journal = {Phys. Lett. A},
	pages = {2851--2856},
	publisher = {Elsevier},
	title = {{Dark and composite rogue waves in the coupled Hirota equations}},
	volume = {378},
	year = {2014}
}

@article{huang2016rational,
	author = {Huang, Xin},
	journal = {Phys. Lett. A},
	pages = {2136--2141},
	publisher = {Elsevier},
	title = {{Rational solitary wave and rogue wave solutions in coupled defocusing Hirota equation}},
	volume = {380},
	year = {2016}
}

@article{jiang2023asymptotic,
	author = {Jiang, Ziwei and Ling, Liming},
	journal = {Commun. Theor. Phys.},
	pages = {115005},
	publisher = {IOP Publishing},
	title = {{Asymptotic analysis of multi-valley dark soliton solutions in defocusing coupled Hirota equations}},
	volume = {75},
	year = {2023}
}

@article{kang2019construction,
	author = {Kang, Zhou-Zheng and Xia, Tie-Cheng},
	journal = {Chin. Phys. Lett.},
	pages = {110201},
	publisher = {IOP Publishing},
	title = {{Construction of multi-soliton solutions of the N-coupled Hirota equations in an optical fiber}},
	volume = {36},
	year = {2019}
}

@article{liu2016mixed,
	author = {Liu, Lei and Tian, Bo and Sun, Wen-Rong and Wu, Xiao-Yu},
	journal = {Comput. Math. Appl.},
	pages = {807--819},
	publisher = {Elsevier},
	title = {{Mixed-type soliton solutions for the \(N\)-coupled Hirota system in an optical fiber}},
	volume = {72},
	year = {2016}
}

@article{pan2024super,
	author = {Pan, Liuyi and Wang, Lei and Liu, Lei},
	journal = {Phys. Rev. A},
	pages = {023523},
	publisher = {APS},
	title = {{Super-regular breathers induced by the higher-order effects in coupled Hirota equations}},
	volume = {110},
	year = {2024}
}

@article{wang2014generalized,
	author = {Wang, Xin and Li, Yuqi and Chen, Yong},
	journal = {Wave Motion},
	pages = {1149--1160},
	publisher = {Elsevier},
	title = {{Generalized Darboux transformation and localized waves in coupled Hirota equations}},
	volume = {51},
	year = {2014}
}

@article{wang2014rogue,
	author = {Wang, Xin and Chen, Yong},
	journal = {Chin. Phys. B},
	pages = {070203},
	publisher = {IOP Publishing},
	title = {{Rogue-wave pair and dark-bright-rogue wave solutions of the coupled Hirota equations}},
	volume = {23},
	year = {2014}
}

@article{wang2021analytical,
	author = {Wang, Pan and Ma, Tian-Ping and Qi, Feng-Hua},
	journal = {Appl. Math. Comput.},
	pages = {126495},
	publisher = {Elsevier},
	title = {{Analytical solutions for the coupled Hirota equations in the firebringent fiber}},
	volume = {411},
	year = {2021}
}

@article{xie2020elastic,
	author = {Xie, Xi-Yang and Liu, Xiao-Bing},
	journal = {Appl. Math. Lett.},
	pages = {106291},
	publisher = {Elsevier},
	title = {{Elastic and inelastic collisions of the semirational solutions for the coupled Hirota equations in a birefringent fiber}},
	volume = {105},
	year = {2020}
}

@article{sasa1991new,
	author = {Sasa, N. and Satsuma, J.},
	journal = {J. Phys. Soc. Jpn.},
	pages = {409--417},
	title = {{New-type of solutions for a higher-order nonlinear evolution equation}},
	volume = {60},
	year = {1991}
}

@article{trippenbach1998effects,
	author = {Trippenbach, Marek and Band, YB},
	journal = {Phys. Rev. A},
	pages = {4791},
	publisher = {APS},
	title = {{Effects of self-steepening and self-frequency shifting on short-pulse splitting in dispersive nonlinear media}},
	volume = {57},
	year = {1998}
}

@article{kodama1987nonlinear,
	author = {Kodama, Yuji and Hasegawa, Akira},
	journal = {IEEE J. Quantum Electron.},
	pages = {510--524},
	publisher = {IEEE},
	title = {{Nonlinear pulse propagation in a monomode dielectric guide}},
	volume = {23},
	year = {1987}
}

@article{mihalache1993inverse,
	author = {Mihalache, Dumitru and Torner, L and Moldoveanu, F and Panoiu, N-C and Truta, N},
	journal = {Phys. Rev. E},
	pages = {4699},
	publisher = {APS},
	title = {{Inverse-scattering approach to femtosecond solitons in monomode optical fibers}},
	volume = {48},
	year = {1993}
}

@article{ohta2010dark,
	author = {Ohta, Y.},
	journal = {AIP Conf. Proc.},
	organization = {American Institute of Physics},
	pages = {114--121},
	title = {{Dark soliton solution of Sasa-Satsuma equation}},
	volume = {1212},
	year = {2010}
}

@article{wu2022multi,
	author = {Wu, Chengfa and Wei, Bo and Shi, Changyan and Feng, Bao-Feng},
	journal = {Proc. R. Soc. A},
	pages = {20210711},
	publisher = {The Royal Society},
	title = {{Multi-breather solutions to the Sasa--Satsuma equation}},
	volume = {478},
	year = {2022}
}

@article{bandelow2012sasa,
	author = {Bandelow, Uwe and Akhmediev, Nail},
	journal = {Phys. Rev. E},
	pages = {026606},
	publisher = {APS},
	title = {{Sasa-Satsuma equation: Soliton on a background and its limiting cases}},
	volume = {86},
	year = {2012}
}

@article{chen2013twisted,
	author = {Chen, Shihua},
	journal = {Phys. Rev. E},
	pages = {023202},
	publisher = {APS},
	title = {{Twisted rogue-wave pairs in the Sasa-Satsuma equation}},
	volume = {88},
	year = {2013}
}

@article{akhmediev2015rogue,
	author = {Akhmediev, N and Soto-Crespo, Jos{\'e} M and Devine, Natasha and Hoffmann, NP},
	journal = {Physica D},
	pages = {37--42},
	publisher = {Elsevier},
	title = {{Rogue wave spectra of the Sasa--Satsuma equation}},
	volume = {294},
	year = {2015}
}

@article{feng2022higher,
	author = {Feng, Bao-Feng and Shi, Changyan and Zhang, Guangxiong and Wu, Chengfa},
	journal = {J. Phys. A},
	pages = {235701},
	publisher = {IOP Publishing},
	title = {{Higher-order rogue wave solutions of the Sasa--Satsuma equation}},
	volume = {55},
	year = {2022}
}

@article{xu2013single,
	author = {Xu, Tao and Xu, Xiang-Min},
	journal = {Phys. Rev. E},
	pages = {032913},
	publisher = {APS},
	title = {{Single-and double-hump femtosecond vector solitons in the coupled Sasa-Satsuma system}},
	volume = {87},
	year = {2013}
}

@article{zhang2025dark,
  title={{Dark Soliton and Breather Solutions to the Coupled Sasa--Satsuma Equation}},
  author={Zhang, Guangxiong and Shi, Changyan and Wu, Chengfa and Feng, Bao-Feng},
  journal={J. Nonlinear Sci.},
  volume={35},
  pages={7},
  year={2025},
  publisher={Springer}
}

@article{wang2020riemann,
  title="{Riemann--Hilbert approach and N-soliton solutions for a new two-component Sasa--Satsuma equation}",
  author={Wang, Jia and Su, Ting and Geng, Xianguo and Li, Ruomeng},
  journal={Nonlinear Dyn.},
  volume={101},
  pages={597--609},
  year={2020},
  publisher={Springer}
}

@article{geng2021darboux,
  title="{Darboux transformation of a two-component generalized Sasa--Satsuma equation and explicit solutions}",
  author={Geng, Xianguo and Li, Yihao and Wei, Jiao and Zhai, Yunyun},
  journal={Math. Methods Appl. Sci.},
  volume={44},
  pages={12727--12745},
  year={2021},
  publisher={Wiley Online Library}
}

@article{shi2025general,
  title={{General soliton solutions to the coupled Hirota equation via the Kadomtsev--Petviashvili reduction}},
  author={Shi, Changyan and Liu, Bingyuan and Feng, Bao-Feng},
  journal={Chaos, Solitons \& Fractals},
  volume={197},
  pages={116400},
  year={2025},
  publisher={Elsevier}
}

@article{xu2014soliton,
  title={{Soliton and breather solutions of the Sasa--Satsuma equation via the Darboux transformation}},
  author={Xu, Tao and Wang, Danhua and Li, Min and Liang, Huan},
  journal={Phys. Scr.},
  volume={89},
  pages={075207},
  year={2014},
  publisher={IOP Publishing}
}

@article{wen2023sasa,
  title={{The Sasa-Satsuma equation on a non-zero background: the inverse scattering transform and multi-soliton solutions}},
  author={Wen, Lili and Fan, Engui and Chen, Yong},
  journal={Acta Math. Sci.},
  volume={43},
  pages={1045--1080},
  year={2023},
  publisher={Springer}
}

@article{xu2018initial,
  title={{The initial-boundary value problem for the Sasa-Satsuma equation on a finite interval via the Fokas method}},
  author={Xu, Jian and Zhu, Qiaozhen and Fan, Engui},
  journal={J. Math. Phys.},
  volume={59},
  year={2018},
  publisher={AIP Publishing}
}

@article{jiang2013bright,
  title={{Bright hump solitons for the higher-order nonlinear Schr{\"o}dinger equation in optical fibers}},
  author={Jiang, Yan and Tian, Bo and Li, Min and Wang, Pan},
  journal={Nonlinear Dyn.},
  volume={74},
  pages={1053--1063},
  year={2013},
  publisher={Springer}
}

@article{wu2024general,
  title={{General rogue wave solutions to the Sasa--Satsuma equation}},
  author={Wu, Chengfa and Zhang, Guangxiong and Shi, Changyan and Feng, Bao-Feng},
  journal={IMA J. Appl. Math.},
  volume={89},
  pages={953--975},
  year={2024},
  publisher={Oxford Academic}
}

@article{xu2015anti,
  title={{Anti-dark and Mexican-hat solitons in the Sasa-Satsuma equation on the continuous wave background}},
  author={Xu, Tao and Li, Min and Li, Lu},
  journal={EPL},
  volume={109},
  pages={30006},
  year={2015},
  publisher={IOP Publishing}
}

@article{bandelow2012persistence,
  title={{Persistence of rogue waves in extended nonlinear Schr{\"o}dinger equations: Integrable Sasa--Satsuma case}},
  author={Bandelow, Uwe and Akhmediev, Nail},
  journal={Phys. Lett. A},
  volume={376},
  pages={1558--1561},
  year={2012},
  publisher={Elsevier}
}

@article{manakov1974theory,
  title="{On the theory of two-dimensional stationary self-focusing of electromagnetic waves}",
  author={Manakov, Sergei V},
  journal={Sov. Phys. JETP},
  volume={38},
  pages={248--253},
  year={1974}
}

@article{gelash2023vector,
  title="{Vector breathers in the Manakov system}",
  author={Gelash, Andrey and Raskovalov, Anton},
  journal={Stud. Appl. Math.},
  volume={150},
  pages={841--882},
  year={2023},
  publisher={Wiley Online Library}
}

@book{maimistov2013nonlinear,
  title="{Nonlinear optical waves}",
  author={Maimistov, AI and Basharov, AM},
  volume={104},
  year={2013},
  publisher={Springer Science \& Business Media}
}

@article{lu2014bright,
  title={{Bright-soliton collisions with shape change by intensity redistribution for the coupled Sasa--Satsuma system in the optical fiber communications}},
  author={L{\"u}, Xing},
  journal={Commun. Nonlinear Sci. Numer. Simul.},
  volume={19},
  pages={3969--3987},
  year={2014},
  publisher={Elsevier}
}

@article{liu2018vector,
  title={{Vector bright soliton interactions of the coupled Sasa--Satsuma equations in the birefringent or two-mode fiber}},
  author={Liu, Lei and Tian, Bo and Yin, Hui-Min and Du, Zhong},
  journal={Wave Motion},
  volume={80},
  pages={91--101},
  year={2018},
  publisher={Elsevier}
}

@article{liu2023riemann,
  title={{Riemann--Hilbert problems and soliton solutions for a generalized coupled Sasa--Satsuma equation}},
  author={Liu, Yaqing and Zhang, Wen-Xin and Ma, Wen-Xiu},
  journal={Commun. Nonlinear Sci. Numer. Simul.},
  volume={118},
  pages={107052},
  year={2023},
  publisher={Elsevier}
}

@article{liu2018dark,
  title={{Dark-bright solitons and semirational rogue waves for the coupled Sasa-Satsuma equations}},
  author={Liu, Lei and Tian, Bo and Yuan, Yu-Qiang and Du, Zhong},
  journal={Phys. Rev. E},
  volume={97},
  pages={052217},
  year={2018},
  publisher={APS}
}

@article{zhang2025rogue,
  title={{Rogue wave solutions to the coupled Sasa--Satsuma equation}},
  author={Zhang, Guangxiong and Chen, Xiyao and Feng, Bao-Feng and Wu, Chengfa},
  journal={Physica D},
  volume={474},
  pages={134549},
  year={2025},
  publisher={Elsevier}
}

@article{zhao2014localized,
  title={{Localized waves on continuous wave background in a two-mode nonlinear fiber with high-order effects}},
  author={Zhao, Li-Chen and Yang, Zhan-Ying and Ling, Liming},
  journal={J. Phys. Soc. Jpn.},
  volume={83},
  pages={104401},
  year={2014},
  publisher={The Physical Society of Japan}
}

@article{hu2022initial,
  title={{The initial-boundary value problems of the new two-component generalized Sasa--Satsuma equation with a 4$\times$ 4 matrix Lax pair}},
  author={Hu, Beibei and Zhang, Ling and Lin, Ji},
  journal={Anal. Math. Phys.},
  volume={12},
  pages={109},
  year={2022},
  publisher={Springer}
}

@article{zhao2025nonlinear,
  title={{Nonlinear behaviors of coupled Hirota equations in the presence of multiple higher-order poles}},
  author={Zhao, Xiaodan and Wang, Lei},
  journal={Nonlinear Dyn.},
  volume={113},
  pages={28039--28053},
  year={2025},
  publisher={Springer}
}

@article{zhao2024two,
  title={{A two-component Sasa--Satsuma equation: Large-time asymptotics on the line}},
  author={Zhao, Xiaodan and Wang, Lei},
  journal={J. Nonlinear Sci.},
  volume={34},
  pages={38},
  year={2024},
  publisher={Springer}
}

@article{zhao2025dynamics,
  title={{Dynamics of Multiple Higher-Order Pole Solutions of A Two-component Sasa--Satsuma Equation based on Riemann--Hilbert Approach and PINN algorithm}},
  author={Zhao, Xiaodan and Pan, Liuyi and Wang, Lei and Liu, Nan},
  journal={Commun. Nonlinear Sci. Numer. Simul.},
  pages={109346},
  year={2025},
  publisher={Elsevier}
}

@mastersthesis{shi2024study,
  title={{A Study on a Vector Complex Modified Korteweg-De Vries Equation}},
  author={Shi, Changyan},
  year={2024},
  school={The University of Texas Rio Grande Valley}
}

@article{weng2021rational,
  title="{Rational vector rogue waves for the n-component Hirota equation with non-zero backgrounds}",
  author={Weng, Weifang and Zhang, Guoqiang and Wang, Li and Zhang, Minghe and Yan, Zhenya},
  journal={Physica D},
  volume={427},
  pages={133005},
  year={2021},
  publisher={Elsevier}
}

@article{wei2022vector,
  title="{Vector multi-pole solutions in the r-coupled Hirota equation}",
  author={Wei, Yun-Chun and Zhang, Hai-Qiang},
  journal={Wave Motion},
  volume={112},
  pages={102959},
  year={2022},
  publisher={Elsevier}
}

@article{zhang2017general,
  title={{General N-dark vector soliton solution for multi-component defocusing Hirota system in optical fiber media}},
  author={Zhang, Hai-Qiang and Yuan, Sha-Sha},
  journal={Commun. Nonlinear Sci. Numer. Simul.},
  volume={51},
  pages={124--132},
  year={2017},
  publisher={Elsevier}
}

@article{kim1998conservation,
  title={{Conservation laws in higher-order nonlinear Schr{\"o}dinger equations}},
  author={Kim, Jongbae and Park, Q-Han and Shin, HJ},
  journal={Phys. Rev. E},
  volume={58},
  pages={6746},
  year={1998},
  publisher={APS}
}

@article{xu2018breathers,
  title={{Breathers and solitons on two different backgrounds in a generalized coupled Hirota system with four wave mixing}},
  author={Xu, Han-Xiang and Yang, Zhan-Ying and Zhao, Li-Chen and Duan, Liang and Yang, Wen-Li},
  journal={Phys. Lett. A},
  volume={382},
  pages={1738--1744},
  year={2018},
  publisher={Elsevier}
}

@article{shi2026soliton,
  title={{Soliton solutions to the coupled Sasa-Satsuma equation under mixed boundary conditions}},
  author={Shi, Changyan and Chen, Xiyao and Zhang, Guangxiong and Wu, Chengfa and Feng, Bao-Feng},
  journal={arXiv preprint arXiv:2603.08686},
  year={2026}
}

\appendix
\section{Results on vector Hirota equation}\label{section:append_a}
In this section, we present bilinear form and soliton solutions to the vector Hirota equation \eqref{m-Hirota}. Soliton solutions in this section are derived from the KP-Toda hierarchy listed in \ref{section:append_b}. Detailed proof of \cref{theorem:bright_m-Hirota} and \cref{theorem:dark_m-Hirota} can be found in Ref.~\cite{shi2024study}, and proof of \cref{theorem:bright_dark_m-Hirota} is similar to Ref.~\cite{shi2026soliton}.

\begin{theorem}[Bright soliton solution to Eq.~\eqref{m-Hirota}]\label{theorem:bright_m-Hirota}
Under the transformation
\[
    u_k= \frac{g_k}{f}
\]
equation \eqref{m-Hirota} is bilinearized into
\begin{align}
    &(D_x^3 - D_t)g_k \cdot f=3\sum_{l=1}^M s_{kl} g_l^* ,\label{BL_bright_1}\\
    &D_x^2 f\cdot f - 2 \sum_{l=1}^M  |g_l|^2=0,\label{BL_bright_2}\\
    &D_x g_k \cdot g_l = s_{kl} f.\label{BL_bright_3}
\end{align}
where \(k, l =1, 2, \ldots, M\), \(s_{kl} = -s_{lk}\).
In this case, \eqref{m-Hirota} admits the bright soliton solutions given by
\begin{align}
    f=|M|,\quad
    g_k=\begin{vmatrix}
        M & \Phi \\ -\left(\bar{\Psi}^{(k)}\right)^T & 0
        \end{vmatrix},\quad
\end{align}
where \(M\) is an \(N\times N\) matrix, \(\Phi \), \(\bar{\Psi} \), are \(N\)-component row vectors whose elements are defined respectively as
\begin{align}
    &m_{ij}=\frac{1}{p_i+p_j^*}\left(e^{\xi_i+\xi_j^*}- \sum_{n=1}^M \varepsilon_n \left(C_i^{(n)}\right)^* C_j^{(n)} \right),\quad \xi_i=p_i x + p_i^3 t+\xi_{i0},\\
    &\Phi = \left( e^{\xi _{1}},e^{\xi _2},\ldots, e^{\xi _{N}}\right)^T,\quad
    \bar{\Psi}^{(k)}=\left(C_1^{(k)}, C_2^{(k)},\ldots, C_N^{(k)}\right)^T,
\end{align}
Here, \(p_i\), \(\xi_{i0}\), \(C_i^{(k)}\) are complex parameters.
\end{theorem}

\begin{theorem}[Dark soliton solution to Eq.~\eqref{m-Hirota}]\label{theorem:dark_m-Hirota}
Under transformation
\begin{equation*}
    u_k=\rho_k \frac{h_k}{f} e^{\i \left(\alpha_k x - \left(\alpha_k^3+3\varepsilon_k\alpha_k\left(\sum\limits_{l=1}^M \rho_l^2\right)+3\sum\limits_{l=1}^M \varepsilon_l\rho_l^2 \alpha_l\right) t\right)},
\end{equation*}
equation \eqref{m-Hirota} is bilinearized into
\begin{align}
    &\left[D_x^3 - D_t +3\i \alpha_k D_x^2-3 \left(\alpha_k^2 +2\sum_{l=1}^M \varepsilon_l \rho_l^2\right)D_x - 3\i \varepsilon_k \alpha_k \sum_{l=1}^M \rho_l^2 + 3\i \sum_{l=1}^M \varepsilon_l \rho_l^2 \alpha_l\right]h_k\cdot f \nonumber\\
    &\quad = - 3\i \sum_{l=1}^M \varepsilon_l (\alpha_k-\alpha_l) \rho_l^2 r_{kl} h_l^* , \label{BL_dark_1}\\
    &\left(D_x^2 - 2\sum_{l=1}^M \varepsilon_l \rho_l^2\right) f \cdot f + 2\sum_{l=1}^M \varepsilon_l \rho_l^2 |h_l|^2=0,\label{BL_dark_2}\\
    &\left[D_x  + \i (\alpha_k-\alpha_l)\right]h_k \cdot h_l=\i (\alpha_k-\alpha_l) r_{kl} f ,\label{BL_dark_3}
\end{align}
where \(k, l =1, 2, \ldots, M\). And \eqref{m-Hirota} admits the dark soliton solutions given by
\begin{equation}
    f=\tau_{\mathbf{0}}, \quad h_k=\tau_{\mathbf{e}_k},
\end{equation}
where \(\tau_{\mathbf{n}}\) is an \(N\times N\) determinant defined as
\begin{equation}
    \tau_{\mathbf{n}}=\det\left(\delta_{ij}d_i e^{-\xi_i-\eta_j} + \frac{1}{p_i+q_j} \prod_{n=1}^M \left(-\frac{p_i - \i \alpha_n}{q_j + \i \alpha_n}\right)^{k_n}\right),
\end{equation}
with \(\xi_i=p_i (x - 3 \sum_{l=1}^M\varepsilon_l \rho_l^2 t) + p_i^3 t + \xi_{i 0}\), \(\eta_i=q_i (x - 3 \sum_{l=1}^M \varepsilon_l \rho_l^2 t) + q_i^3 t + \xi_{i 0}\). Where \(\mathbf{n} = (k_1, k_2, \dots, k_M) \in \mathbb{Z}^M\) and \(\mathbf{0}, \mathbf{e}_k\) are zero vector and standard unit vector in \(\mathbb{Z}^M\). \(\xi_{i 0},\ \alpha_1,\ \alpha_2,\ \rho_1,\ \rho_2\) are real parameters, \(p_i\), \(q_i\) are complex parameters. For each \(h = 0, 1, \ldots, \lfloor N/2 \rfloor\), the parameters satisfy the following complex conjugate relation
\begin{align}\label{dark_complex_relation}
    \begin{split}
        p_i = q_i^*,\ p_{N+1-i} = q_{N+1-i}^*,\ \text{and}\ d_i,d_{N+1-i},\xi_{i,0},\xi_{N+1-i,0} \in \mathbb{R},& \quad \text{for } i \in \{\mathbb{Z} | 1 \leq i \leq h\},\\
        p_i = q_{N+1-i}^*,\ p_{N+1-i} = q_i^*,\ d_i = d_{N+1-i} \in \mathbb{R},\ \xi_{i,0} = \xi_{N+1-i,0} \in \mathbb{R},& \quad \text{for } i \in  \{\mathbb{Z} | h+1 \leq i \leq \lceil N/2 \rceil\}\,.
    \end{split}
\end{align}
Moreover, these parameters also need to satisfy the equation \(G(p_i,q_i) = 0\), for \(i = 1, 2, \ldots, N\), where \(G(p,q)\) defined as
\begin{align}
    G(p,q) &= \sum_{l=1}^M \frac{\varepsilon_l\rho_l^2}{(p-\i\alpha_l)(q+\i\alpha_l)}-1.
\end{align}
\end{theorem}

\begin{theorem}[Bright-dark soliton solution to Eq.~\eqref{m-Hirota}]\label{theorem:bright_dark_m-Hirota}
Under transformation
\begin{align*}
    u_k &= \frac{g_k}{f}\exp\left( -3 \i \sum\limits_{i=m+1}^M \rho_i^2\varepsilon_i\alpha_i t \right), \quad &\text{for} \quad k = 1,\ldots, m,\\
    u_l &= \rho_l \frac{h_l}{f} \exp\left(\i \left(\alpha_l x - \left(\alpha_l^3 + 3\sum\limits_{i=m+1}^M \varepsilon_i\alpha_i \rho_i^2 + 3\varepsilon_l\sum\limits_{i=m+1}^M \alpha_i \rho_i^2 \right) t\right)\right), \quad &\text{for} \quad l = m+1,\ldots, M,
\end{align*}
equation \eqref{m-Hirota} is bilinearized into
\begin{align}
    &\left[D_x^3 - D_t - 6\left(\sum_{l=m+1}^{M} \varepsilon_l \rho_l^2 \right)D_x  + 3\i \left(\sum_{l=m+1}^M \varepsilon_l \alpha_l \rho_l^2\right) \right]g_k \cdot f\nonumber\\
    &\quad = -3\sum_{i=1}^{m} \varepsilon_i g_i^* s_{ki} + 3\i\sum_{i=m+1}^M \varepsilon_i \alpha_i \rho_i^2 h_i^* r_{ki},\label{BL_bright_dark_1}\\
    &\left[D_x^3 - D_t +3\i \alpha_l D_x^2-3 \left(\alpha_l^2 +2\sum_{i=m+1}^M \varepsilon_i \rho_i^2\right)D_x - 3\i \varepsilon_l \alpha_l \sum_{i=m+1}^M \rho_i^2 + 3\i \sum_{i=m+1}^M \varepsilon_i \alpha_i \rho_i^2 \right]h_l \cdot f \nonumber\\
    &\quad = -3\i\sum_{i=1}^m \varepsilon_i \alpha_i g_i^* r_{li} - 3\i \sum_{i=m+1}^M \varepsilon_i (\alpha_l-\alpha_i) \rho_i^2 h_i^* r_{li}  , \label{BL_bright_dark_2}\\
    &\left(D_x^2 - 2\sum_{l=m+1}^M \varepsilon_l \rho_l^2\right) f \cdot f +  2\sum_{k=1}^m \varepsilon_k |g_k|^2 + 2\sum_{l=m+1}^M \varepsilon_l \rho_l^2 |h_l|^2=0,\label{vcmKdVDE_BL_2}\\
    &D_x g_k \cdot g_i = s_{ki} f, \quad \text{for} \quad i=1,\ldots,m, \label{BL_bright_dark_3}\\
    &(D_x - \i \alpha_i)g_k \cdot h_i = -\alpha_i r_{ki} f, \quad \text{for} \quad i=m+1,\ldots,M, \label{BL_bright_dark_4}\\
    &\left[D_x  + \i (\alpha_l-\alpha_i)\right]h_l \cdot h_i=\i (\alpha_l-\alpha_i) r_{li} f, \quad \text{for} \quad i=m+1,\ldots,M, \label{BL_bright_dark_5}
\end{align}
where \(k =1, 2, \ldots, m\), \(l=m+1,\ldots, M\), and \(s_{ki} = -s_{ik}\) for \(i=1,\ldots, m\), \(r_{li}=r_{il}\) for \(i=1,\ldots, M\).
In this case, Eq.~\eqref{m-Hirota} admits the following \(m\)-bright-\((M-m)\)-dark soliton solution, where \(0<m<M\), and the bright soliton solution \(u_k\) and dark soliton solution \(u_l\) are given by
\begin{equation}
    f = \left|M_\mathbf{0}\right|,\quad g_k = \begin{vmatrix}
        M_\mathbf{0} & \Phi \\
        -\left(\bar{\Psi}^{(k)}\right)^T & 0
    \end{vmatrix},\quad h_l = \left|M_{\mathbf{e}_l}\right|
\end{equation}
where \(M_{\mathbf{e}_k}\) is \(N\times N\) matrix, \(\Phi\) and \(\bar{\Psi}^{(k)}\) are \(N\)-component vectors whose elements are defined as
\begin{align}
    &\left(M_{\mathbf{e}_k}\right)_{ij} = \frac{1}{p_i+p_j^*}e^{\xi_i+\xi_j^*}\prod_{n=m+1}^M\left(-\frac{p_i - \i \alpha_n}{q_j + \i \alpha_n}\right)^{k_n} + \frac{\sum
    \limits_{l=1}^{m}\varepsilon_l \left(C_i^{(l)}\right)^* C_j^{(l)}}{(p_i+p_j^*) \left(\sum\limits_{l=m+1}^M \dfrac{ \varepsilon_l \rho_l^2}{(p_i - \i \alpha_l)(p_j^* + \i \alpha_l)} - 1\right)},\\
    &\xi_i = p_i \left(x - 3\sum_{l=m+1}^n \varepsilon_l \rho_l^2 t\right) + p_i^3 t +\xi_{i0}, \\
    &\Phi = \left(e^{\xi _{1}}, e^{\xi _2},\ldots, e^{\xi _{N}}\right)^T, \quad \Psi^{(k)} = \left(C_1^{(k)}, C_2^{(k)},\ldots, C_N^{(k)}\right)^T.
\end{align}
Here, \(p_i, \xi_{i0}, C_i^{(k)}\) are complex parameters and \(\alpha_l\) is a real number.
\end{theorem}

\section{Corresponding bilinear equations and \(\tau\)-functions from KP-Toda hierarchy}\label{section:append_b}
From the KP-Toda hierarchy, we have the following lemmas.

\begin{lemma}
The bilinear equations
\begin{align}
    &\left(D_{x_1}^3+3D_{x_1}D_{x_2}-4D_{x_3}\right)g_k \cdot f=0,\label{KP_bright_1}\\
    &D_{y_1^{(k)}}D_{x_1}f\cdot f=-2g_k \bar{g}_k,\label{KP_bright_2}\\
    &D_{x_1}g_k\cdot g_l=s_{kl}f,\label{KP_bright_3}\\
    &D_{y_1^{(k)}}\left(D_{x_1}^2-D_{x_2}\right)g_k \cdot f=-4s_{kl}\bar{g}_l,\label{KP_bright_4}
\end{align}
where \(k,l=1,\ldots, M\), are satisfied by the following \(\tau\) functions \(f,\ s_{kl},\ g_k,\ \bar{g_k}\),
\begin{align}
    &f=|M|,\label{tau_bright_begin}\\
    &g_k=\begin{vmatrix}
        M & \Phi \\ -\left(\bar{\Psi}^{(k)}\right)^T & 0
    \end{vmatrix},
    \quad
    \bar{g}_k=\begin{vmatrix}
        M & \Psi^{(k)} \\ -\bar{\Phi}^T & 0
    \end{vmatrix},
    \\
    & s_{kl}=\begin{vmatrix}
        M & \Phi & \partial_{x_1} \Phi \\
        -\left(\bar{\Psi}^{(l)}\right)^T & 0 & 0 \\
        -\left(\bar{\Psi}^{(k)}\right)^T & 0 & 0
    \end{vmatrix},\label{tau_bright_end}
\end{align}
where \(M\) is a \(N\times N\) matrix, \(\Phi \), \(\bar{\Phi}\), \(\Psi^{(k)} \), and \(\bar{\Psi}^{(k)}\), are \(N\)-component row vectors whose elements are defined respectively as
\begin{align}
    &m_{ij}=\frac{1}{p_i+\bar{p}_j}e^{\xi_i+\bar{\xi}_j}+\sum_{n=1}^{M}\frac{\tilde{C}_i^{(n)} \bar{C}_j^{(n)}}{q_i^{(n)}+\bar{q}_j^{(n)}}e^{\eta_i^{(n)}+\bar{\eta}_j^{(n)}},\\
    &\Phi =\left( e^{\xi _{1}},e^{\xi _2},\ldots, e^{\xi _{N}}\right)^T, \quad \bar{\Phi} =\left( e^{\bar{\xi}_{1}},e^{\bar{\xi}_2},\ldots ,e^{\bar{\xi}_{N}}\right)^T, \quad\\
    &\Psi^{(k)} =\left(\tilde{C}_1^{(k)} e^{\eta _{1}^{(k)}}, \tilde{C}_2^{(k)} e^{\eta _2^{(k)}},\ldots, \tilde{C}_N^{(k)} e^{\eta _{N}^{(k)}}\right)^T,\\
    &\bar{\Psi}^{(k)}=\left(\bar{C}_1^{(k)} e^{\bar{\eta}_{1}^{(k)}}, \bar{C}_2^{(k)} e^{\bar{\eta}_2^{(k)}},\ldots, \bar{C}_N^{(k)} e^{\bar{\eta}_{N}^{(k)}}\right)^T,\\
    &\xi_i=p_i x_1 + p_i^2 x_2 + p_i^3 x_3+\xi_{i0},\quad \bar{\xi}_i=\bar{p}_i x_1 - \bar{p}_i^2 x_2 + \bar{p}_i^3 x_3+\bar{\xi}_{i0},\\
    &\eta_i^{(k)}=q_i y_1^{(k)},\quad \bar{\eta}_i^{(k)}=\bar{q}_i y_1^{(k)}.
\end{align}
Note that with above defined \(\tau\) function, we have \(s_{kl}=-s_{lk}\) by exchanging two rows in a determinant. And \(D_{x_1}g_k\cdot g_l=-D_{x_1}g_l\cdot g_k\) by the definition of \(D\)-operator.
In particular, when \(k=l\), we have \(s_{kk}=0\) and \(D_{x_1}g_k\cdot g_k=0.\)
\end{lemma}

\begin{lemma}
The following bilinear equations
\begin{align}
    & \left(D_{x_{-1}^{(k)}} D_x-2\right) \tau_{\mathbf{n}} \cdot \tau_{\mathbf{n}}=-2 \tau_{\mathbf{n} + \mathbf{e}_k} \tau_{\mathbf{n} - \mathbf{e}_k}, \label{KP_dark_1} \\
    & \left(D_x^2-D_y+2 a_k D_x\right) \tau_{\mathbf{n} + \mathbf{e}_k} \cdot \tau_{\mathbf{n}}=0, \label{KP_dark_2}\\
    & \left(D_x^3+3 D_x D_y-4 D_t+3 a_k\left(D_x^2+D_y\right)+6 a_k^2 D_x\right) \tau_{\mathbf{n} + \mathbf{e}_k} \cdot \tau_{\mathbf{n}}=0, \label{KP_dark_3}\\
    & \left(D_{x_{-1}^{(l)}}\left(D_x^2-D_y+2 a_k D_x\right)-4\left(D_x+a_k-a_l\right)\right) \tau_{\mathbf{n} + \mathbf{e}_k} \cdot \tau_{\mathbf{n}} + 4(a_k-a_l) \tau_{\mathbf{n} + \mathbf{e}_k + \mathbf{e}_l} \cdot \tau_{\mathbf{n} - \mathbf{e}_l}=0, \label{KP_dark_4}\\
    & \left(D_x+a_k-a_l\right) \tau_{\mathbf{n} + \mathbf{e}_k} \cdot \tau_{\mathbf{n} + \mathbf{e}_l}=(a_k-a_l) \tau_{\mathbf{n} + \mathbf{e}_k + \mathbf{e}_l}  \tau_{\mathbf{n}}, \label{KP_dark_5}
\end{align}
where \(k,l = 1, \ldots, M\), \(\mathbf{n} \in \mathbb{Z}^M\) and \(\mathbf{e}_k\) is the \(k\)-th standard unit vector in \(\mathbb{Z}^M\), is satisfied by the \(\tau\) function defined as
\begin{equation}\label{tau_dark}
    \tau_{\mathbf{n}}=\det\left(m_{i j}^{\mathbf{n}}\right)_{1 \leq i, j \leq N},
\end{equation}
where \(\mathbf{n} = (k_1, k_2, \ldots, k_M) \in \mathbb{Z}^M\), \(N\) and \(M\) are positive integer. And the matrix element is defined as
\begin{align*}
    & m_{i j}^{\mathbf{n}}=c_{i j} + \frac{e^{\xi_i+\eta_j}}{p_i+q_j} \prod_{n=1}^M \left(-\frac{p_i-a_n}{q_j+a_n}\right)^{k_n}, \\
    & \xi_i=p_i x+p_i^2 y+p_i^3 t+ \sum_{n=1}^M \left(\frac{1}{p_i-a_n} x_{-1}^{(n)}\right)+\xi_{i 0}, \\
    & \eta_i=q_i x-q_i^2 y+q_i^3 t+\sum_{n=1}^M \left(\frac{1}{q_i+a_n} x_{-1}^{(n)}\right)+\eta_{i 0}.
\end{align*}
Here \(c_{i j},\ p_i,\ q_j,\ \xi_{i 0},\ \eta_{j 0}\), and \(a_n\) are constants.
\end{lemma}

\begin{lemma}
Denote index sets \(I_1 = \{i\in\mathbb Z|1\leq i \leq m\}\), \(I_2 = \{i\in \mathbb Z|1\leq i \leq M-m\}\). Denote arbitrary vector from \(\mathbb{Z}^{M-m}\) by \(\mathbf{n}=(k_1, k_2, \ldots, k_{M-m})\), and denote \(\mathbf{e}_j\) to be the \(j\)-th standard unit vector in \(\mathbb{Z}^{M-m}\).

For \(k,i \in I_1\), \(l \in I_2\), we have the following bilinear equations about \(\tau\)-functions \(\tau^{(k)}_{\mathbf n}\) and \(\tau^{(0)}_{\mathbf n}\),
\begin{align}
    &\left(D_{x_1}^3+3D_{x_1}D_{x_2}-4D_{x_3}\right)\tau^{(k)}_{\mathbf n} \cdot \tau^{(0)}_{\mathbf n}=0,\label{KP_bright_dark_1}\\
    &D_{y_1^{(i)}}\left(D_{x_1}^2-D_{x_2}\right)\tau^{(k)}_{\mathbf n} \cdot \tau^{(0)}_{\mathbf n}=-4\tau^{(k,i)}_{\mathbf n}\bar{\tau}^{(i)}_{\mathbf n},\label{KP_bright_dark_2}\\
    &\left(D_{x_{-1}^{(l)}}\left(D_{x_1}^2 - D_{x_2}\right) - 4 (D_{x_1} -a_l )\right)\tau^{(k)}_{\mathbf n}\cdot \tau^{(0)}_{\mathbf n} -4 a_l \tau^{(k)}_{\mathbf n + \mathbf e_l} \tau^{(0)}_{\mathbf n - \mathbf e_l} = 0,\label{KP_bright_dark_3}\\
    &D_{y_1^{(k)}}D_{x_1}\tau^{(0)}_{\mathbf n} \cdot \tau^{(0)}_{\mathbf n}=-2\tau^{(k)}_{\mathbf n} \bar\tau^{(k)}_{\mathbf n}.\label{KP_bright_dark_4}
\end{align}

For \(k\in I_1\), \(l,j \in I_2\), we have the following bilinear equations about \(\tau\)-functions \(\tau^{(0)}_{\mathbf{n} + \mathbf{e}_l}\) and \(\tau^{(0)}_{\mathbf n}\)
\begin{align}
    &\left(D_{x_1}^2-D_{x_2}+2 a_l D_{x_1}\right) \tau^{(0)}_{\mathbf{n} + \mathbf{e}_l} \cdot \tau^{(0)}_{\mathbf{n}}=0, \label{KP_bright_dark_5}\\
    &\left(D_{x_1}^3+3 D_{x_1} D_{x_2}-4 D_{x_3}+3 a_l\left(D_{x_1}^2+D_{x_2}\right)+6 a_l^2 D_{x_1}\right) \tau^{(0)}_{\mathbf{n} + \mathbf{e}_l} \cdot \tau^{(0)}_{\mathbf{n}}=0, \label{KP_bright_dark_6}\\
    &\left(D_{x_{-1}^{(j)}}\left(D_{x_1}^2-D_{x_2}+2 a_l D_{x_1}\right)-4\left(D_{x_1}+a_l-a_j\right)\right) \tau^{(0)}_{\mathbf{n} + \mathbf{e}_l} \cdot \tau^{(0)}_{\mathbf{n}}+ 4(a_l-a_j) \tau^{(0)}_{\mathbf{n} + \mathbf{e}_l + \mathbf{e}_j} \cdot \tau^{(0)}_{\mathbf{n} - \mathbf{e}_j} = 0,\label{KP_bright_dark_7}\\
    &\left(D_{y_1^{(k)}}\left(D_{x_1}^2 - D_{x_2} + 2a_l D_{x_1}\right)\right) \tau_{\mathbf n + \mathbf e_l}^{(0)} \cdot \tau_{\mathbf n}^{(0)} +4a_l \tau_{\mathbf n + \mathbf e_l}^{(k)} \bar{\tau}_{\mathbf n}^{(k)}= 0, \label{KP_bright_dark_8}\\
    &\left(D_{x_{-1}^{(l)}} D_{x_1}-2\right) \tau^{(0)}_{\mathbf{n}} \cdot \tau^{(0)}_{\mathbf{n}}=-2 \tau^{(0)}_{\mathbf{n} + \mathbf{e}_l} \tau^{(0)}_{\mathbf{n} - \mathbf{e}_l}. \label{KP_bright_dark_9}
\end{align}

For \(k,i\in I_1\), \(l,j \in I_2\), we have the following bilinear equations
\begin{align}
    &D_{x_1}\tau^{(k)}_{\mathbf n}\cdot \tau^{(i)}_{\mathbf n}=\tau^{(k,i)}_{\mathbf n} \tau^{(0)}_{\mathbf n}, \label{KP_bright_dark_10}\\
    &\left(D_{x_1}+a_l\right)\tau^{(0)}_{\mathbf n + \mathbf e_l}\cdot \tau^{(k)}_{\mathbf n} = a_l\tau^{(k)}_{\mathbf n + \mathbf e_l} \tau^{(0)}_{\mathbf n},\label{KP_bright_dark_11}\\
    &\left(D_{x_1}+a_l-a_j\right)\tau^{(0)}_{\mathbf{n} + \mathbf{e}_l} \cdot \tau^{(0)}_{\mathbf{n} + \mathbf{e}_j}=(a_l-a_j) \tau^{(0)}_{\mathbf{n} + \mathbf{e}_l + \mathbf{e}_j}  \tau^{(0)}_{\mathbf{n}}. \label{KP_bright_dark_12}
\end{align}

Above bilinear equations \eqref{KP_bright_dark_1}-\eqref{KP_bright_dark_12} are satisfied by the following \(\tau\)-functions
\begin{align}\label{tau_bright_dark}
    \begin{split}
        &\tau_{\mathbf n}^{(0)} = \left|M_{\mathbf n}\right|,\\
        &\tau_{\mathbf n}^{(k)} = \begin{vmatrix}
            M_{\mathbf n} & \Phi_{\mathbf n} \\
            -\left(\bar{\Psi}^{(k)}\right)^T & 0
        \end{vmatrix},
        \quad
        \bar{\tau}_{\mathbf n}^{(k)} = \begin{vmatrix}
            M_{\mathbf n} & \Psi^{(k)} \\
            -\left(\bar{\Phi}_{\mathbf n}\right)^T & 0
        \end{vmatrix},
        \\
        &\tau_{\mathbf n}^{(k,i)} = \begin{vmatrix}
            M_{\mathbf n} & \Phi_{\mathbf n} & \partial_{x_1} \Phi_{\mathbf n} \\
            -\left(\bar{\Psi}^{(i)}\right)^T & 0 & 0 \\
            -\left(\bar{\Psi}^{(k)}\right)^T & 0 & 0
        \end{vmatrix}
    \end{split}
\end{align}
where \(M_{\mathbf n}\) is a \(N \times N\) matrix, \(\Phi_{\mathbf n}\), \(\bar{\Phi}_{\mathbf n}\), \(\Psi^{(k)}\), \(\bar{\Psi}^{(k)}\) are \(N\)-component vectors whose elements are defined as
\begin{align}
    &m_{ij}^{\mathbf n}=\frac{e^{\xi_i+\bar{\xi}_j}}{p_i+\bar{p}_j}\prod_{n=1}^{M-m}\left(-\frac{p_i - a_n}{\bar{p}_j + a_n}\right)^{k_n}  + \sum_{n=1}^m \frac{\tilde{C}_i^{(n)} \bar{C}_j^{(n)}}{q_i^{(n)}+\bar{q}_j^{(n)}}e^{\eta_i^{(n)}+\bar{\eta}_j^{(n)}},\\
    &\Phi_{\mathbf n} = \left(e^{\xi _{1}}\prod_{n=1}^{M-m} \left(1-\frac{p_1}{a_n}\right)^{k_n}, e^{\xi_{2}}\prod_{n=1}^{M-m} \left(1-\frac{p_2}{a_n}\right)^{k_n},\ldots, e^{\xi _{N}}\prod_{n=1}^{M-m} \left(1-\frac{p_N}{a_n}\right)^{k_n}\right)^T, \\
    &\bar{\Phi}_{\mathbf n} = \left(e^{\bar\xi _{1}}\prod_{n=1}^{M-m} \left(1+\frac{\bar p_1}{a_n}\right)^{k_n}, e^{\bar\xi _{2}}\prod_{n=1}^{M-m} \left(1+\frac{\bar p_2}{a_n}\right)^{k_n},\ldots , e^{\bar\xi _{N}}\prod_{n=1}^{M-m} \left(1+\frac{\bar p_N}{a_n}\right)^{k_n}\right)^T, \quad\\
    &\Psi^{(k)} =\left(\tilde{C}_1^{(k)} e^{\eta _{1}^{(k)}}, \tilde{C}_2^{(k)} e^{\eta _2^{(k)}},\ldots, \tilde{C}_N^{(k)} e^{\eta _{N}^{(k)}}\right)^T,\\
    &\bar{\Psi}^{(k)}=\left(\bar{C}_1^{(k)} e^{\bar{\eta}_{1}^{(k)}}, \bar{C}_2^{(k)} e^{\bar{\eta}_2^{(k)}},\ldots, \bar{C}_N^{(k)} e^{\bar{\eta}_{N}^{(k)}}\right)^T,\\
    &\xi_i=p_i x_1 + p_i^2 x_2 + p_i^3 x_3 + \sum_{n=1}^{M-m}\frac{1}{p_i - a_n} x_{-1}^{(n)} +\xi_{i0},\\
    &\bar{\xi}_i=\bar{p}_i x_1 - \bar{p}_i^2 x_2 + \bar{p}_i^3 x_3 + \sum_{n=1}^{M-m}\frac{1}{\bar p_i + a_n} x_{-1}^{(n)} +\bar{\xi}_{i0},\\
    &\eta_i^{(k)}=q_i^{(k)} y_1^{(k)},\quad \bar{\eta}_i^{(k)}=\bar{q}_i^{(k)} y_1^{(k)}.
\end{align}
\end{lemma}
\end{document}